\documentclass{jfm}

\usepackage{amsmath,amssymb}
\usepackage{mathrsfs}
\usepackage{xcolor}
\usepackage{subfigure}
\usepackage{bbm}
\usepackage{bm}
\usepackage{booktabs}
\usepackage{tabularx}

\usepackage[colorlinks=true, linkcolor=blue, urlcolor=blue, citecolor=blue]{hyperref}
\usepackage{natbib}

\begin{document}

\newtheorem{lemma}{Lemma}
\newtheorem{corollary}{Corollary}

\shorttitle{Higher-order moment theories for granular gases} 
\shortauthor{V. K. Gupta et al.} 

\title{Higher-order moment theories for dilute granular gases of smooth hard-spheres}

\author
 {
 {Vinay Kumar Gupta\aff{1},
 \corresp{
 \email{vinay.libra.gupta@gmail.com 
 }
 }}
 Priyanka Shukla\aff{2} 
 \and 
 Manuel Torrilhon\aff{3}
 }

\affiliation
{
\aff{1}
SRM Research Institute and Department of Mathematics, 
\\
SRM University, Chennai 603203, India
\aff{2}
Department of Mathematics,
Indian Institute of Technology Madras, Chennai 600036, India
\aff{3}
 Center for Computational Engineering Science, 
Department of Mathematics, 
\\
RWTH Aachen University, Schinkelstr.~2, D-52062 Aachen, Germany
}

\maketitle

\begin{abstract}
Grad's method of moments is employed to develop higher-order Grad moment equations---up to first 26-moments---for dilute granular gases within the framework of the (inelastic) Boltzmann equation. The homogeneous cooling state of a freely cooling granular gas is investigated with the Grad 26-moment equations in a semi-linearized setting and it is shown that the granular temperature in the homogeneous cooling state still decays according to Haff's law while the other higher-order moments decay on a faster time scale. The nonlinear terms of fully contracted fourth moment are also considered and, by exploiting the stability analysis of fixed points, it is shown that these nonlinear terms have negligible effect on Haff's law. Furthermore, an even larger Grad moment system which includes the fully contracted sixth moment is also  scrutinized and the stability analysis of fixed points is again exploited to conclude that even the inclusion of scalar sixth order moment into the Grad moment system has negligible effect on Haff's law. The constitutive relations for the stress and heat flux (i.e., the Navier--Stokes and Fourier relations) are derived through the Grad 26-moment equations and compared with those obtained via CE expansion and via computer simulations. The linear stability of the homogeneous cooling state is analyzed through the Grad 26-moment system and various sub-systems by decomposing them into longitudinal and transverse systems. It is found that one eigenmode in both longitudinal and transverse systems in the case of inelastic gases is unstable. By comparing the eigenmodes from various theories, it is established that the 13-moment eigenmode theory predicts that the unstable heat mode of the longitudinal system remains unstable for all wavenumbers below a certain coefficient of restitution while any other higher-order moment theory shows that this mode becomes stable above some critical wavenumber for all values of the coefficient of restitution. In particular, the Grad 26-moment theory leads to a smooth profile for the critical wavenumber in contrast to the other considered theories. Furthermore, the critical system size obtained through the Grad 26-moment and existing theories are also in excellent agreement.
%
\end{abstract}
\section{Introduction}
A conglomeration of discrete macroscopic particles characterized by dissipative collisions is termed as a granular material. Granular materials are prevalent in various industries---for instance, in chemical, agriculture and food industries---as well as in nature---for instance, in asteroid belt, sand dunes, debris, etc. Under substantially strong driving forces (e.g., vibration, shearing, etc.), granular materials are in rapid flow regime, in which they exhibit fluid-like behavior and often referred to as \emph{granular gases} \citep{Campbell1990,Goldhirsch2003}. In the rapid flow regime, the particles of a granular material move randomly and, similar to a monatomic gas, it can be assumed that the collisions among them are binary and instantaneous. However, in contrast to a monatomic gas, the collisions between any two particles of a granular material are inherently inelastic, and thereby energy is dissipated during each collision. The inelastic collisions in granular materials lead to many interesting phenomena---for instance, standing wave patterns \citep{MUS1995,UMS1996}, clustering 
\citep{KWG1997}, fingering \citep{PDS1997},
mixing and segregation \citep{OK2000, Mullin2000, BEKR2003}, shear banding \citep{MDKENJ2000, SA2009},
jamming \citep{CJN2005},
density waves \citep{LSG2002, ACG2009}, 
etc. Nevertheless, the mechanics of granular material is still not well-understood, although there have been significant developments in the last couple of decades. 

The analogy between granular and molecular fluids has motivated several researchers to devise theoretical methods---for studying granular fluids---based on kinetic theory within the framework of the Boltzmann equation, see e.g., \cite{JR1985,JR1985PoF,GS1995,BMD1996, SGJFM1998,BDKS1998,GS2003,BP2004, BST2004,KSG2014}. 
Kinetic theory for granular flows was first introduced in the two seminal papers by \cite{JS1983} and \cite{LSJC1984} where the former deals with the theory for nearly elastic granular flows while the latter with arbitrary inelasticity.
The reported works on the kinetic theory for granular fluids also attempted to extend the two well-known approximation methods in kinetic theory for monatomic gases, namely the Chapman--Enskog (CE) expansion \citep{CC1970} and Grad's method of moments \citep{Grad1949}. 
Despite the fact that the higher-order approximations (Burnett and beyond) resulting from the CE expansion lead to unstable equations while the Grad's method of moments always yields linearly stable sets of equations in the case of monatomic gases \citep{Bobylev1982}, the former has extensively been studied even for granular gases. 
\citet{GS1995} applied the CE expansion to obtain Euler-like hydrodynamic equations for rough granular flows. 
The papers by \cite{BDKS1998} and \cite{GD1999} apply the CE expansion to dilute and dense granular gases, respectively, to derive the first order (Navier--Stokes and Fourier) constitutive relations for the system of mass, momentum and energy balance equations. Both the works assume that the space and time dependence of the distribution function can be expressed completely in terms of the hydrodynamic fields and represent the distribution function in a formal series of a uniformity parameter which measures the strength of spatial gradients of the hydrodynamic fields. 
\citet{SGN1996} performed a generalized expansion on the distribution function in powers of two small parameters, namely the Knudsen number and the degree of inelasticity $\epsilon=1-e^2$ with $e$ being the coefficient of (normal) restitution and determined the constitutive relations up to  Burnett order for the system of mass, momentum and energy balance equations specialized to a simply sheared two-dimensional hard-sphere granular gas via the CE expansion. Subsequently, with the same method, \citet{SGJFM1998} \citep[see also][]{Gupta2011} determined the constitutive relations up to Burnett order for a smooth hard-sphere granular gas in three dimensions. However, both these works restrict the coefficient of restitution for being very close to $1$ in order to get the zeroth order solution in the expansion as Maxwellian, similar to a monatomic gas.
\citet{Lutsko2005} employed the CE expansion to obtain the constitutive relations for dense granular gases with arbitrary energy loss models. His model produced the homogeneous cooling state (HCS) solution at the zeroth order in the expansion while Navier--Stokes and Fourier constitutive relations at first order. Following the approach of expanding the distribution function in powers of uniformity parameter, \citet{KGS2014} recently obtained the Burnett order constitutive relations for a smooth granular gas described with (inelastic) Maxwell interaction potential via the CE expansion. It is worth to point out that, unlike \cite{SGJFM1998}, their approach does not restrict the coefficient of restitution for being close to $1$. Recently, \citet{KSG2014} employed the CE expansion to obtain the transport coefficients up to Navier--Stokes order for a granular gas of rough hard-spheres. 

Despite the success of the Grad's method of moments over the CE expansion in the case of monatomic gases \citep[see e.g.,][and references therein]{TorrilhonARFM}, the former in the case of granular flows has received much less attention than the latter which has been investigated extensively. Nevertheless, over the last few years, researchers have shown interest in exploring the former as well \cite[see e.g.,][]{KM2011, Garzo2013,SahaAlam2014}.
%
The pioneering work on Grad's method of moments for granular flows is due to \citeauthor{JR1985} who extended the method to granular flows and derived the Grad 13-moment (G13) equations for a dense and smooth granular gas in \cite{JR1985} and, subsequently, the Grad 16-moment equations for a dense and rough two-dimensional granular gas in \cite{JR1985PoF}---the extra field variables in the latter are due to rotational motion of rough particles on collision. 
\cite{BST2004} further extended the Grad's method of moments to weakly inelastic granular flows with variable coefficient of restitution, in one-dimension though. As it is well-established that the scalar fourth moment is necessary for the proper description of a granular gas, \citet{RC2002} included the scalar fourth moment and---for a granular gas---obtained Grad moment equations for $9$ hydrodynamic fields in two dimensions (corresponding to $14$ field variables in three dimensions), and applied them to investigate two problems, namely the HCS and a granular system steadily heated by two parallel walls. \citet{KM2011} presented the Grad 14-moment (G14) theory for dilute granular gases and by employing this theory, they investigated the HCS, found the constitutive relations for five moment equations, and performed eigenmode analysis on thirteen field theory, where they assumed a constant value for the scalar fourth moment. \citet{Garzo2013} demonstrated the G14 theory for moderately dense granular flows, although the work aimed at finding the Navier--Stokes level constitutive relations through the G14 equations. Recently, \citet{SahaAlam2014} developed the Grad's method of moments based on anisotropic Gaussian and employed it to investigate the non-Newtonian stress, collisional dissipation and heat flux in a sheared two-dimensional granular flow.

The main reasons---among others---that the Grad's method of moments for granular flows is receiving less attention could be that $(i)$ the production terms, which emanate through the Boltzmann collision integral, in the moment equations are very difficult to evaluate for a general interaction potential, $(ii)$ how many and which moments one should consider for describing a process is not known \emph{a priori}, and $(iii)$ the boundary conditions associated with the Grad moment equations are unclear, which is apparently also the case for higher-order equations resulting from the CE expansion. Over the last couple of years, the two of the present authors have developed a computational methodology, which can compute the production terms associated with the Grad moment equations for $(i)$ a monatomic gas, $(ii)$ a mixture of monatomic gases, and $(iii)$ a dilute granular gas, all interacting with a general interaction potential \citep{GT2012,GT2015CAMWA,Gupta2015}. The \emph{order of magnitude} method developed by \citet{Struchtrup2004,Struchtrup2005}---which regularizes the original G13 equations---identifies the required moments for describing a process in rarefied monatomic gases in a systematic way, although the method requires the Grad 26-moment (G26) equations which at present are not available for granular gases. 
The boundary conditions associated with the Grad moment equations for monatomic gases are typically obtained using the Maxwell accommodation model \citep{Maxwell1879}; however, those for granular gases are yet to be explored.

Within moment approximations the number of moments is crucial and has a strong influence on the results, especially when considering only few moments. It is known form studies in monatomic gases that, while moment approximations do converge to the solution of the Boltzmann equation for large number of moments, when using small systems oscillatory convergence pattern makes it difficult to judge the approximation quality, see \cite{Torrilhon2015}. 
Consequently, when using moment approximations, one should always compare a set of successive approximations with different number of moments. The result of a 14-moment theory alone is not conclusive and must be cross-checked with larger systems. A similar argument applies to using linearized models, where it is important to check the possible influence of non-linearity even if the outcome confirms its negligibility. 
Another important reason to consider moment equations beyond 14 moments is the long-term perspective of establishing a complete set of predictive moment equations for granular flows. For monatomic gases, the regularized 13-moment (R13) equations provide such a model. The R13 system is derived from the G26 model and only on that level relevant phenomena like Knudsen layers or non-gradient transport can be predicted. 

In this paper, we derive the G26 equations for dilute granular gases through the Grad's method of moments, although we neither consider the regularization of the Grad moment equations nor the required boundary conditions---these problems will be considered elsewhere in future. The fully nonlinear production terms associated with an even larger Grad moment system---which includes the fully contracted sixth moment as a field variable along with the $26$ moments and is referred to as the system of Grad 27-moment (G27) equations here---are computed with computer algebra software \textsc{Mathematica}\textsuperscript{\textregistered} and presented in appendix~\ref{App:ProdTerms} for hard-sphere interaction potential. The production terms for the G26 equations can readily be found from those for the G27 equations by simply discarding the terms containing the  fully contracted sixth moment. Here, it is noteworthy to point out that the present work does not have any restriction on the coefficient of restitution other than being a constant and is, consequently, expected to be applicable not only for nearly elastic granular gases but also for those having large inelasticity. With the semi-linearized G26 and G27 equations, we investigate the HCS---which has also been studied a lot theoretically, numerically as well as experimentally---of a freely cooling granular gas to show that the decay of granular temperature closely follows Haff's law \citep{Haff1983} while the other higher-order moments relax on a faster time scale than the granular temperature. We further study the effects of nonlinear terms of scalar fourth moment and that of linear terms of scalar sixth moment on Haff's law. Following the approach of \cite{Garzo2013}, we determine the constitutive relations for the stress and heat flux for dilute granular gases through the G26 equations and compare them with those obtained via CE expansion and via computer simulations. We further investigate the linear stability of HCS by scrutinizing the eigenmodes of longitudinal and transverse problems associated with the G26 system and other sub-systems. Similar problems related to the stability of eigenmodes have been investigated earlier by \citet{BMD1996} for five moment theory and by \citet{KM2011} for thirteen moment theory. We find that one eigenmode from each Grad moment theory considered in this paper is unstable for inelastic gases while others are stable, while all the eigenmodes are stable for monatomic (elastic) gases. Moreover, the thirteen moment theory of \cite{KM2011} shows that the unstable heat mode of the longitudinal system remains unstable below a certain value of the coefficient of restitution for all wavenumbers while the present work reveals that the unstable heat mode of the longitudinal system associated with all moment theories expect for the G13 theory becomes stable above a critical wavenumber for all values of the coefficient of restitution. 
By investigating the critical wavenumbers of the longitudinal and transverse systems from the G26 theory, we also study the critical system size, beyond which the system becomes unstable, and compare the results with the existing theories and simulations.
The findings of the paper will be useful in better understanding of granular gases, in developing new mathematical models---such as regularized moment equations---and boundary conditions, and in capturing some intriguing features of granular gases theoretically.
  
%
%
%
%
%
%
%

The rest of the paper is structured as follows. 
Kinetic theory for granular gases is briefly reviewed in \S\,\ref{Sec:reviewKT}. 
Grad's method of moments is outlined and applied to derive the Grad moment systems of various orders---in particular, the system of G26 equations---for granular gases in \S\,\ref{Sec:GradMethod}.
The HCS of a freely cooling granular gas is investigated through the G26 and G27 equations in \S\,\ref{Sec:HCS} in order to discern the effects of nonlinear terms of scalar fourth moment and of linear terms of scalar sixth moment on HCS. The constitutive relations for the stress and heat flux are computed through the G26 equations in \S\,\ref{Sec:NSF_rels}. The linear stability of HCS is analyzed in \S\,\ref{Sec:Stability} through the G26 equations by decomposing them into longitudinal and transverse systems. The conclusions of the paper are given in  \S\,\ref{Sec:Conclusion}. 
\section{Short review of kinetic theory} \label{Sec:reviewKT}
We consider a dilute granular gas composed of smooth-inelastic-identical hard-spheres of mass $m$ and diameter $d$. The binary collision between two such spheres having pre-collisional velocities $\bm{c}$ and $\bm{c}_1$ leads to the following velocity transformation after the collision \citep[see e.g.,][]{BP2004,RaoNott2008}:
\begin{align}
\label{VelTran}
\left.
\begin{aligned}
\bm{c}^{\prime}&=\bm{c}-\frac{1+e}{2}(\hat{\bm{k}}\cdot\bm{g})\hat{\bm{k}},
\\
\bm{c}_1^{\prime}&=\bm{c}_1+\frac{1+e}{2}(\hat{\bm{k}}\cdot\bm{g})\hat{\bm{k}},
\end{aligned}
\right\}
\end{align} 
where $\bm{c}^{\prime}$ and $\bm{c}_1^{\prime}$ are the post-collisional velocities of the respective spheres, $\bm{g}=\bm{c}-\bm{c}_1$ is the relative velocity, $\hat{\bm{k}}$ is the unit vector directed from the center of one sphere to that of other at the time of collision, and $e$ is the coefficient of normal restitution (also, referred to as the coefficient of restitution). The coefficient of restitution $e$, in principle, is not a constant and usually depends on impact velocity \citep{BSSS1999,BP2004}. Nevertheless, for simplicity, we assume that $e$ is constant with $0\leq e \leq 1$ in this work. The two limiting cases of $e=0$ and $e=1$ correspond to \emph{sticky} and \emph{perfectly elastic} collisions, respectively.

In kinetic theory, the state of a dilute granular gas can be described with single-particle velocity distribution function $f\equiv f(t, \bm{x}, \bm{c})$ which obeys the (inelastic) Boltzmann equation \citep{GS1995,SGJFM1998,BP2004}
\begin{equation}
\label{BE}
\frac{\partial f}{\partial t}+c_i\,\frac{\partial f}{\partial x_i}+F_i\,\frac{\partial f}{\partial c_i}=d^2\int\limits_{\mathbb{R}^3} \!\int\limits_{S^2} \!\left(\frac{1}{e^2}f^{\prime\prime} f_1^{\prime\prime}
-f  f_1\right)(\hat{\bm{k}}\cdot \bm{g})\,\Theta(\hat{\bm{k}}\cdot \bm{g})
\,\mathrm{d}\hat{\bm{k}}\,\mathrm{d}\bm{c}_1,
\end{equation} 
where $\Theta$ is the Heaviside step function, $\bm{F}$ is the external force per unit mass and usually do not depend on $\bm{c}$, $f_1\equiv f(t,\bm{x},\bm{c}_1)$, $f^{\prime\prime}\equiv f(t, \bm{x},\bm{c}^{\prime\prime})$, $f_1^{\prime\prime}\equiv f(t, \bm{x},\bm{c}_1^{\prime\prime})$, the integration limits of $\hat{\bm{k}}$ extend over the unit sphere $S^2$ and Einstein summation convention is assumed over repeated indices. The velocities $\bm{c}^{\prime\prime}$ and $\bm{c}_1^{\prime\prime}$ are the pre-collisional velocities in an inverse collision and are related to the $\bm{c}$ and $\bm{c}_1$ as follows \citep[see e.g.,][]{BP2004,RaoNott2008}.
\begin{align}
\label{VelTranInverse}
\left.
\begin{aligned}
\bm{c}^{\prime\prime}&=\bm{c}-\frac{1+e}{2e}(\hat{\bm{k}}\cdot\bm{g})\hat{\bm{k}},
\\
\bm{c}_1^{\prime\prime}&=\bm{c}_1+\frac{1+e}{2e}(\hat{\bm{k}}\cdot\bm{g})\hat{\bm{k}}.
\end{aligned}
\right\}
\end{align} 
Owing to brevity, hereafter, we shall omit the limits of integrations. Nevertheless, the integration over any velocity space will stand for the volume integral over $\mathbb{R}^3$ and that over $\hat{\bm{k}}$ will stand for the volume integral over the unit sphere $S^2$. The right-hand side (RHS) of \eqref{BE} is referred to as the (inelastic) Boltzmann collision operator.

The hydrodynamic variables---number density $n\equiv n(t, \bm{x})$, macroscopic velocity $\bm{v}\equiv \bm{v}(t, \bm{x})$, \emph{granular} temperature $T\equiv T(t, \bm{x})$---are directly related to the velocity distribution function as follows.
\begingroup
\allowdisplaybreaks
\begin{align}
n(t, \bm{x})&=\int\! f(t, \bm{x}, \bm{c})\,\mathrm{d}\bm{c},
\\
n(t, \bm{x})\, \bm{v}(t, \bm{x})&=\int\! \bm{c}\,f(t, \bm{x}, \bm{c})\,\mathrm{d}\bm{c},
\\
\label{granTemp}
\frac{3}{2}n(t, \bm{x})\, T(t, \bm{x})&=\frac{1}{2}m\int\! C^2\,f(t, \bm{x}, \bm{c})\,\mathrm{d}\bm{c},
\end{align}
\endgroup
where $\bm{C}(t, \bm{x}, \bm{c})=\bm{c}-\bm{v}(t, \bm{x})$ is the peculiar velocity. Here, the definition of the granular temperature---which is a measure of fluctuating kinetic energy---is adopted following the references \cite{BP2004, Garzo2013}, although some authors---for instance, \cite{LSJC1984,Campbell1990,SahaAlam2014}---also refer $\theta = T/m$ as granular temperature (defined as fluctuating kinetic energy per unit mass). It should be noted that in the case of monatomic gases, $T$ in \eqref{granTemp} is replaced with $k_B T_{\textrm{th}}$, where $k_B$ denotes the Boltzmann constant and $T_{\textrm{th}}$ is the thermodynamic temperature.

The governing equations for the hydrodynamic variables can be derived from the Boltzmann equation \eqref{BE} as follows. Let us first introduce the moments. For a particle property, $\psi\equiv\psi(t, \bm{x}, \bm{c})$, its average value $\langle \psi \rangle\equiv \langle \psi \rangle(t, \bm{x})$ is defined in terms of the distribution function $f$ as
\begin{align}
\label{moment}
\langle \psi\rangle (t, \bm{x}) = \frac{1}{n(t, \bm{x})} \int\!\psi(t, \bm{x}, \bm{c})\, f(t, \bm{x}, \bm{c})\,\mathrm{d}\bm{c}.
\end{align} 
The integral $\int\!\psi f\,\mathrm{d}\bm{c}$ on the RHS of \eqref{moment} is referred to as the \emph{moment} of the velocity distribution function with respect to $\psi$. Clearly,
\begin{align}
1=\langle 1\rangle,\quad \bm{v}=\langle \bm{c}\rangle\quad\textrm{and}\quad T=\left\langle \frac{1}{3} m C^2\right\rangle.
\end{align}
The governing equation for the moment $\int\!\psi f\,\mathrm{d}\bm{c}$---often, referred to as the moment equation for $n\langle \psi \rangle$ or the transfer equation for property $\psi$---is obtained by multiplying the Boltzmann equation \eqref{BE} with $\psi$ and integrating over the velocity space $\bm{c}$. The transfer equation for the quantity $\psi$ reads
\begin{align}
\label{TransferEqn}
\frac{\mathrm{D}}{\mathrm{D} t}\!\int\!\psi f\,\mathrm{d}\bm{c}+\underline{\frac{\partial}{\partial x_i}\!\int\!\psi\, C_i f\,\mathrm{d}\bm{c}}+\frac{\partial v_i}{\partial x_i}\!\int\!\psi f\,\mathrm{d}\bm{c} -\int\!\left(\frac{\mathrm{D}\psi}{\mathrm{D}t} +C_i\frac{\partial \psi}{\partial x_i}+F_i\frac{\partial \psi}{\partial c_i}\right) f\,\mathrm{d}\bm{c}
=\mathcal{P}(\psi),
\end{align} 
where $\frac{\mathrm{D}}{\mathrm{D} t}\equiv\frac{\partial}{\partial t}+\bm{v}\cdot\bm{\nabla}$ is the material derivative, the underline denotes the flux term, and 
\begin{align}
\label{prodtermpsi}
\mathcal{P}(\psi)&=\frac{d^2}{2}\iiint\!\big(\psi^{\prime}+\psi_1^{\prime}-\psi-\psi_1\big)f  f_1(\hat{\bm{k}}\cdot\bm{g})\,\Theta(\hat{\bm{k}}\cdot\bm{g})
\,\mathrm{d}\hat{\bm{k}}\,\mathrm{d}\bm{c}\,\mathrm{d}\bm{c}_1
\nonumber\\
&=d^2\iiint\!\big(\psi^{\prime}-\psi\big)f  f_1(\hat{\bm{k}}\cdot\bm{g})\,\Theta(\hat{\bm{k}}\cdot\bm{g})
\,\mathrm{d}\hat{\bm{k}}\,\mathrm{d}\bm{c}\,\mathrm{d}\bm{c}_1,
\end{align}
with $\psi^{\prime}\equiv \psi(t,\bm{x},\bm{c}^{\prime})$ etc.,
is the rate of change of $\langle \psi \rangle$ per unit volume due to collisions and referred to as the \emph{production term} or the (inelastic) \emph{Boltzmann collision integral} corresponding to moment $\int\!\psi f\,\mathrm{d}\bm{c}$. While writing \eqref{prodtermpsi}, the symmetry properties of the Boltzmann collision operator \citep{BP2004,RaoNott2008} have been employed. 
It should be noted from the transfer equation \eqref{TransferEqn} that irrespective of the value of $\psi$ chosen for defining a moment by \eqref{moment}, the transfer equation \eqref{TransferEqn} will always contain an additional moment of one more order in the flux term. On substituting $\psi$ in \eqref{TransferEqn} with $1$, $c_i$ and $\frac{1}{3}m\, C^2$ successively, we get the mass, momentum and energy balance equations, respectively, which read
\begin{align}
\label{massBal}
\frac{\mathrm{D} n}{\mathrm{D}t}+ n \frac{\partial v_i}{\partial x_i}=0,
\end{align}
%
\begin{align}
\label{momBal}
\frac{\mathrm{D} v_i} {\mathrm{D}t} + \frac{1}{m\,n} \left[\frac{\partial \sigma_{ij}}{\partial x_j} + \frac{\partial (n\, T)}{\partial x_i}\right] - F_i=0,
\end{align}
%
\begin{align}
\label{energyBal}
\frac{\mathrm{D} T} {\mathrm{D}t} + \frac{2}{3\,n} \left[ \frac{\partial q_i}{\partial x_i} +\sigma_{ij}\frac{\partial v_i}{\partial x_j}+n \,T\frac{\partial v_i}{\partial x_i}\right] = - \zeta\,T.
\end{align}
In \eqref{momBal} and \eqref{energyBal}, $\sigma_{ij}\equiv \sigma_{ij}(t, \bm{x})$ and $q_i\equiv q_i(t, \bm{x})$ are the component of stress tensor and heat flux, respectively, and---following the notations of \cite{Struchtrup2005}---these are given by
\begin{align}
\label{stressHFdef}
\sigma_{ij}=m\int\!\, C_{\langle i}C_{j\rangle} f\,\mathrm{d}\bm{c} \quad\textrm{and}\quad q_i=\frac{1}{2}m\int\!\, C^2 C_i f\,\mathrm{d}\bm{c},
\end{align}
where the angle brackets around the indices denote the symmetric and traceless part of the tensor \citep{Struchtrup2005}, and $\zeta$ in \eqref{energyBal} is the cooling rate due to inelastic collisions:
\begin{align}
\label{CoolingRate}
\zeta=-\frac{m\,d^2}{6\,n\, T}
\iiint\!\Big((C^{\prime})^2+(C_1^{\prime})^2-C^2-C_1^2\Big)f  f_1(\hat{\bm{k}}\cdot\bm{g})\,\Theta(\hat{\bm{k}}\cdot\bm{g})
\,\mathrm{d}\hat{\bm{k}}\,\mathrm{d}\bm{c}\,\mathrm{d}\bm{c}_1.
\end{align}
Note that the RHSs of \eqref{massBal} and \eqref{momBal} vanish due to conservation of mass and momentum; however, the energy is not conserved due to dissipative collisions resulting into non vanishing RHS in \eqref{energyBal}. It is worth pointing out that the stress and heat flux appearing in \eqref{momBal} and \eqref{energyBal}---in general---have non-vanishing collisional contributions in addition to the usual kinetic contribution given by \eqref{stressHFdef} \citep{BDS1997,GD1999,RaoNott2008,Garzo2013}. However, for dilute granular gases, the kinetic contributions to the stress and heat flux dominate over their respective collisional transfer contributions, and therefore the collisional transfer contributions to the stress and heat flux can be neglected for dilute granular gases \citep{BP2004}. In any case, since \eqref{momBal} and \eqref{energyBal} are obtained here by taking the velocity moments of the distribution function---which is essentially the idea of Grad's method of moments too, only kinetic 
contributions to the stress and heat flux will emerge in the equations. The collisional contributions to stress and heat flux, in the case of dense granular gases, must be computed separately \citep{Garzo2013}.  


Obviously, the system of equations \eqref{massBal}--\eqref{energyBal} in hydrodynamic variable $n$, $v_i$ and $T$ is not closed, since it contains the unknowns $\sigma_{ij}$, $q_i$ and $\zeta$, and in order to deal with this system further one must close it. Here, we employ Grad's method of moments \citep{Grad1949} in order to obtain a closed system of equations.
%
\section{Grad's method of moments} \label{Sec:GradMethod}
The function $\psi$ in \eqref{moment} can be chosen in infinitely many ways. Therefore, the transfer equation \eqref{TransferEqn} leads to an infinite hierarchy of moment equations. However, in practice, only a finite number of moment equations---obtained by truncating the infinite hierarchy of moment equations at a certain level---are used. Nonetheless, this truncated system of equation is not closed due to the (underlined) flux term in \eqref{TransferEqn}. In order to obtain a closed (finite) system of moment equations, \citet{Grad1949} expanded the velocity distribution function $f$ in a finite linear combination of the $N$-dimensional Hermite polynomials \citep{Grad1949a} in peculiar velocity and computed the unknown coefficients in the expansion in terms of the considered moments by satisfying their definitions with the approximated distribution function. This method of obtaining a closed set of moment equations is referred to as \emph{Grad's method of moments} and its details can be found in \cite{
Grad1949} and in many standard textbooks, see e.g., \cite{Struchtrup2005,Kremer2010}.

On including the governing equations for stress ($\sigma_{ij}$) and heat flux ($q_i$) into the system of mass, momentum and energy balance equations \eqref{massBal}--\eqref{energyBal}, one obtains the well-known 13-moment equations (in three dimensions). Here, we want to derive and explore the G26 equations for dilute granular flows. To this end, let us first introduce the general form of a moment. The typical form of $\psi$ is $m\, C^{2a}C_{\langle i_1}C_{i_2}\dots C_{i_n\rangle}$, where $a,n\in\mathbb{N}_0$ and the angle brackets around the indices again denote the symmetric and traceless part of the corresponding quantity \citep{Struchtrup2005}. Thus, a general moment is given by
\begin{align}
\label{genmoment}
u_{i_1 i_2 \dots i_n}^{a}=m\int\!C^{2a}C_{\langle i_1}C_{i_2}\dots C_{i_n\rangle}\, f\,\mathrm{d}\bm{c}, \quad a,n\in\mathbb{N}_0.
\end{align} 
Clearly,
\begin{align}
u^{0}=m\, n=\rho,\quad u_i^{0}=0,\quad u^{1}=3\, n\,T = 3\, \rho \, \theta,\quad u_{ij}^{0}=\sigma_{ij},\quad u_i^{1}=2\,q_i,
\end{align}
where $\rho=m\, n$ is the mass density and $\theta = T/m$. 
\subsection{26-moment equations}
In addition to the well-known 13 moments, the system of 26-moment equations include full third rank tensor and one- and full-traces of the fourth moment (i.e., the 26-moment equations include the moments $n$, $v_i$, $T$, $\sigma_{ij}$, $q_i$, $u_{ijk}^0$, $u_{ij}^1$, $u^2$). The system of G26 equations is obtained by substituting $\psi$ in \eqref{TransferEqn} with $1$, $c_i$, $\frac{1}{3}m\, C^2$, $m\,C_{\langle i} C_{j\rangle}$, $\frac{1}{2}m\,C^2C_i$, $m\,C_{\langle i} C_j C_{k\rangle}$, $m\,C^2 C_{\langle i} C_{j\rangle}$ and $m\,C^4$ successively. The system of 26-moment equations consists of the  mass, momentum and energy balance equations \eqref{massBal}--\eqref{energyBal} and other higher-order moment equations, which on using the abbreviations
\begin{align}
\label{abbreviations}
m_{ijk}=u_{ijk}^0,\quad R_{ij}=u_{ij}^1-7\theta\sigma_{ij}, \quad w=u^2-15\rho\theta^2,
\end{align}
read
\begin{align}
\label{eqn:stress}
\frac{\mathrm{D}\sigma_{ij}}{\mathrm{D}t}+\frac{\partial m_{ijk}}{\partial x_k}+\frac{4}{5}\frac{\partial q_{\langle i}}{\partial x_{j\rangle}}+\sigma_{ij}\frac{\partial v_k}{\partial x_k}+2\sigma_{k\langle i}\frac{\partial v_{j\rangle}}{\partial x_k}+2\rho\theta\frac{\partial v_{\langle i}}{\partial x_{j\rangle}}=\mathcal{P}_{ij}^0,
\end{align}
\begin{align}
\label{eqn:HF}
\frac{\mathrm{D} q_i}{\mathrm{D} t} + \frac{1}{2} \frac{\partial R_{ij}}{\partial x_j} +\frac{1}{6}\frac{\partial w}{\partial x_i}+\theta\frac{\partial \sigma_{ij}}{\partial x_j}+\frac{5}{2}\sigma_{ij}\frac{\partial \theta}{\partial x_j} +\frac{5}{2}\rho\theta\frac{\partial \theta}{\partial x_i}+ m_{ijk}\frac{\partial v_j}{\partial x_k}
\nonumber\\
- \sigma_{ij}\frac{1}{\rho}\left(\frac{\partial \sigma_{jk}}{\partial x_k}+\theta\frac{\partial \rho}{\partial x_j}\right) 
+\frac{7}{5}q_i\frac{\partial v_j}{\partial x_j}+\frac{7}{5}q_j\frac{\partial v_i}{\partial x_j}+\frac{2}{5}q_j\frac{\partial v_j}{\partial x_i} = \frac{1}{2} \mathcal{P}_i^{1},
\end{align}
\begin{align}
\label{eqn:mijk}
\frac{\mathrm{D}m_{ijk}}{\mathrm{D}t}+\frac{\partial u_{ijkl}^0}{\partial x_l}+\frac{3}{7}\frac{\partial R_{\langle ij}}{\partial x_{k\rangle}}+3\theta\frac{\partial \sigma_{\langle ij}}{\partial x_{k\rangle}}-3\frac{1}{\rho}\sigma_{\langle ij}\left(\frac{\partial \sigma_{k\rangle l}}{\partial x_l}+\theta\frac{\partial \rho}{\partial x_{k\rangle}}\right)
\nonumber\\
+m_{ijk}\frac{\partial v_l}{\partial x_l}+3m_{l\langle ij}\frac{\partial v_{k\rangle}}{\partial x_l}+\frac{12}{5}q_{\langle i}\frac{\partial v_{j}}{\partial x_{k\rangle}}=\mathcal{P}_{ijk}^0,
\end{align}
\begin{align}
\label{eqn:Rij}
&\frac{\mathrm{D} R_{ij}}{\mathrm{D} t} +2 u_{ijkl}^{0}\frac{\partial v_k}{\partial x_l}+\frac{\partial u_{ijk}^{1}}{\partial x_k} +\frac{2}{5}\frac{\partial u_{\langle i}^{2}}{\partial x_{j\rangle}} +R_{ij}\frac{\partial v_k}{\partial x_k}-\frac{28}{5}\theta\frac{\partial q_{\langle i}}{\partial x_{j\rangle}}- \frac{28}{5} q_{\langle i} \frac{\partial \theta}{\partial x_{j\rangle}}
\nonumber\\
&+4\theta \sigma_{k\langle i}\frac{\partial v_k}{\partial x_{j\rangle}}+4\theta\sigma_{k\langle i}\frac{\partial v_{j\rangle}}{\partial x_k}-\frac{8}{3}\theta\sigma_{ij}\frac{\partial v_k}{\partial x_k} -\frac{14}{3}\frac{1}{\rho}\sigma_{ij}\frac{\partial q_k}{\partial x_k}-\frac{14}{3}\frac{1}{\rho}\sigma_{ij}\sigma_{kl} \frac{\partial v_k}{\partial x_l}
\nonumber\\
&-7\theta\frac{\partial m_{ijk}}{\partial x_k}+\frac{6}{7}R_{\langle ij}\frac{\partial v_{k\rangle}}{\partial x_k}+\frac{4}{5}R_{k\langle i}\frac{\partial v_k}{\partial x_{j\rangle}}+2R_{k\langle i}\frac{\partial v_{j\rangle}}{\partial x_k}+\frac{14}{15}w\frac{\partial v_{\langle i}}{\partial x_{j\rangle}}
\nonumber\\
&- 2 m_{ijk}\frac{\partial \theta}{\partial x_k} - 2\frac{1}{\rho}  m_{ijk}\left( \frac{\partial \sigma_{kl}}{\partial x_l}
+ \theta \frac{\partial \rho}{\partial x_k}\right) - \frac{28}{5} \frac{1}{\rho}q_{\langle i} \left(\frac{\partial \sigma_{j\rangle k}}{\partial x_k}+\theta \frac{\partial \rho}{\partial x_{j\rangle}} \right)
\nonumber\\
&=\mathcal{P}_{ij}^{1} -7\theta\mathcal{P}_{ij}^{0}- \frac{14}{3} \frac{1}{\rho}\sigma_{ij} \left( \frac{1}{2} \mathcal{P}^1\right),
\end{align}
\begin{align}
\label{eqn:Delta}
&\frac{\mathrm{D}w}{\mathrm{D}t}+\frac{\partial u_i^2}{\partial x_i}-20\theta\frac{\partial q_i}{\partial x_i}-8q_i\frac{\partial \theta}{\partial x_i}+\frac{7}{3}w\frac{\partial v_i}{\partial x_i}+4R_{ij}\frac{\partial v_i}{\partial x_j}
\nonumber\\
&+8\theta\sigma_{ij}\frac{\partial v_i}{\partial x_j}-8\frac{1}{\rho}q_i\left(\frac{\partial \sigma_{ij}}{\partial x_j}+\theta\frac{\partial \rho}{\partial x_i}\right)
=\mathcal{P}^2 - 20 \theta \left( \frac{1}{2} \mathcal{P}^1\right),
\end{align}
where
\begin{align}
\label{genProdTerm}
\mathcal{P}_{i_1 i_2 \dots i_n}^a=m\, d^2\iiint&\!\Big((C^{\prime})^{2a}C_{\langle i_1}^{\prime}C_{i_2}^{\prime}\dots C_{i_n\rangle}^{\prime}-C^{2a}C_{\langle i_1}C_{i_2}\dots C_{i_n\rangle}\Big)
\nonumber\\
&\times f  f_1(\hat{\bm{k}}\cdot\bm{g})\,\Theta(\hat{\bm{k}}\cdot\bm{g})
\,\mathrm{d}\hat{\bm{k}}\,\mathrm{d}\bm{c}\,\mathrm{d}\bm{c}_1
\end{align}
are the production terms. Notice that $\mathcal{P}^0=\mathcal{P}_i^0=0$ due to mass and momentum conservation, and $\mathcal{P}^1=-3\,n\,T\,\zeta$.

In \eqref{abbreviations}, $w$ has been defined as the difference of the full trace of fourth moment to its value computed with the Maxwellian distribution function
\begin{align}
\label{Maxwellian}
f_M\equiv f_M(t, \bm{x}, \bm{c})=n\left(\frac{1}{2\,\pi\,  \theta}\right)^{\!3/2}\exp{\left(-\frac{C^2}{2\,\theta}\right)},
\end{align}
and $R_{ij}$ is defined as the difference of the one trace of fourth moment to its value computed with the Grad 13-moment (G13) distribution function so that the quantities $m_{ijk}$, $R_{ij}$ and $w$ vanish for the G13 theory. It may be noted that the value of the full trace of fourth moment $u^2$ when computed either with Maxwellian distribution function or with the G13 distribution function is $15 \rho \theta^2$.
\subsection{Grad 26-moment closure}
Clearly, the system of 26-moment equations (eqs.~\eqref{massBal}--\eqref{energyBal} and \eqref{eqn:stress}--\eqref{eqn:Delta}) is not closed, since it contains the unknown higher-order moments $u_{ijkl}^0$, $u_i^2$ and $u_{ijk}^1$, and on top of that the production terms are also not known. The system is closed with the Grad distribution function based on the 26 moments considered, which is referred to as the G26 distribution function and reads
\begin{align}
\label{G26DisFunc}
f_{|\textrm{G26}}
&=f_M\left[1+\frac{1}{2}\frac{\sigma_{ij}}{\rho\,\theta^2}C_iC_j +\frac{1}{5}\frac{q_i}{\rho\,\theta^2}C_i\left(\frac{C^2}{\theta}-5\right)+\frac{1}{6}\frac{ m_{ijk}}{\rho\,\theta^3} C_{i}C_jC_{k} \right.
\nonumber\\
&\left.\hspace{13.7mm}+\frac{1}{28}\frac{R_{ij}}{\rho \,\theta^3} C_{i}C_{j} \left(\frac{C^2}{\theta}-7\right)+\frac{1}{8}\frac{w}{\rho \,\theta^2}\left(1-\frac{2}{3}\frac{C^2}{\theta}+\frac{1}{15}\frac{C^4}{\theta^2}\right)\right].
\end{align}
Insertion of the G26 distribution function \eqref{G26DisFunc} into the definitions of higher-order moments $u_{ijkl}^0$, $u_i^2$ and $u_{ijk}^1$, and into the production terms \eqref{genProdTerm} expresses them in terms of the considered 26 moments. The unknown higher-order moments $u_{ijkl}^0$, $u_i^2$ and $u_{ijk}^1$ turn into
\begin{align}
\label{closure}
u_{ijkl|\textrm{G26}}^0=0,\quad u_{i|\textrm{G26}}^2=28\,\theta\,q_i, \quad\textrm{and}\quad u_{ijk|\textrm{G26}}^1=9\,\theta\, m_{ijk},
\end{align}
where the subscript ``{\scriptsize $|\textrm{G26}$}" just denotes that these moments are evaluated with the G26 distribution function \eqref{G26DisFunc}. Unfortunately, the production terms \eqref{genProdTerm} are still not easy to evaluate by hand. The two authors of the present paper implemented the strategy for computing the production terms into computer algebra software  {\textsc{Mathematica}}\textsuperscript{\textregistered} and obtained the fully nonlinear production terms associated with the G26 equations for dilute granular gases of smooth hard-spheres. The details of the computation can be found in \cite{GT2012} and the source code for the computation is provided as supplementary material with the present paper. Interested readers are also referred to \cite{GT2015PRSA,Gupta2015} for deriving the higher-order moment equations for monatomic gas mixtures and to \cite{GT2015CAMWA,Gupta2015} for learning the computation of their associated production terms, which might be useful in developing higher-order moment theories for granular gas mixtures.
For the sake of completeness, we provide the production terms associated with the G26 equations (eqs.~\eqref{massBal}--\eqref{energyBal} and \eqref{eqn:stress}--\eqref{eqn:Delta}) in appendix \ref{App:ProdTerms}; the production terms associated with the G26 equations are \eqref{P1}--\eqref{P2} on taking $\Xi=0$. Notice from \eqref{P1} that the cooling rate computed via the G26 equations is given by
\begin{align}
\label{coolingRateG26}
\zeta=\frac{5}{12}(1-e^2)\,\nu
\bigg[1+\frac{1}{80}\frac{w}{\rho\,\theta^2} \underline{+ \frac{1}{25600}\frac{w^2}{\rho^2\theta^4} + \frac{1}{40}\frac{\sigma_{ij}\sigma_{ij}}{\rho^2\theta^2}+\frac{1}{200}\frac{q_i q_i}{\rho^2\theta^3}}
\nonumber\\
\underline{ +\frac{1}{1680}\frac{m_{ijk} m_{ijk}}{\rho^2\theta^3} +\frac{3}{31360}\frac{R_{ij} R_{ij}}{\rho^2\theta^4} -\frac{1}{560}\frac{\sigma_{ij} R_{ij}}{\rho^2\theta^3}} \bigg],
\end{align}
where 
\begin{align}
\label{CollFreq}
\nu=\frac{16}{5}\sqrt{\pi}\,n\, d^2\sqrt{\theta}
\end{align}
is the collision frequency. From \eqref{coolingRateG26}, one can see that
the cooling rate computed via the G26 equations is also proportional to $(1-e^2)$ as obtained in previous studies, e.g., \cite{SGJFM1998,vNE1998,BP2004,KM2011}, and it vanishes identically for monatomic gases ($e=1$) resulting into the conservation of energy. Furthermore, on dropping the underlined nonlinear terms in \eqref{CoolingRate}, the expression for the dissipation matches with that in \cite{KM2011}. 
\subsection{Various Grad moment systems}\label{Subsec:various}
Equations \eqref{massBal}--\eqref{energyBal} and \eqref{eqn:stress}--\eqref{eqn:Delta} along with \eqref{P1}--\eqref{P2} form the system of G26 equations. Various Grad moment systems may be obtained from the system of G26 equations.
\begin{enumerate}\itemsep1.5ex
\item The system of well-known G13 equations contains the balance equations for variables $n, v_i, T, \sigma_{ij}, q_i$, i.e., the system of G13 equations consists of equations \eqref{massBal}--\eqref{energyBal}, \eqref{eqn:stress} and \eqref{eqn:HF} along with \eqref{P1}--\eqref{Pi1} and $m_{ijk}=R_{ij}=w=0$.
\item The system of G14 equations contains the balance equations for variables $n, v_i, T, \sigma_{ij}, q_i, w$, i.e., the system of G14 equations includes equations \eqref{massBal}--\eqref{energyBal}, \eqref{eqn:stress}, \eqref{eqn:HF} and \eqref{eqn:Delta} along with \eqref{P1}--\eqref{Pi1}, \eqref{P2} and $m_{ijk}=R_{ij}=0$.
\item The system of G20 equations contains the balance equations for variables $n, v_i, T, \sigma_{ij}, q_i, m_{ijk}$, i.e., the system of G20 equations includes equations \eqref{massBal}--\eqref{energyBal} and \eqref{eqn:stress}, \eqref{eqn:HF}, \eqref{eqn:mijk} along with \eqref{P1}--\eqref{Pijk0} and $R_{ij}=w=0$.
\item The system of G21 equations contains the balance equations for variables $n, v_i, T, \sigma_{ij}, q_i, m_{ijk}, w$, i.e., the system of G21 equations includes equations \eqref{massBal}--\eqref{energyBal}, \eqref{eqn:stress}--\eqref{eqn:mijk} and \eqref{eqn:Delta} along with \eqref{P1}--\eqref{Pijk0}, \eqref{P2} and $R_{ij}=0$.
\end{enumerate}
It is commonly accepted that the dependence of heat flux on the density gradient in addition to the temperature gradient is akin to inclusion of the full trace of the fourth moment into the moment system \citep{KM2011,Garzo2013}. In that sense, the G13 and G20 theories for granular flows may not lead to meaningful results. Nevertheless, we shall also include them in this study for comparison purposes.

\section{Homogeneous cooling state of a freely cooling granular gas}\label{Sec:HCS}
The state of a granular gas when in the absence of any external forces its granular temperature decays continuously but its spatial homogeneity is maintained is termed as the \emph{homogeneous cooling state} \citep{BP2004}. In this section, we study the HCS of a granular gas with Grad moment equations. We assume a spatially homogeneous state (i.e., $\frac{\partial(\cdot)}{\partial x_i}=0$) without any external force acting on the particles of the granular gas (i.e., $\bm{F}=\bm{0}$).  Furthermore, following \cite{KM2011}, in the production terms (eqs.~\eqref{P1}--\eqref{P2}), we shall retain only those nonlinear terms which are the product of the scalar fourth moment ($w$) and a vectorial or tensorial moment, and all other nonlinear terms are simply ignored. This means that we are focusing our attention on the early evolution stage of homogeneously cooling granular gas. 

Often, It is more convenient to use a dimensionless variable $\Delta:=\frac{w}{\rho\,\theta^2}$ instead of the field variable $w$ in the Grad moment systems, where the governing equations for the former can be obtained by \eqref{massBal}, \eqref{energyBal} and \eqref{eqn:Delta}. Therefore, in the following, we shall use this new variable $\Delta$ while writing the Grad moment system.
The G26 equations in the zero external force case and with aforementioned simplification of the production terms---in terms of $\Delta$---read
\begin{align}
\label{massBalSimp}
\frac{\mathrm{D} n}{\mathrm{D}t}+ n \frac{\partial v_i}{\partial x_i}=0,
\end{align}
\begin{align}
\label{momBalSimp}
\frac{\mathrm{D} v_i} {\mathrm{D}t} + \frac{1}{m\,n} \left[\frac{\partial \sigma_{ij}}{\partial x_j} + \frac{\partial (n\, T)}{\partial x_i}\right]=0,
\end{align}
\begin{align}
\label{energyBalSimp}
\frac{\mathrm{D} T} {\mathrm{D}t} + \frac{2}{3\,n} \left[\frac{\partial q_i}{\partial x_i} +\sigma_{ij}\frac{\partial v_i}{\partial x_j}+n \,T\frac{\partial v_i}{\partial x_i}\right] = - \zeta \, T,
\end{align}
\begin{align}
\label{eqn:stressSimp}
&\frac{\mathrm{D}\sigma_{ij}}{\mathrm{D}t}+\frac{\partial m_{ijk}}{\partial x_k}+\frac{4}{5}\frac{\partial q_{\langle i}}{\partial x_{j\rangle}}+\sigma_{ij}\frac{\partial v_k}{\partial x_k}+2\sigma_{k\langle i}\frac{\partial v_{j\rangle}}{\partial x_k}+2\rho\theta\frac{\partial v_{\langle i}}{\partial x_{j\rangle}}
\nonumber\\
&=-\frac{(1+e)(3-e)}{4}\nu \left[\left(1-\frac{1}{480}\Delta \right) \sigma_{ij}
+\frac{1}{28}\left(1+\frac{1}{160}\Delta \right) \frac{R_{ij}}{\theta}\right],
\end{align}
\begin{align}
\label{eqn:HFSimp}
&\frac{\mathrm{D} q_i}{\mathrm{D} t} + \frac{1}{2} \frac{\partial R_{ij}}{\partial x_j} + \frac{1}{6}\rho\theta^2 \frac{\partial \Delta}{\partial x_i} + \frac{1}{6}\Delta \left(\theta^2\frac{\partial \rho}{\partial x_i}+2\rho\theta \frac{\partial \theta}{\partial x_i}\right)
+\frac{5}{2}\rho\theta\frac{\partial \theta}{\partial x_i}+\frac{5}{2}\sigma_{ij}\frac{\partial \theta}{\partial x_j} \nonumber\\
& 
+\theta\frac{\partial \sigma_{ij}}{\partial x_j} + m_{ijk}\frac{\partial v_j}{\partial x_k} - \sigma_{ij}\frac{1}{\rho}\left(\frac{\partial \sigma_{jk}}{\partial x_k}+\theta\frac{\partial \rho}{\partial x_j}\right)
+\frac{7}{5}q_i\frac{\partial v_j}{\partial x_j}+\frac{7}{5}q_j\frac{\partial v_i}{\partial x_j}+\frac{2}{5}q_j\frac{\partial v_j}{\partial x_i} 
\nonumber\\
&= -\frac{(1+e)}{48}\nu \left[(49-33e)+\frac{(19-3e)}{480}\Delta \right]q_i,
\end{align}
\begin{align}
\label{eqn:mijkSimp}
&\frac{\mathrm{D}m_{ijk}}{\mathrm{D}t} + \frac{3}{7}\frac{\partial R_{\langle ij}}{\partial x_{k\rangle}}+3\theta\frac{\partial \sigma_{\langle ij}}{\partial x_{k\rangle}}-3\frac{1}{\rho}\sigma_{\langle ij}\left(\frac{\partial \sigma_{k\rangle l}}{\partial x_l}+\theta\frac{\partial \rho}{\partial x_{k\rangle}}\right) +m_{ijk}\frac{\partial v_l}{\partial x_l}
\nonumber\\
&+3m_{l\langle ij}\frac{\partial v_{k\rangle}}{\partial x_l}+\frac{12}{5}q_{\langle i}\frac{\partial v_{j}}{\partial x_{k\rangle}}
=-\frac{3(1+e)(3-e)}{8}\nu \left(1-\frac{1}{1120}\Delta \right) m_{ijk} ,
\end{align}
\begin{align}
\label{eqn:RijSimp}
&\frac{\mathrm{D} R_{ij}}{\mathrm{D} t} + R_{ij}\frac{\partial v_k}{\partial x_k}+\frac{28}{5}\theta\frac{\partial q_{\langle i}}{\partial x_{j\rangle}}+ \frac{28}{5} q_{\langle i} \frac{\partial \theta}{\partial x_{j\rangle}} +4\theta \sigma_{k\langle i}\frac{\partial v_k}{\partial x_{j\rangle}}
\nonumber\\
& +4\theta\sigma_{k\langle i}\frac{\partial v_{j\rangle}}{\partial x_k}-\frac{8}{3}\theta\sigma_{ij}\frac{\partial v_k}{\partial x_k} -\frac{14}{3}\frac{1}{\rho}\sigma_{ij}\frac{\partial q_k}{\partial x_k}-\frac{14}{3}\frac{1}{\rho}\sigma_{ij}\sigma_{kl} \frac{\partial v_k}{\partial x_l}
\nonumber\\
&+2\theta\frac{\partial m_{ijk}}{\partial x_k}+\frac{6}{7}R_{\langle ij}\frac{\partial v_{k\rangle}}{\partial x_k}+\frac{4}{5}R_{k\langle i}\frac{\partial v_k}{\partial x_{j\rangle}}+2R_{k\langle i}\frac{\partial v_{j\rangle}}{\partial x_k}+\frac{14}{15}\rho\theta^2\Delta\frac{\partial v_{\langle i}}{\partial x_{j\rangle}}
\nonumber\\
&+7 m_{ijk}\frac{\partial \theta}{\partial x_k} - 2\frac{1}{\rho}  m_{ijk}\left( \frac{\partial \sigma_{kl}}{\partial x_l}
+ \theta \frac{\partial \rho}{\partial x_k}\right) - \frac{28}{5} \frac{1}{\rho}q_{\langle i} \left(\frac{\partial \sigma_{j\rangle k}}{\partial x_k}+\theta \frac{\partial \rho}{\partial x_{j\rangle}} \right)
\nonumber\\
&= -\frac{(1+e)}{12}\nu\bigg[\frac{1}{28} \bigg\{(436-267 e+66 e^2-30 e^3)-\frac{(52-27 e+66 e^2-30 e^3)}{480}\Delta
\bigg\}R_{ij}
&\nonumber\\
&\quad-\bigg\{(11-2e-22 e^2+10 e^3)
+\frac{(202 - 207 e - 66 e^2 + 30 e^3)}{480} \Delta \bigg\}\,\theta\,\sigma_{ij}\bigg],
\end{align}
%
\begin{align}
\label{eqn:DeltaSimp}
&\frac{\mathrm{D}\Delta}{\mathrm{D}t}+8\left(1-\frac{1}{6}\Delta\right) \frac{1}{\rho\,\theta}\left(\frac{\partial q_i}{\partial x_i} + \sigma_{ij}\frac{\partial v_i}{\partial x_j}\right) \nonumber\\
&
+ \frac{1}{\rho\,\theta^2}\left[ 20 q_i\frac{\partial \theta}{\partial x_i}+4R_{ij}\frac{\partial v_i}{\partial x_j}-8\frac{1}{\rho}q_i\left(\frac{\partial \sigma_{ij}}{\partial x_j}+\theta\frac{\partial \rho}{\partial x_i}\right)\right]
\nonumber\\
&
=\frac{5(1+e)}{4}\nu\left[(1-e)(1-2 e^2)-\frac{(81-17 e+30 e^2-30 e^3)}{240}\Delta \underline{+ \frac{(1-e)}{120}\Delta^2}\right]
\end{align}
with
\begin{align}
\zeta = \frac{5}{12}(1-e^2)\,\nu\left(1+\frac{1}{80}\Delta\right).
\end{align}
Notice that \eqref{eqn:DeltaSimp} contains $\Delta^2$ term, even though we have dropped all the nonlinear terms including $\Delta^2$ terms in the RHSs of \eqref{energyBalSimp} and \eqref{eqn:Delta} owing to aforementioned simplification.

%

We now introduce the following scaling
\begin{align}
\label{scaling}
\left.
\begin{gathered}
n_{\ast}=\frac{n}{n_0},\quad v_i^{\ast}=\frac{v_i}{\sqrt{\theta_0}},\quad T_{\ast}=\frac{T}{T_0},\quad \sigma_{ij}^{\ast}=\frac{\sigma_{ij}}{n_0 T_0},\quad q_i^{\ast}=\frac{q_i}{n_0 T_0\sqrt{\theta_0}}, 
\\
m_{ijk}^{\ast}=\frac{m_{ijk}}{n_0 T_0\sqrt{\theta_0}}, \quad R_{ij}^{\ast}=\frac{R_{ij}}{n_0 T_0\theta_0},\quad w_{\ast}=\frac{w}{n_0 T_0\theta_0},\quad t_{\ast}=\nu_0 t
\\
\textrm{with}\quad \nu_0=\frac{16}{5}\sqrt{\pi}\,n_0\, d^2\sqrt{\theta_0}\,, \quad \theta_0=\frac{T_0}{m}, \quad n_0 = n(0) \quad \textrm{and} \quad T_0 = T(0).
\end{gathered}
\right\}
\end{align}
With the scaling \eqref{scaling} and noting that $\Delta = w_{\ast} / (n_{\ast}T_{\ast}^2)$,  the G26 equations in the HCS (i.e., $\frac{\partial(\cdot)}{\partial x_i}=0$, $\bm{F}=\bm{0}$) reduce to
\begin{flalign}
\label{HCS:massBal}
&\frac{\mathrm{d} n_{\ast}}{\mathrm{d}t_{\ast}}=0,&
\end{flalign}
\begin{flalign}
\label{HCS:momBal}
&\frac{\mathrm{d} v_i^{\ast}} {\mathrm{d}t_{\ast}}=0,&
\end{flalign}
\begin{flalign}
\label{HCS:energyBal}
&\frac{\mathrm{d} T_{\ast}} {\mathrm{d}t_{\ast}} = -\frac{5}{12}(1-e^2)\, n_{\ast}T_{\ast}^{3/2}\left(1+\frac{1}{80}\Delta\right),&
\end{flalign}
\begin{flalign}
\label{HCS:stress}
&\frac{\mathrm{d}\sigma_{ij}^{\ast}}{\mathrm{d}t_{\ast}}=-\frac{(1+e)(3-e)}{4}n_{\ast}\sqrt{T_{\ast}}\bigg[\bigg(1-\frac{1}{480}\Delta\bigg)\sigma_{ij}^{\ast}+\frac{1}{28}\bigg(1+\frac{1}{160}\Delta\bigg)\frac{R_{ij}^{\ast}}{T_{\ast}}\bigg],&
\end{flalign}
\begin{flalign}
\label{HCS:HF}
&\frac{\mathrm{d} q_i^{\ast}}{\mathrm{d} t_{\ast}} =-\frac{(1+e)}{48}n_{\ast}\sqrt{T_{\ast}}\bigg[(49-33e)+\frac{(19-3e)}{480}\Delta\bigg]q_i^{\ast},&
\end{flalign}
\begin{flalign}
\label{HCS:mijk}
&\frac{\mathrm{d}m_{ijk}^{\ast}}{\mathrm{d}t_{\ast}}=-\frac{3(1+e)(3-e)}{8}n_{\ast}\sqrt{T_{\ast}}\bigg(1-\frac{1}{1120}\Delta\bigg)m_{ijk}^{\ast},&
\end{flalign}
\begin{flalign}
\label{HCS:Rij}
\frac{\mathrm{d} R_{ij}^{\ast}}{\mathrm{d} t_{\ast}} 
=&-\frac{(1+e)}{12} n_{\ast}\sqrt{T_{\ast}} \bigg[\frac{1}{28} \bigg\{(436 - 267 e + 66 e^2 - 30 e^3)
\nonumber\\
&-\frac{(52 - 27 e + 66 e^2 - 30 e^3)}{480}\Delta\bigg\}R_{ij}^{\ast}&
\nonumber\\
&-\bigg\{(11-2 e-22 e^2+10 e^3)+\frac{(202-207 e-66 e^2+30e^3)}{480}\Delta\bigg\}T_{\ast}\sigma_{ij}^{\ast}\bigg],&
\end{flalign}
\begin{align}
\label{HCS:DeltaBal}
&\frac{\mathrm{d}\Delta}{\mathrm{d}t_{\ast}}
=\frac{5(1+e)}{4}n_{\ast} \sqrt{T_{\ast}} \left[(1-e)(1-2 e^2)-\frac{(81-17 e+30 e^2-30 e^3)}{240}\Delta \underline{+ \frac{(1-e)}{120}\Delta^2}\right].
\end{align}
%
\subsection{Haff's law}\label{Subsec:HaffLaw}
By following heuristic approach, \cite{Haff1983} discovered that in a freely cooling granular gas (with constant coefficient of restitution), the decay rate of granular temperature is given by
\begin{align}
\label{Haffheuristic}
\frac{\mathrm{d}T}{\mathrm{d} t} \propto - \bar{n} \, d^2(1-e^2) T^{3/2},
\end{align}
where $\bar{n}$ is the average number density. The solution of \eqref{Haffheuristic} is given by
\begin{align}
\label{HaffLawheuristic}
T(t) = \frac{T(0)}{(1+t/\tau_\circ)^2},
\end{align}
where $\tau_\circ^{-1} \propto \bar{n} \, d^2(1-e^2) \sqrt{T_\circ}$ is an inverse time scale, see e.g., \cite{Haff1983,BP2004}. Equation \eqref{HaffLawheuristic} is termed as Haff's law for evolution of the granular temperature of a freely cooling granular gas (with a constant coefficient of restitution).

On comparing \eqref{HCS:energyBal} and \eqref{Haffheuristic}, one readily perceives that Haff's law can be obtained from \eqref{HCS:energyBal}, provided $\Delta$ is constant or, in other words, 
\begin{align}
\label{HaffLawCond}
\frac{\mathrm{d}\Delta}{\mathrm{d}t_\ast}=0.
\end{align}
With a constant value of $\Delta$ ($=\alpha$, let us say)---which is obtained from condition \eqref{HaffLawCond}---equations \eqref{HCS:massBal} and \eqref{HCS:energyBal} along with the initial conditions $n_\ast(0)=T_\ast(0)=1$ yield Haff's law for evolution of the dimensionless granular temperature $T_\ast$:
\begin{align}
\label{eq:HaffLaw}
T_\ast (t_\ast) = \frac{1}{(1+t_\ast/\tau_0)^2},
\end{align}
where
\begin{align}
\label{tau0inverse}
\tau_0^{-1} = \frac{5}{24}(1-e^2)\left(1+\frac{1}{80}\alpha\right)
\end{align} 
is the inverse of a (dimensionless) time scale $\tau_0$.

It is worthwhile to note that, in the light of condition \eqref{HaffLawCond}, the constant $\alpha$ in \eqref{tau0inverse} is an \emph{equilibrium point} \citep{Strogatz1994} of differential equation \eqref{HCS:DeltaBal}.
Here, we shall discard the underlined term (proportional to $\Delta^2$) in \eqref{HCS:DeltaBal}, although we shall analyze the effect of this term in \S\,\ref{Subsec:HaffLawNL}. On discarding the underlined term in \eqref{HCS:DeltaBal}, one readily obtains the equilibrium point---which is the constant $\alpha$ in \eqref{tau0inverse}---as
\begin{align}
\label{alpha}
\alpha = 15 \, a_2, \quad\textrm{where}\quad a_2:=\frac{16\,(1-e)\,(1-2e^2)}{81 - 17e + 30 e^2 - 30 e^3}
\end{align}
is same as the coefficient of second Sonine polynomial $S_2(v^2)$  while performing the CE expansion on the inelastic Boltzmann equation, see e.g., \cite{vNE1998,BP2004}.

\subsection{Relaxation of moments in the homogeneous cooling state}\label{Relax}
It is clear from \eqref{HCS:massBal} and \eqref{HCS:momBal} that the number density $n_\ast$ remains constant while the macroscopic velocity $v_i^{\ast}$ remains zero (since there is no macroscopic velocity in the initial state) during the homogeneous cooling. Furthermore, equations \eqref{HCS:stress}--\eqref{HCS:DeltaBal} are coupled and, therefore need to be solved numerically for further analysis. It may also be noticed from the structure of \eqref{HCS:HF} and \eqref{HCS:mijk} that vanishing initial conditions on $q_i^\ast$ and $m_{ijk}^\ast$ will result into vanishing solution for these quantities. Similarly, vanishing initial conditions on both $\sigma_{ij}^\ast$ and $R_{ij}^\ast$ will also result into zero solution for them; however, owing to the coupling on the RHSs of \eqref{HCS:stress} and \eqref{HCS:Rij}, a non-vanishing initial condition on any one of them will trigger the non-vanishing solution for both of them. On the contrary, vanishing initial condition on $\Delta$ results into non-vanishing 
solution for it for all values of the coefficient of  restitution except for $e=1$ and $e=1/\sqrt{2}$. In the following analysis, we take the initial conditions as $n_\ast(0)=T_\ast(0)=\sigma_{ij}^{\ast}(0)=q_i^{\ast}(0)=m_{ijk}^{\ast}(0)=R_{ij}^{\ast}(0)=1$, $v_i^{\ast}(0)=0$ and $\Delta(0) = 15$.

\begin{figure}
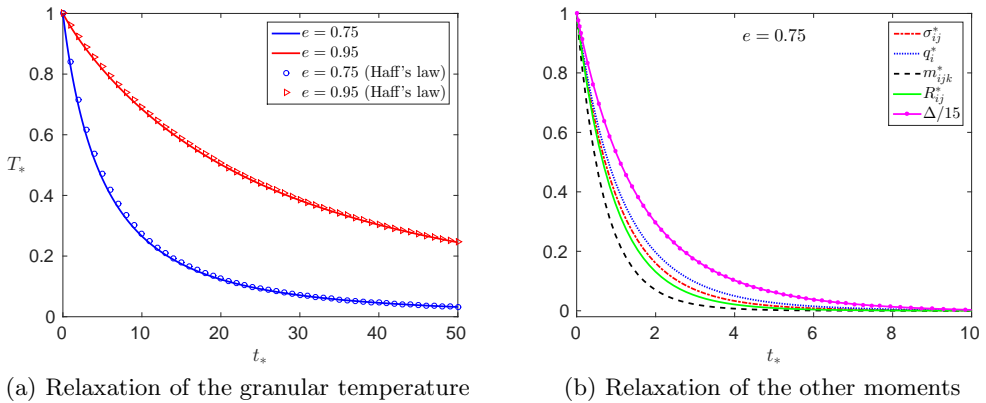

\begin{center}
\subfigure[Relaxation of the granular temperature]{
\label{fig:TempRelax}
\includegraphics[height=48mm]{plots/Fig_Haff_law.eps}
}
\qquad
\subfigure[Relaxation of the other moments]{
\label{fig:OtherMomRelax}
\includegraphics[height=48mm]{plots/Fig_moments_relax_e0p75.eps}
}
\caption{\label{fig:RelaxHCS} \small{Relaxation of various moments in the HCS: 
(a) relaxation of the granular temperature $T_\ast$; the lines depict solutions obtained by solving the system \eqref{HCS:massBal}--\eqref{HCS:DeltaBal} (without the underlined term in \eqref{HCS:DeltaBal}) while the symbols delineate Haff's law \eqref{eq:HaffLaw}; the blue and red colors correspond to coefficient of restitution $e=0.75$ and $e=0.95$, respectively, and (b) relaxation of the other moments---$\sigma_{ij}^\ast$, $q_i^\ast$, $m_{ijk}^\ast$, $R_{ij}^\ast$ and $(\Delta/15)$---for coefficient of restitution $e=0.75$. Initial conditions are taken as $n_\ast(0)=T_\ast(0)=\sigma_{ij}^{\ast}(0)=q_i^{\ast}(0)=m_{ijk}^{\ast}(0)=R_{ij}^{\ast}(0)=1$, $v_i^{\ast}(0)=0$ and $\Delta(0) = 15$.
}}

\end{center}
\end{figure}

Figure~\ref{fig:RelaxHCS} illustrates the numerical solution for the (dimensionless) granular temperature and other higher-order moments obtained by solving the system \eqref{HCS:massBal}--\eqref{HCS:DeltaBal} (without the underlined term in \eqref{HCS:DeltaBal}) along with the aforementioned initial conditions. The solutions for the number density and velocity are not shown since they are just constants.
Figure~\ref{fig:TempRelax} exhibits the granular temperature for two values of the coefficient of  restitution $e=0.75$ (in blue color) and $e=0.95$ (in red color). The solid lines depict the numerical solution for the granular temperature obtained by solving the system \eqref{HCS:massBal}--\eqref{HCS:DeltaBal} (without the underlined term in \eqref{HCS:DeltaBal}) along with the aforementioned initial conditions while the symbols denote the corresponding granular temperature computed via Haff's law~\eqref{eq:HaffLaw}. Clearly, the numerical results (denoted by solid lines) are in good agreement with those obtained via Haff's law (denoted by symbols). Moreover, it is also clear that the granular temperature relaxes faster with decreasing the coefficient of restitution. This is due to the fact that more inelastic particles dissipate more energy during the collision in comparison to the less inelastic ones resulting into the faster decay of the granular temperature for the former in comparison to the latter.
The relaxation of other moments---$\sigma_{ij}^\ast$, $q_i^\ast$, $m_{
ijk}^\ast$, $R_{ij}^\ast$ and $(\Delta/15)$---with time $t_\ast$ for $e=0.75$ is displayed in figure~\ref{fig:OtherMomRelax}. 
It turns out that all these moments decay with time much faster than the granular temperature.
Among themselves, $m_{ijk}^\ast$ decays with time $t_\ast$ faster than any other moment, followed by $R_{ij}^\ast$, $\sigma_{ij}^\ast$, $q_i^\ast$ and $\Delta$ (in the order of fast to slow). That the stress (dashed red line) decays faster than the heat flux (dotted blue line) which decays faster than $\Delta/15$ (magenta line with symbols) is in agreement with the findings of \cite{KM2011}.

\subsection{Effect of nonlinear terms of dimensionless scalar moment ($\Delta$) on Haff's law}\label{Subsec:HaffLawNL}
In \S\,\ref{Subsec:HaffLaw}, we obtained Haff's law by dropping all the nonlinear terms of non-equilibrium moments ($\sigma_{ij}^{\ast}$, $q_i^{\ast}$, $m_{ijk}^{\ast}$, $R_{ij}^{\ast}$, $\Delta$) in \eqref{HCS:energyBal} and \eqref{HCS:DeltaBal}. In this subsection, we shall investigate the effect of nonlinear terms of (dimensionless) scalar moment ($\Delta$) on Haff's law in detail.
\subsubsection{Case 1: Effect of $\Delta^2$ term present in \eqref{HCS:DeltaBal}}
The constant $\alpha$ in characteristic time scale $\tau_0$ for the temperature decay via Haff's law \eqref{eq:HaffLaw} was obtained by discarding $\Delta^2$ term in \eqref{HCS:DeltaBal}. We now consider \eqref{HCS:DeltaBal} without dropping any term. Equation \eqref{HCS:DeltaBal} has two equilibrium points for $e \neq 1$. Nevertheless, a simple stability analysis of these equilibrium points \citep[see Chapter 2 of][]{Strogatz1994} shows that only one equilibrium point is stable whereas the other one is unstable. In the limiting case of $e\to1$, the stable equilibrium point tends to zero while the unstable equilibrium point tends to infinity. This makes sense as \eqref{HCS:DeltaBal} in the case of $e=1$ has only one equilibrium point (which is zero). Furthermore, in the limiting case of $e\to1$, the stable equilibrium point leads to infinite relaxation time $\tau_0$ which is meaningful for monatomic gases (i.e., for $e=1$); however the unstable equilibrium point in this limit leads to a finite relaxation time which 
is meaningless. Owing to these reasons, we shall neglect the unstable equilibrium point. 

\begin{figure}
\begin{center}
\includegraphics[height=48mm]{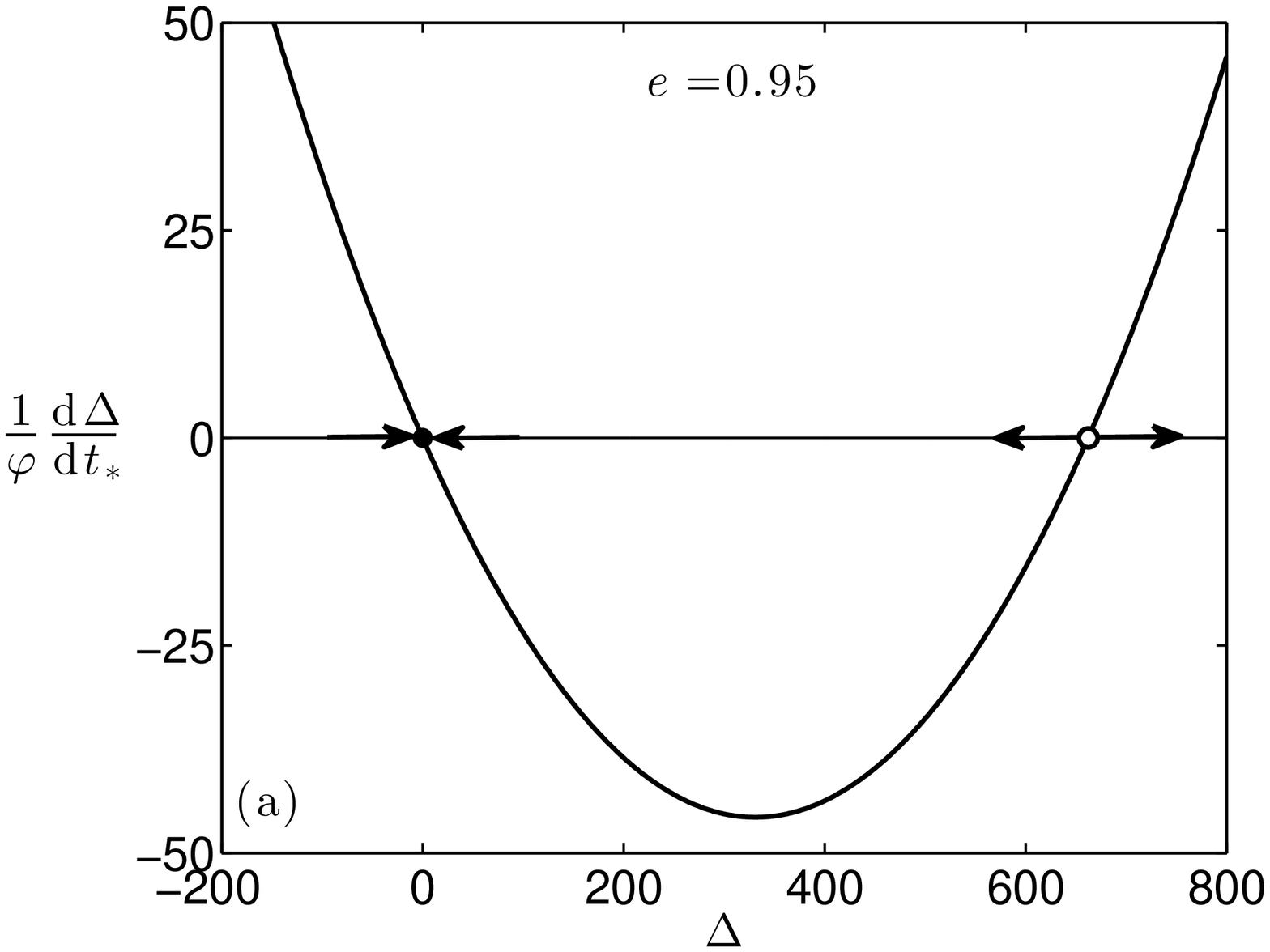}
\qquad
\includegraphics[height=48mm]{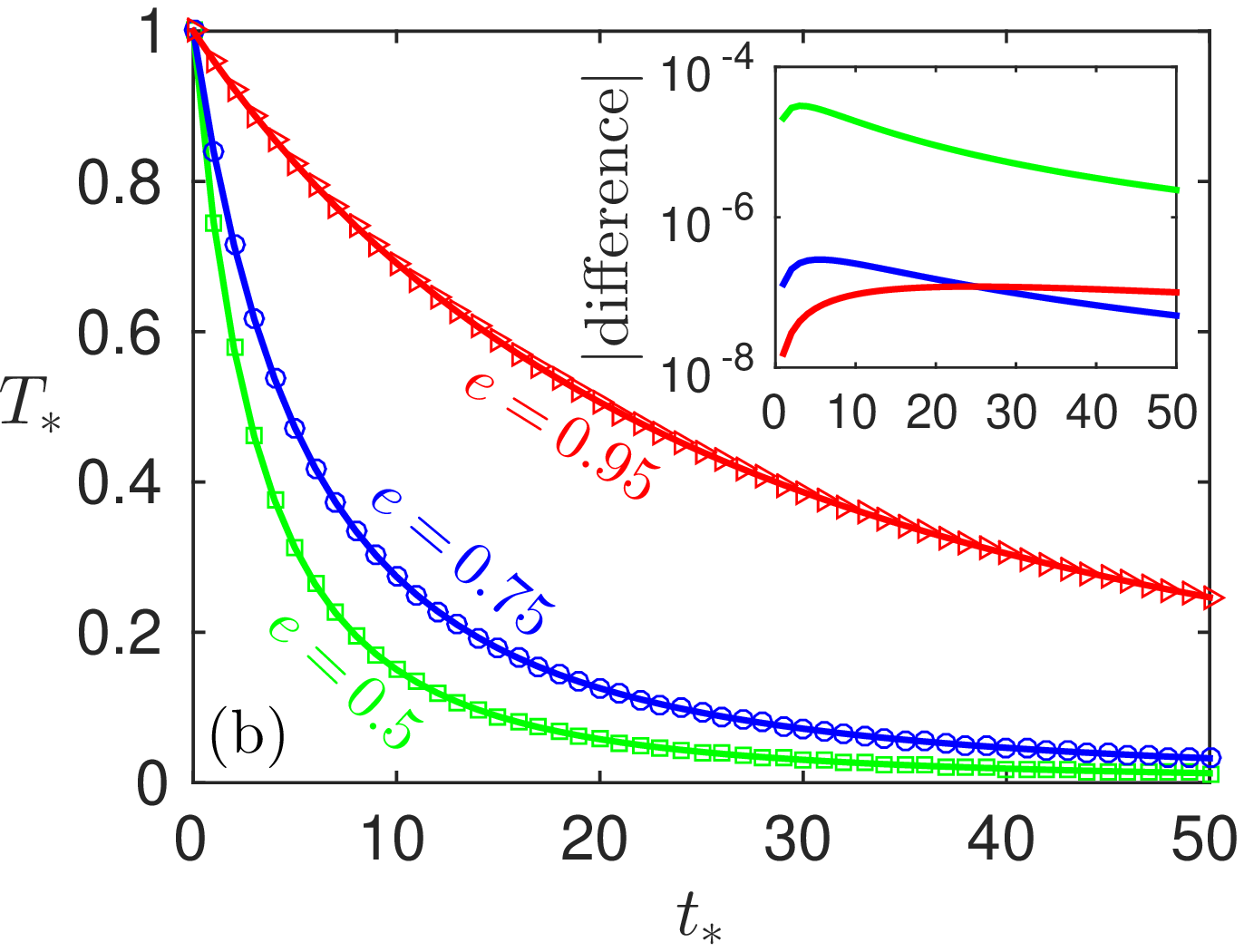}
\caption{\label{fig:TempRelaxQuad}Stability of roots and the temperature decay: (a) phase portrait showing the stable (black filled circle) and unstable (white filled circle) equilibrium points of \eqref{HCS:DeltaBal}; here, $\varphi = 5\,(1+e)\,n_\ast\sqrt{T_\ast}/4>0$ (cf.~\eqref{HCS:DeltaBal}), thus does not play any role in analyzing the stability of equilibrium points, and (b) temperature decay via Haff's law \eqref{eq:HaffLaw} with the inverse time scale \eqref{tau0inverse} 
for $e=0.5,0.75,0.95$ shown by green, blue and red colors respectively; the solid lines represent Haff's law when the constant $\alpha$ in \eqref{tau0inverse} is taken as the stable equilibrium point of \eqref{HCS:DeltaBal} (marked by black filled circle in the phase portrait) while the symbols delineate that for $\alpha$ given in \eqref{alpha}$_1$; the inset shows the absolute difference between the temperature values in these two cases.
}
\end{center}
\end{figure}

Figure~\ref{fig:TempRelaxQuad}(a) portrays the stable (black filled circle) and unstable (white filled circle) equilibrium points  for $e=0.95$ on a phase portrait. We consider only the stable equilibrium point (black filled circle) and neglect the unstable equilibrium point (white filled circle). The main panel of figure~\ref{fig:TempRelaxQuad}(b) illustrates the temperature decay  via Haff's law \eqref{eq:HaffLaw} with the inverse time scale \eqref{tau0inverse} for three values of the coefficient of restitution $e=0.5,0.75,0.95$, shown by green, blue and red colors, respectively. The solid lines depict Haff's law when the constant $\alpha$ in \eqref{tau0inverse} is taken as the stable equilibrium point of \eqref{HCS:DeltaBal} while the symbols show that when $\alpha$ is taken as the single equilibrium point \eqref{alpha}$_1$ (i.e., in the linear case). In the inset of figure~\ref{fig:TempRelaxQuad}(b), we also plot the corresponding absolute difference between the temperature values obtained in both the cases. It turns out that the difference is many order of magnitude smaller than the original temperature values. Moreover, the difference is even smaller for large values of the coefficient of restitution. Thus we conclude that the $\Delta^2$ term present in \eqref{HCS:DeltaBal} do not play any significant role on Haff's law.
\subsubsection{Case 2: Effect of $\Delta^2$ terms present on the RHSs of \eqref{HCS:energyBal} and \eqref{eqn:Delta}}
To investigate the effect of nonlinear terms of scalar moment (i.e., $\Delta^2$ terms) on Haff's law one needs to include $\Delta^2$ terms in the production terms of the energy balance equation \eqref{energyBal} and $w$ balance equation \eqref{eqn:Delta}. On including $\Delta^2$ terms along with linear terms on the RHSs of \eqref{energyBal} and \eqref{eqn:Delta}, and introducing the scaling \eqref{scaling}, the new energy and $\Delta$ balance equations in the HCS read
\begin{align}
\label{HCS:energyBalwithDel2term}
\frac{\mathrm{d} T_{\ast}} {\mathrm{d}t_{\ast}} &= -\frac{5}{12}(1-e^2)\, n_{\ast}T_{\ast}^{3/2}\left(1 + \frac{1}{80}\Delta + \frac{1}{25600}\Delta^2\right),
\end{align}
\begin{align}
\label{Cubic}
\frac{\mathrm{d}\Delta}{\mathrm{d}t_\ast} &= \frac{5(1+e)}{4}n_{\ast} \sqrt{T_{\ast}} \left[(1-e)(1-2 e^2) -\frac{(81-17 e+30 e^2-30 e^3)}{240}\Delta \right.\nonumber\\
&\left.\quad + \frac{(1873 - 2001 e+30 e^2-30 e^3)}{230400} \Delta^2 + \frac{(1-e)}{38400}\Delta^3\right].
\end{align}
Notice again that although we included the nonlinear terms proportional to $\Delta^2$ on the RHSs of \eqref{energyBal} and \eqref{eqn:Delta}, the RHS of \eqref{Cubic} contains $\Delta^3$ terms as well.
Similar to above, for obtaining Haff's law form \eqref{HCS:energyBalwithDel2term}, $\Delta$ must be a constant or, in other words, condition \eqref{HaffLawCond} must be fulfilled. With a constant value of $\Delta$ ($=\beta$, let us say), equations~\eqref{HCS:massBal} and \eqref{HCS:energyBalwithDel2term} along with the initial conditions $n_\ast(0)=T_\ast(0)=1$ again yield Haff's law \eqref{eq:HaffLaw}. However, the inverse of dimensionless characteristic time scale in this case reads
\begin{align}
\label{tau0inverseNL}
\tau_0^{-1} = \frac{5}{24}(1-e^2)\left(1+\frac{1}{80}\beta+\frac{1}{25600}\beta^2\right).
\end{align} 
Again, in the light of condition \eqref{HaffLawCond}, the constant $\beta$ in \eqref{tau0inverseNL} is an equilibrium point of differential equation \eqref{Cubic}.

\begin{figure}
\begin{center}
\includegraphics[height=48mm]{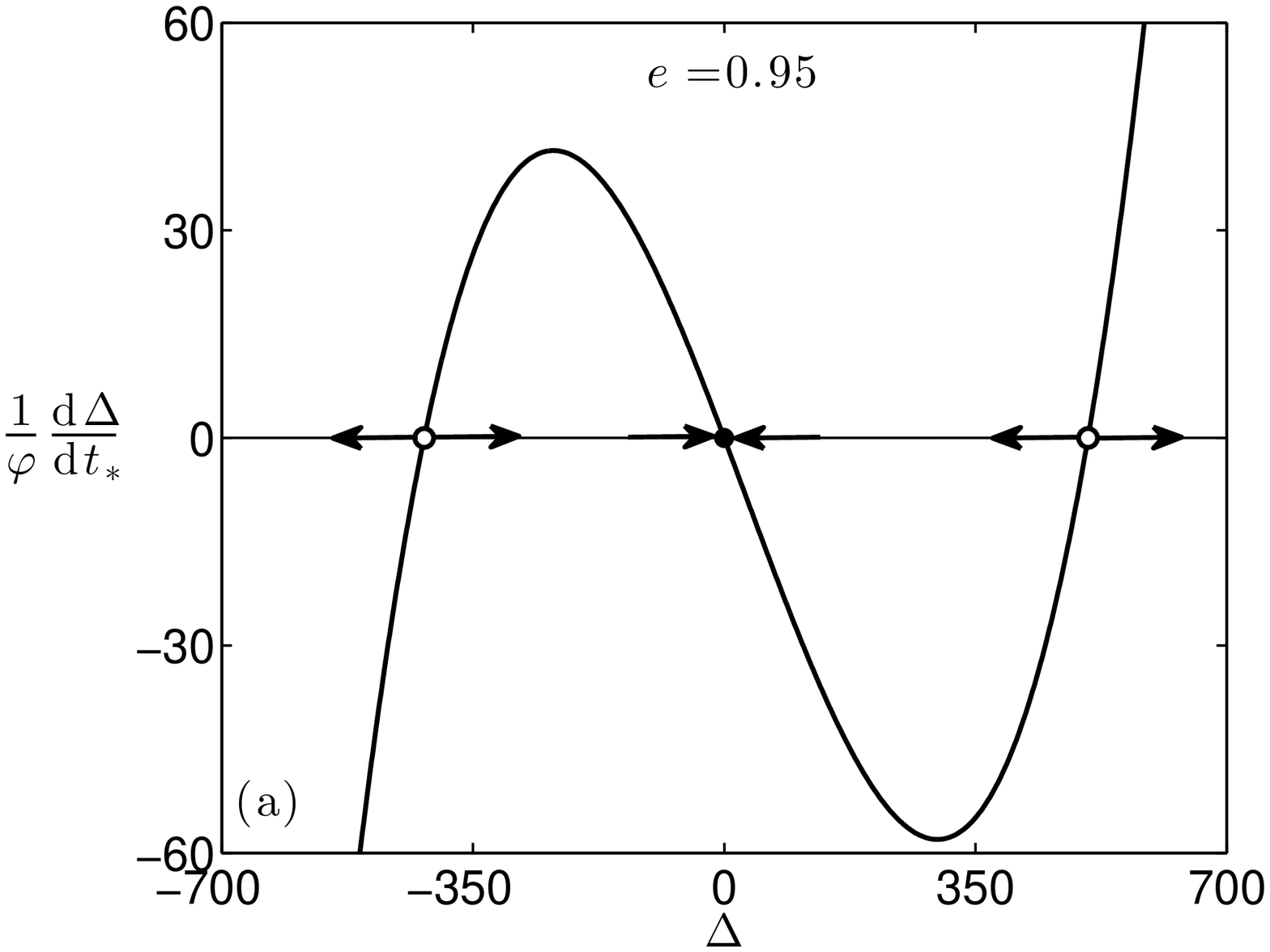}
\qquad
\includegraphics[height=48mm]{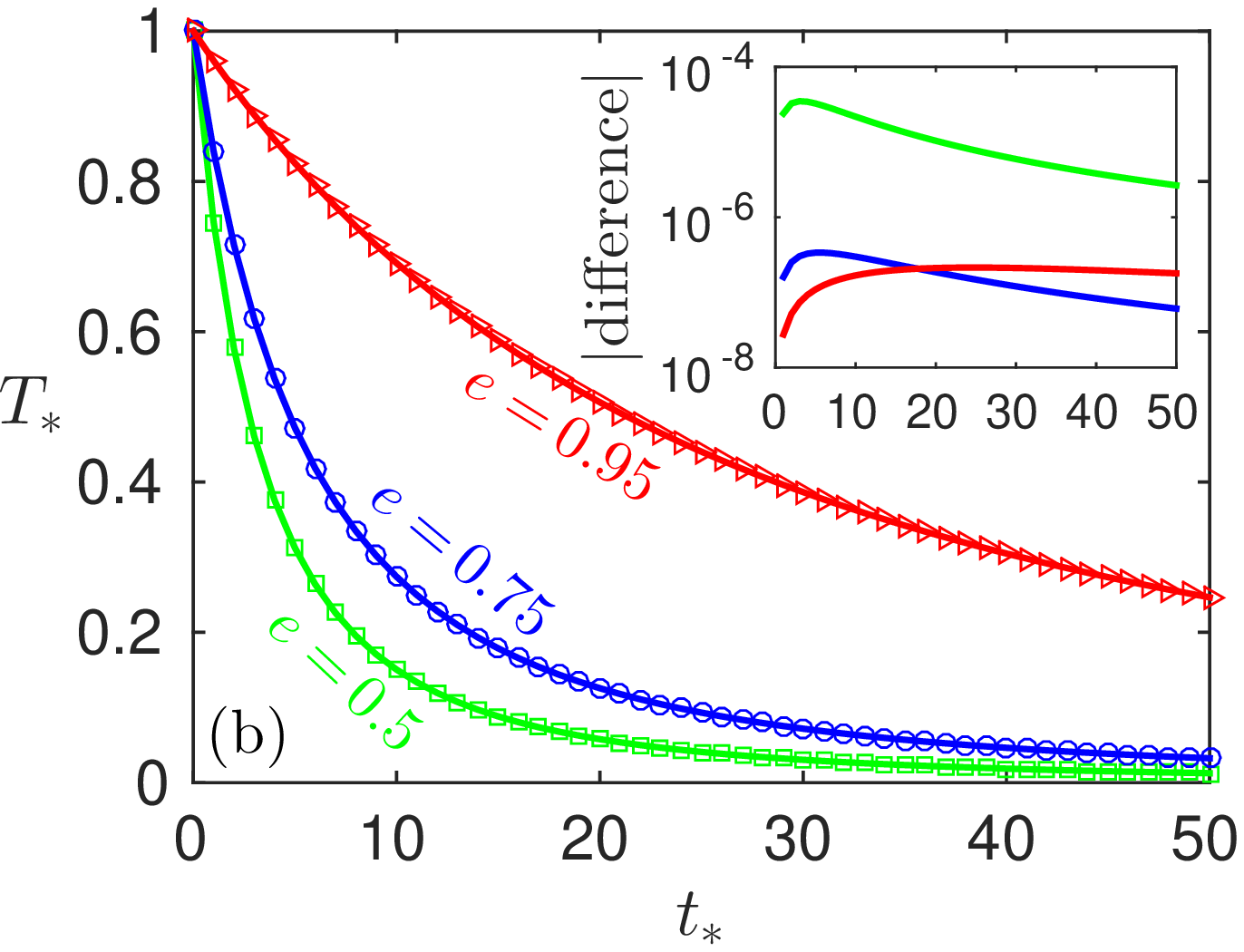}
\caption{\label{fig:TempRelaxCubic}Stability of roots and  the temperature decay: (a) phase portrait showing the stable (black filled circle) and unstable (white filled circles) equilibrium points of \eqref{Cubic}; here, $\varphi = 5\,(1+e)\,n_\ast\sqrt{T_\ast}/4>0$ (cf.~\eqref{HCS:DeltaBal}), thus does not play any role in analyzing the stability of equilibrium points, and (b) temperature decay via Haff's law \eqref{eq:HaffLaw} for $e=0.5,0.75,0.95$ shown by green, blue and red colors respectively; the solid lines represent Haff's law \eqref{eq:HaffLaw} with the inverse time scale \eqref{tau0inverseNL} when the constant $\beta$ in \eqref{tau0inverseNL} is taken as the stable equilibrium point of  \eqref{Cubic} (marked by black filled circle in the phase portrait) while the symbols delineate Haff's law \eqref{eq:HaffLaw} with the inverse time scale \eqref{tau0inverse} and the constant $\alpha$ as given in \eqref{alpha}$_1$; the inset shows the absolute difference between the temperature values in these two cases.
}
\end{center}
\end{figure}

Equation \eqref{Cubic} has three equilibrium points for $e \neq 1$. Again, a simple stability analysis of these equilibrium points \citep[see Chapter 2 of][]{Strogatz1994} shows that only one equilibrium point is stable whereas the other two are unstable. Figure~\ref{fig:TempRelaxCubic}(a) displays the stable (black filled circle) and unstable equilibrium points (white filled circles) for $e=0.95$ on a phase portrait.
In the limiting case of $e\to1$, the stable equilibrium point tends to zero, the left unstable point tends to $-480$ and the right unstable equilibrium point tends to infinity. That one equilibrium point of \eqref{Cubic} in the limiting case of $e \to 1$ tends to infinity makes sense as \eqref{Cubic} in the case of $e=1$ has only two equilibrium points. Furthermore, in the limiting case of $e\to1$, the stable equilibrium point and the left unstable  equilibrium point lead to infinite relaxation time $\tau_0$ which is meaningful for monatomic gases (i.e., for $e=1$); however the right unstable equilibrium point in this limit leads to a vanishing relaxation time which is meaningless, therefore we neglect the right unstable equilibrium point. Although, the left unstable equilibrium point lead to meaningful infinite relaxation time $\tau_0$ in the limit $e \to 1$, its value in the limit $e \to 1$ itself is not meaningful because the equilibrium points are the steady state solutions of \eqref{Cubic} and the distribution function for monatomic gases (i.e., for $e=1$) in the steady state is Maxwellian, and consequently, by definition, $w$ (and hence $\Delta$) must vanish in the limit $e \to 1$ and in the steady state.

Therefore, we again consider only the stable equilibrium point (marked by black filled circle) and neglect the unstable equilibrium points (marked by white filled circles). The main panel of figure~\ref{fig:TempRelaxCubic}(b) illustrates the temperature decay via Haff's law \eqref{eq:HaffLaw} for three values of the coefficient of restitution $e=0.5,0.75,0.95$, shown by green, blue and red colors, respectively. The solid lines depict the results obtained with the inverse time scale \eqref{tau0inverseNL} and the constant $\beta$ in \eqref{tau0inverseNL} as the stable equilibrium point of \eqref{Cubic} (marked by black filled circle in the phase portrait) whereas the symbols delineate the corresponding results obtained with the inverse time scale \eqref{tau0inverse} and the constant $\alpha$ as given in \eqref{alpha}$_1$  (i.e., in the linear case). In the inset of figure~\ref{fig:TempRelaxCubic}(b), we also plot the corresponding absolute difference between the temperature values obtained in both the cases. There is practically no difference in both the results as the difference is again many order of 
magnitude smaller than the original temperature values. 

Thus, we conclude that the nonlinear terms of (dimensionless) scalar moment $\Delta$ do not play any significant role on Haff's law.

\subsection{Effect of higher-order scalar moment on Haff's law}
In this subsection, we shall analyze effect of the sixth order scalar moment $u^3$, which is the full trace of the sixth order moment, on Haff's law. Nevertheless, for convenience, we introduce a new dimensionless variable $\Xi$ which is related to the sixth order scalar moment $u^3$ via
\begin{align}
\label{XiDefinition}
\Xi = \frac{m}{\rho\,\theta^3} \int\!\big(f - f_M \big) \big( C^6 - 21 \, \theta \, C^4 \big)\,\mathrm{d}\bm{c}
= \frac{u^3}{\rho\,\theta^3} - 21 \, \Delta - 105
\end{align}
so that $\Xi$ vanishes if it is computed with $f$ as either Maxwellian or G13 or G26 distribution function. 

We now include the governing equation for the moment $\Xi$ to our existing 26-moment system (eqs.~\eqref{massBal}--\eqref{energyBal} and \eqref{eqn:stress}--\eqref{eqn:Delta}) and close this system with the Grad distribution function (which we shall name as the G27 distribution function)
\begin{align}
\label{G27DisFunc}
f_{|\textrm{G27}}
&=f_M\left[1+\frac{1}{2}\frac{\sigma_{ij}}{\rho\,\theta^2}C_iC_j +\frac{1}{5}\frac{q_i}{\rho\,\theta^2}C_i\left(\frac{C^2}{\theta}-5\right)+\frac{1}{6}\frac{ m_{ijk}}{\rho\,\theta^3} C_{i}C_jC_{k} \right.
\nonumber\\
&\left.\hspace{13.7mm}+ \frac{1}{28}\frac{R_{ij}}{\rho \,\theta^3} C_{i}C_{j} \left(\frac{C^2}{\theta}-7\right)+\frac{1}{8}\Delta
\left(1-\frac{2}{3}\frac{C^2}{\theta}+\frac{1}{15}\frac{C^4}{\theta^2}\right) \right.
\nonumber\\
&\left.\hspace{13.7mm}- \frac{1}{48}\Xi
\left(1-\frac{C^2}{\theta} + \frac{1}{5}\frac{C^4}{\theta^2} - \frac{1}{105}\frac{C^6}{\theta^3}\right)\right].
\end{align}
We shall refer to the system of these 27-moment equations closed with the G27 distribution function as the system of G27 equations.

\begin{figure}
\begin{center}
\includegraphics[height=48mm]{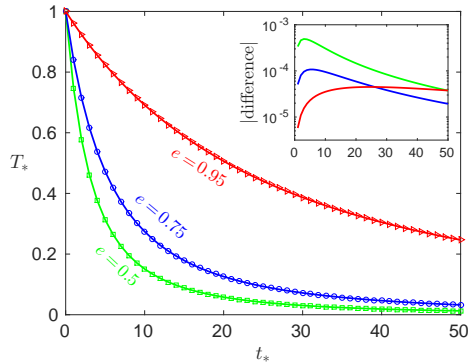}
\caption{\label{fig:TempRelaxSixthMoment}Temperature decay via Haff's law \eqref{eq:HaffLaw} for $e=0.5,0.75,0.95$ shown by green, blue and red colors respectively; the solid lines depict the results obtained with the inverse time scale \eqref{tau0inverseHighOrder} and the constants $(\varkappa_1,\varkappa_2)$ in \eqref{tau0inverseHighOrder} as the unique equilibrium point of \eqref{HCS:DeltaG27} and \eqref{HCS:XiG27} whereas the symbols show Haff's law \eqref{eq:HaffLaw} with the inverse time scale \eqref{tau0inverse} and the constant $\alpha$ given in \eqref{alpha}$_1$; the inset shows the absolute difference between the temperature values in these two cases.
}
\end{center}
\end{figure}

For our purposes, it is not necessary to derive the full G27 equations but we do need the production terms for the energy, $\Delta$ and $\Xi$ balance equations. All the production terms associated with the G27 equations are given in appendix~\ref{App:ProdTerms}. Note that the production terms for the $\Delta$ and $\Xi$ balance equations are given by
\begin{align*}
\frac{1}{\rho\,\theta^2}\mathcal{P}^2 - \frac{2}{3}\frac{(\Delta+15)}{\rho\,\theta}\mathcal{P}^1
\quad\textrm{and}\quad 
\frac{1}{\rho\,\theta^3}\mathcal{P}^3 - \frac{21}{\rho\,\theta^2}\mathcal{P}^2 - \frac{(\Xi-7\Delta-105)}{\rho\,\theta}\mathcal{P}^1,
\end{align*}
respectively. In the HCS, the dimensionless energy, $\Delta$ and $\Xi$ balance equations from the system of G27 equations---on employing \eqref{scaling} and 
and discarding all the nonlinear terms of non-equilibrium scalar, vector or tensor moments---read 
%
\begin{flalign}
\label{HCS:energyBalG27}
&\frac{\mathrm{d} T_{\ast}} {\mathrm{d}t_{\ast}} = -\frac{5}{12}(1-e^2)\, n_{\ast}T_{\ast}^{3/2}\left(1 + \frac{1}{80} \Delta -  \frac{1}{6720} \Xi \right),&
\end{flalign}
\begin{flalign}
\label{HCS:DeltaG27}
\frac{\mathrm{d}\Delta}{\mathrm{d}t_\ast}&= \frac{5}{4}(1+e)\,n_{\ast} \sqrt{T_{\ast}} \bigg[(1-e)(1-2 e^2) - \frac{(81-17 e+30 e^2-30 e^3)}{240} \Delta&
\nonumber\\
&\quad - \frac{(191-127 e+10 e^2-10 e^3)}{6720} \Xi 
\bigg],&
\end{flalign}
\begin{flalign}
\label{HCS:XiG27}
\frac{\mathrm{d}\Xi}{\mathrm{d}t_\ast}&= - \frac{15}{16}(1+e)\,n_{\ast} \sqrt{T_{\ast}} \bigg[(1 - e) (3 - 12 e^2 + 8 e^4) &
\nonumber\\
&\quad- \frac{(1111 - 727 e - 2012 e^2 + 1500 e^3 - 280 e^4 + 280 e^5)}{240} \Delta&
\nonumber\\
&\quad+\frac{(2673 - 241 e + 5196 e^2 - 3660 e^3 + 280 e^4 - 280 e^5)}{6720} \Xi\bigg].&
\end{flalign}
%
Clearly, for obtaining Haff's law form \eqref{HCS:energyBalG27}, $\Delta$ and $\Xi$ 
must be constants or, in other words, conditions 
\begin{align}
\label{HaffLawCondG27}
\frac{\mathrm{d} \Delta}{\mathrm{d} t_\ast} = \frac{\mathrm{d} \Xi}{\mathrm{d} t_\ast} = 0
\end{align}
must be fulfilled. With constant values of $\Delta$ ($=\varkappa_1$, let us say) and $\Xi$ ($=\varkappa_2$, let us say), eqs.~\eqref{HCS:massBal} and \eqref{HCS:energyBalG27} along with the initial conditions $n_\ast(0)=T_\ast(0)=1$ again yield Haff's law \eqref{eq:HaffLaw}. However, the inverse of dimensionless characteristic time scale now reads
\begin{align}
\label{tau0inverseHighOrder}
\tau_0^{-1} = \frac{5}{24}(1-e^2)\left(1 + \frac{1}{80}\varkappa_1 - \frac{1}{6720} \varkappa_2\right).
\end{align}
In the light of conditions \eqref{HaffLawCondG27}, the constants ($\varkappa_1, \varkappa_2$) in \eqref{tau0inverseHighOrder} are the equilibrium points ($\Delta_{\mathrm{eq}}, \Xi_{\mathrm{eq}}$) of the system of two first-order ordinary differential equations \eqref{HCS:DeltaG27} and \eqref{HCS:XiG27} \citep{Strogatz1994}.
The system of equations~\eqref{HCS:DeltaG27} and \eqref{HCS:XiG27} has a unique equilibrium point. Figure~\ref{fig:TempRelaxSixthMoment} illustrates the temperature decay via Haff's law \eqref{eq:HaffLaw} for three values of the coefficient of restitution $e=0.5,0.75,0.95$ shown by green, blue and red colors, respectively. The solid lines depict the results obtained with the inverse time scale \eqref{tau0inverseHighOrder} and the constants $(\varkappa_1,\varkappa_2)$ in \eqref{tau0inverseHighOrder} as the unique equilibrium point of \eqref{HCS:DeltaG27} and \eqref{HCS:XiG27} whereas the symbols delineate the corresponding results obtained with the inverse time scale \eqref{tau0inverse} and the constant $\alpha$ as given in \eqref{alpha}$_1$. In the inset of figure~\ref{fig:TempRelaxSixthMoment}(b), we also plot the corresponding absolute difference between the temperature values obtained in both the cases.
Again, there is practically no difference in the results as the differences are many order of magnitude smaller than the original temperature values.

Furthermore, in view of \S\,\ref{Subsec:HaffLawNL}, it is expected that the nonlinear terms of scalar moments in \eqref{HCS:energyBalG27}--\eqref{HCS:XiG27} would also have only negligible effect on Haff's law. Thus, it is concluded \emph{empirically} that the nonlinear terms of non-equilibrium scalar moments as well as higher-order scalar moments do not have any significant effect on Haff's law. Therefore, it suffices to consider only those moment systems which consists of the fourth order scalar moment ($\Delta$)---i.e., to consider 14- or 26-moment system. Moreover, in the production terms of these moment systems, it is sufficient to retain only those nonlinear terms which are products of $\Delta$ and a vector or a tensor.
Owing to this reason, equations \eqref{massBalSimp}--\eqref{eqn:DeltaSimp} without the underlined term in \eqref{eqn:DeltaSimp} will be referred to as the G26 equations henceforth.
\section{Constitutive relations for the stress and heat flux: Navier--Stokes, and Fourier laws}
\label{Sec:NSF_rels}
The mass, momentum and energy balance equations \eqref{massBal}--\eqref{energyBal} are not closed since they contain the stress ($\sigma_{ij}$), heat flux ($q_i$) and cooling rate ($\zeta$) as additional unknowns. One of the main goal in the kinetic theory for granular gases is to derive the constitutive relations for these unknowns in order to close the system of mass, momentum and energy balance equations. To the linear approximation in the spatial gradients, these constitutive relations read \citep[see e.g.,][]{JR1985,JR1985PoF,GD1999,Garzo2013}
\begin{align}
\label{NSFconstitutiveRels}
\sigma_{ij} = -2 \eta \frac{\partial v_{\langle i}}{\partial x_{j\rangle}},
\qquad
q_i = - \kappa \frac{\partial T}{\partial x_i} - \lambda \frac{\partial n}{\partial x_i}
\qquad\textrm{and}\qquad
\zeta = \zeta_0 + \zeta_1 \frac{\partial v_i}{\partial x_i}.
\end{align}
Equations \eqref{NSFconstitutiveRels}$_1$ and \eqref{NSFconstitutiveRels}$_2$ are the Navier--Stokes law and Fourier law, respectively. The coefficients $\zeta_0$ and $\zeta_1$ in \eqref{NSFconstitutiveRels}$_3$ are the zeroth- and first-order contributions, respectively, to the cooling rate.
The transport coefficients $\eta$, $\kappa$ are referred to as the \emph{shear viscosity} and \emph{thermal conductivity}, respectively; and the transport coefficient $\lambda$ is a special coefficient for granular gases which vanishes identically for elastic gases. Typically, these transport coefficients are obtained by a formal CE expansion on the Boltzmann equation \citep[see e.g.,][]{BDKS1998,SGJFM1998,Gupta2011}, nevertheless some authors have also computed them via computer-aided kinetic theory \citep{NBSG2007} and via Grad's method of moment, \citep[see e.g.,][]{JR1985,KM2011,Garzo2013}. Although Grad's method of moments does not restrict the coefficient of restitution to be close to $1$, G13 theory of \citet{JR1985} and the Maxwellian iteration procedure on the G14 equations of \citet{KM2011} neglect the effect of cooling rate on the transport coefficients. On the other hand, the work of \citet{Garzo2013} incorporates the effect of cooling rate  on the transport coefficients and, consequently, the transport coefficients obtained in this work are in good agreement with those obtained via the classical CE expansion. In this section, we shall determine these transport coefficients through the G26 equations (eqs.~\eqref{massBalSimp}--\eqref{eqn:DeltaSimp} without the underlined term in \eqref{eqn:DeltaSimp}). To this end, we shall follow the approach of \cite{Garzo2013} so that the effect of cooling rate on the transport coefficients is incorporated.

We are interested in the \emph{linear} approximation in the spatial gradients. To this end, let us first analyze the zeroth-order contributions in the spatial gradients. 
\subsection{Zeroth-order contributions in the spatial gradients}
To zeroth-order in the spatial gradients, $\Delta$ balance equation \eqref{eqn:DeltaSimp} yields
\begin{align}
\label{DeltaZerothOrder}
\Delta = 15 \, a_2.
\end{align}
Consequently, to zeroth-order in the spatial gradients, the mass, momentum and energy balance equations \eqref{massBalSimp}--\eqref{energyBalSimp} reduce to
\begin{align}
\label{zerothorderConservationlaws}
\frac{\partial n}{\partial t} = 0, \qquad \frac{\partial v_i}{\partial t} = 0 \qquad \textrm{and} \qquad \frac{\partial T}{\partial t} = -\zeta_0 T,
\end{align}
with
\begin{align}
\label{zeta0}
\zeta_0 = \nu \, \zeta_0^\ast \qquad\textrm{and}\qquad  \zeta_0^\ast=\frac{5}{12}(1-e^2)
\left(1+\frac{3 \, a_2}{16}\right),
\end{align}
while the balance equations for the higher moments (eqs.~\eqref{eqn:stressSimp}--\eqref{eqn:RijSimp}) simplify to 
\begin{align}
\label{zerothorderEq}
-\nu_\sigma^\ast \, \sigma_{ij} - \nu_{\sigma R}^\ast \frac{R_{ij}}{\theta} = 0, \qquad q_i = 0, \qquad m_{ijk} = 0, \qquad -\nu_R^\ast \, R_{ij} + \nu_{R\sigma}^\ast \,\theta \sigma_{ij} = 0,
\end{align}
respectively, with
\begin{align}
\label{nu*}
\left.
\begin{aligned}
\nu_\sigma^\ast &= \frac{(1+e)(3-e)}{4} \left(1-\frac{a_2}{32}\right),
\\
\nu_{\sigma R}^\ast &= \frac{(1+e)(3-e)}{112} \left(1+\frac{3\,a_2}{32}\right),
\\
\nu_R^\ast &= \frac{(1+e)}{336} \left[(436-267 e+66 e^2-30 e^3)-\frac{(52-27 e+66 e^2-30 e^3)\,a_2}{32}
\right],
\\
\nu_{R\sigma}^\ast  &= \frac{(1+e)}{12} \left[(11-2e-22 e^2+10 e^3)
+\frac{(202 - 207 e - 66 e^2 + 30 e^3)\,a_2}{32}\right].
\end{aligned}
\right\}
\end{align}
Equations~\eqref{zerothorderEq}$_{1,4}$ imply that
\begin{align}
\sigma_{ij} = R_{ij} = 0.
\end{align}
Thus, to zeroth-order in the spatial gradients, $\sigma_{ij}$, $q_i$, $m_{ijk}$ and $R_{ij}$ are zero while $\Delta = 15 \, a_2$. This means that all the non-equilibrium vectorial and tensorial moments are \emph{at least} of linear order in the spatial gradients.
\subsection{First-order contributions in the spatial gradients}
We now investigate the moments up to first-order in the spatial gradients, which will lead to the desired Navier--Stokes and Fourier laws.

To first-order in the spatial gradients, the mass, momentum and energy balance equations \eqref{massBalSimp}--\eqref{energyBalSimp} yield
\begin{align}
\label{FirstOrderConservationEqs}
\frac{\partial n}{\partial t} &= - v_i \frac{\partial n}{\partial x_i} - n \frac{\partial v_i}{\partial x_i}, 
\\
\frac{\partial v_i}{\partial t} &= - v_j \frac{\partial v_i}{\partial x_j} - \frac{1}{m\,n} \frac{\partial (n\, T)}{\partial x_i},
\\ 
\frac{\partial T}{\partial t} &= - v_i \frac{\partial T}{\partial x_i} - \frac{2}{3} T\frac{\partial v_i}{\partial x_i} - \zeta \, T,
\end{align}
where $\zeta$ needs to be considered up to linear order in the spatial gradients. Furthermore, the stress and heat flux balance equations (eqs.~\eqref{eqn:stressSimp} and~\eqref{eqn:HFSimp}), to first-order in the spatial gradients, reduce to
\begin{align}
\label{FirstOrderSigmaEq1}
\frac{\partial \sigma_{ij}}{\partial t} + 2\rho\theta\frac{\partial v_{\langle i}}{\partial x_{j\rangle}}
&= -\nu \left[\nu_\sigma^\ast \, \sigma_{ij}
+\nu_{\sigma R}^\ast \, \frac{R_{ij}}{\theta}\right],
\\
\label{FirstOrderHFEq}
\frac{\partial q_i}{\partial t} + \frac{5}{2} a_2 \left(\theta^2\frac{\partial \rho}{\partial x_i}+2\rho\theta \frac{\partial \theta}{\partial x_i}\right) +\frac{5}{2}\rho\theta\frac{\partial \theta}{\partial x_i} 
&= -\nu \, \nu_q^\ast \, q_i
\end{align}
with
\begin{align}
\label{nuq}
\nu_q^\ast = \frac{(1+e)}{48} \left[(49-33e)+\frac{(19-3e)\,a_2}{32} \right].
\end{align}
Notice that we do not need the balance equations for other moments except \eqref{eqn:RijSimp}, the one for $R_{ij}$, for first-order approximation. Equation \eqref{eqn:RijSimp} is only needed because the contribution of $R_{ij}$ is required in \eqref{FirstOrderSigmaEq1}. To first-order in the spatial gradients, the $R_{ij}$ balance equation \eqref{eqn:RijSimp} reduces to
\begin{align}
\label{FirstOrderRijEq}
&\frac{\partial R_{ij}}{\partial t} + 14 a_2 \rho\theta^2\frac{\partial v_{\langle i}}{\partial x_{j\rangle}}
= -\nu \big[\nu_R^\ast \, R_{ij} - \nu_{R\sigma}^\ast \,\theta \sigma_{ij}\big].
\end{align}
Here, we ignore the time derivative of $R_{ij}$ by assuming that it decreases with time faster than stress and heat flux. With this assumption, equation~\eqref{FirstOrderRijEq} yields
\begin{align}
\label{RijSol}
\frac{R_{ij}}{\theta} = \frac{\nu_{R\sigma}^\ast}{\nu_R^\ast} \sigma_{ij} - \frac{14 a_2 \rho\theta}{\nu \, \nu_R^\ast} \frac{\partial v_{\langle i}}{\partial x_{j\rangle}}.
\end{align}
With \eqref{RijSol}, equation~\eqref{FirstOrderSigmaEq1} turns to
\begin{align}
\label{FirstOrderSigmaEq}
\frac{\partial \sigma_{ij}}{\partial t} + \nu \left(\nu_\sigma^\ast + \frac{\nu_{\sigma R}^\ast \, \nu_{R\sigma}^\ast}{\nu_R^\ast} \right) \sigma_{ij} 
= - 2 \left(1 - 7 a_2 \frac{\nu_{\sigma R}^\ast}{\nu_R^\ast}\right) \rho\theta \frac{\partial v_{\langle i}}{\partial x_{j\rangle}}.
\end{align}
The time derivatives of the stress and heat flux in \eqref{FirstOrderSigmaEq} and \eqref{FirstOrderHFEq} are now calculated as follows. From \eqref{NSFconstitutiveRels}$_{1,2}$, we have 
\begin{align}
\label{timeDerStressHF1}
\left.
\begin{aligned}
\frac{\partial \sigma_{ij}}{\partial t} &= - 2 \frac{\partial v_{\langle i}}{\partial x_{j\rangle}} \frac{\partial \eta}{\partial t} - \underline{2\eta \frac{\partial}{\partial x_{j\rangle}} \frac{\partial v_{\langle i}}{\partial t}},
\\
\frac{\partial q_i}{\partial t} &= - \frac{\partial T}{\partial x_i} \frac{\partial \kappa}{\partial t} - \kappa \underbrace{\frac{\partial}{\partial x_i} \frac{\partial T}{\partial t}} - \frac{\partial n}{\partial x_i} \frac{\partial \lambda}{\partial t} - \underline{\lambda \frac{\partial}{\partial x_i} \frac{\partial n}{\partial t}}.
\end{aligned}
\right\}
\end{align}
For first-order approximation in the spatial gradients, the time derivatives on the RHSs of \eqref{timeDerStressHF1} are required only up to zeroth-order in spatial gradients. Hence, by virtue of \eqref{zerothorderConservationlaws}, the underlined terms in \eqref{timeDerStressHF1} vanish while the underbraced term in \eqref{timeDerStressHF1}$_2$ becomes
\begin{align}
\label{aterm}
\frac{\partial}{\partial x_i} \frac{\partial T}{\partial t} = - \zeta_0 \frac{\partial T}{\partial x_i} - T \zeta_0^\ast \frac{\partial \nu}{\partial x_i} = - \frac{3}{2} \zeta_0 \frac{\partial T}{\partial x_i} - \zeta_0 \frac{T}{n} \frac{\partial n}{\partial x_i}.
\end{align}
By dimensional analysis, it turns out that $\eta \propto \sqrt{T}$, $\kappa \propto \sqrt{T}$ and $\lambda \propto T^{3/2}$. Therefore, to zeroth-order in spatial gradients---on using \eqref{zerothorderConservationlaws}$_3$---the time derivatives of the transport coefficients read
\begin{align}
\label{timeDerTransCoeff}
\frac{\partial \eta}{\partial t} =  - \frac{1}{2} \eta \, \zeta_0, \qquad
\frac{\partial \kappa}{\partial t} =  - \frac{1}{2} \kappa \, \zeta_0 \qquad \textrm{and} \qquad
\frac{\partial \lambda}{\partial t} =  - \frac{3}{2} \lambda \, \zeta_0.
\end{align}
The time derivatives of the stress and heat flux \eqref{timeDerStressHF1} on using \eqref{aterm} and \eqref{timeDerTransCoeff} become
\begin{align}
\label{timeDerStressHF}
\frac{\partial \sigma_{ij}}{\partial t} = \eta \, \zeta_0 \frac{\partial v_{\langle i}}{\partial x_{j\rangle}}
\qquad\textrm{and}\qquad
\frac{\partial q_i}{\partial t} = 2 \kappa \, \zeta_0 \frac{\partial T}{\partial x_i} + \left(\kappa \frac{T}{n} + \frac{3}{2} \lambda \right) \zeta_0 \frac{\partial n}{\partial x_i}.
\end{align}
Inserting the time derivatives of the stress and heat flux from \eqref{timeDerStressHF} into \eqref{FirstOrderSigmaEq} and \eqref{FirstOrderHFEq}, and comparing the coefficients of the spatial gradient of each hydrodynamic variable, one obtains the transport coefficients in the Navier--Stokes and Fourier laws \eqref{NSFconstitutiveRels}$_{1,2}$ as
\begin{align}
\label{transCoeff}
\eta = \eta_0 \, \eta^\ast, \qquad 
\kappa = \kappa_0 \, \kappa^\ast, \qquad \textrm{and} \qquad 
\lambda = \frac{\kappa_0 \, T}{n}\, \lambda^\ast
\end{align}
where
\begin{align}
\eta_0 = \frac{n\,T}{\nu} = \frac{5}{16\sqrt{\pi}} \frac{\sqrt{m\,T}}{d^2}
\qquad \textrm{and} \qquad 
\kappa_0 = \frac{15}{4 \, m} \eta_0 = \frac{75}{64\sqrt{\pi}} \frac{1}{d^2}\sqrt{\frac{T}{m}}
\end{align}
are the elastic values---in first Sonine approximation---of the shear viscosity and thermal conductivity, respectively. They are also referred to as the \emph{Enskog} viscosity and \emph{Enskog} thermal conductivity, respectively \citep{BP2004}. Moreover,
$\eta^\ast$, $\kappa^\ast$ and $\lambda^\ast$ in \eqref{transCoeff} are the reduced shear viscosity, reduced thermal conductivity and reduced coefficient corresponding to $\lambda$. These reduced transport coefficients are given by
\begin{align}
\label{ReducedTransCoeff}
\eta^\ast = \frac{1 - 7 a_2 \frac{\nu_{\sigma R}^\ast}{\nu_R^\ast}}{\nu_\sigma^\ast + \frac{\nu_{\sigma R}^\ast \, \nu_{R\sigma}^\ast}{\nu_R^\ast} - \frac{1}{2} \zeta_0^\ast}, 
\quad
\kappa^\ast = \frac{2}{3} \left[\frac{1 + 2\,a_2}{\nu_q^\ast - 2 \zeta_0^\ast}\right],
\quad
\lambda^\ast = \frac{\kappa^\ast \zeta_0^\ast + \frac{2}{3} a_2}{\nu_q^\ast - \frac{3}{2} \zeta_0^\ast}.
\end{align}
Note that if we take $\nu_{\sigma R}^\ast = 0$, i.e., if we ignore the contribution of $R_{ij}$, the reduced transport coefficients \eqref{ReducedTransCoeff} obtained here agree with those obtained via Grad's method of moments in \cite{Garzo2013} for dilute granular gases and also with those obtained by first-Sonine approximation via the CE expansion in \cite{BDKS1998}. If we further neglect the cooling effect (i.e., $\zeta_0^\ast = 0$) as well, the reduced transport coefficients obtained here concur with those derived in \cite{KM2011}. 

In the elastic case ($e=1$), the reported (exact) value of the reduced shear viscosity $\eta^\ast$ is $1.016034$ which was obtained by \citet{PA1957} by reducing the Boltzmann equation to an ordinary differential equation of order $4$ and subsequently integrating it numerically. \citet{ReineckeKremer1990} obtained almost the same value $\eta^\ast \approx 1.016028$ for the reduced shear viscosity by employing Maxwell iteration procedure on Grad moment equations---at fifth order of approximation. The reduced shear viscosity for $e=1$ obtained from \eqref{ReducedTransCoeff}$_1$ is $\eta^\ast = 205/202 \, (\approx 1.014851)$, which is exactly same as that obtained at second order approximation in \cite{ReineckeKremer1990}, thanks to the coupling between $\sigma_{ij}$ and $R_{ij}$ in their balance equations. Note that the first Sonine approximation \citep{BDKS1998}, modified Sonine approximation \citep{GSM2007}, the G14 theory of \citet{KM2011} and Grad moment theory of \citet{Garzo2013}, they all in the elastic case lead to $\eta^\ast = 1$, which is the value obtained at first order approximation in \cite{ReineckeKremer1990}. 
Furthermore, the reported value of the reduced thermal conductivity $\kappa^\ast$ in the elastic case is $1.025218$ \citep{PA1957}. \citet{ReineckeKremer1990} again obtained almost the same value $\kappa^\ast \approx 1.025197$ at fifth order approximation. However, the reduced thermal conductivity obtained here as well as that obtained by all the four aforementioned approaches is $\kappa^\ast = 1$, which is the value obtained at first order approximation in \cite{ReineckeKremer1990}. The reason for not obtaining an improved value of the reduced thermal conductivity here is that, unlike the RHS of the stress balance equation which is coupled with the one trace of the fourth moment ($R_{ij}$), the RHS of the heat flux balance equation does not have coupling with other vectorial or tensorial moments in its considered form.
\subsection{Comparison with existing theories and computer simulations}
In this subsection, we shall compare the reduced transport coefficients derived here with those obtained by CE expansion as well as with those obtained through computer simulations.

\begin{figure}
\centering
\includegraphics[scale = 0.4]{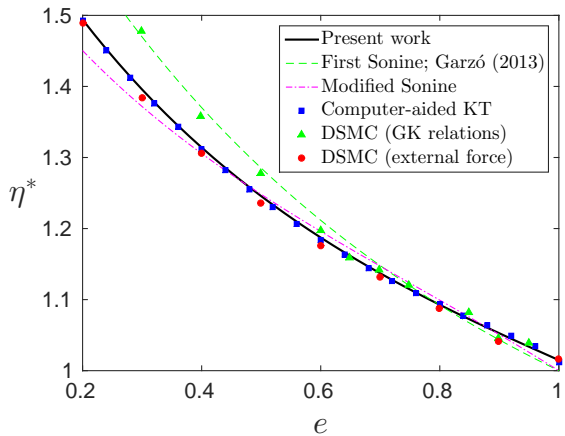}
\caption{Variation of the reduced shear viscosity $\eta^\ast$ with the coefficient of restitution $e$. The solid (black) line represents the results obtained in the present work. The dashed (green) and dash-dotted (magenta) lines delineate the first Sonine approximation \citep{BDKS1998} and modified version of the first Sonine approximation \citep{GSM2007}, respectively. The squares are the results from the theoretical expressions obtained via the computer-aided method devised by \cite{NBSG2007}. The other symbols are the DSMC simulation results of \cite{BRMG2005} obtained using GK relations (triangles) and of \cite{MSG2005} obtained with the implementation of an external force (circles).  
}
\label{fig:viscosity}
\end{figure}

\begin{figure}
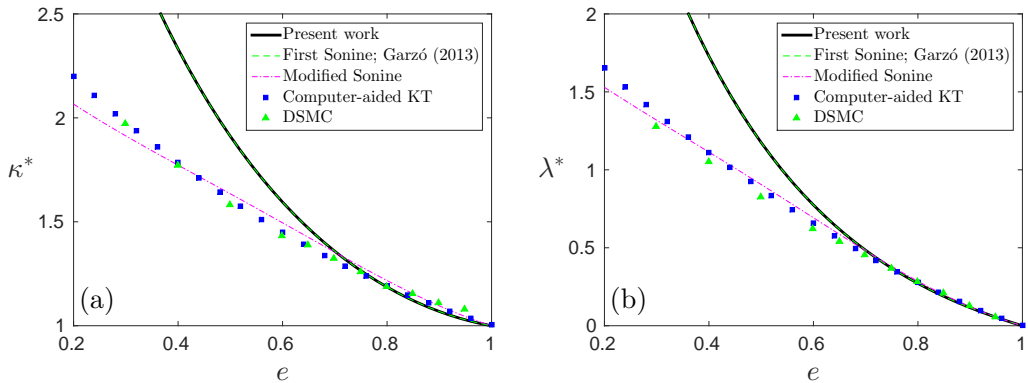

\centering
\includegraphics[width=0.48\textwidth]{plots/ThermalConductivity.eps}
\hfill
\includegraphics[width=0.48\textwidth]{plots/lambda.eps}
\caption{Variation of (a) the reduced thermal conductivity $\kappa^\ast$ and (b) the reduced coefficient $\lambda^\ast$ with the coefficient of restitution $e$. The lines and symbols are same as described in figure~\ref{fig:viscosity}.%
}
\label{fig:kappa_lambda}
\end{figure}

Figure~\ref{fig:viscosity} illustrates the reduced shear viscosity $\eta^\ast$ while figure~\ref{fig:kappa_lambda} illustrates the reduced thermal conductivity $\kappa^\ast$ and reduced coefficient $\lambda^\ast$ for different values of the coefficient of restitution. The solid (black) lines in both the figures represent the results obtained with expressions \eqref{ReducedTransCoeff} in the present work. The dashed (green) lines are the plots for the reduced transport coefficients derived at first Sonine approximation through CE expansion on the Boltzmann equation in \cite{BDKS1998}. As the reduced transport coefficients obtained via Grad's method of moments in \cite{Garzo2013} coincide with those in \cite{BDKS1998}, the dashed (green) lines also display the reduced transport coefficients obtained in \cite{Garzo2013}. The dash-dotted (magenta) lines display the results obtained with the theoretical expressions for the reduced transport coefficients deduced through the \emph{modified} version of the first Sonine approximation in \cite{GSM2007}. The squares are the results obtained with the theoretical expressions derived via the so-called \emph{computer-aided} method devised by \cite{NBSG2007} while the triangles denote the numerical solution of the Boltzmann equation obtained through Green--Kubo (GK) relations by means of the direct-simulation Monte Carlo (DSMC) method \citep{Bird1994} in \cite{BRMG2005}. The circles in figure~\ref{fig:viscosity} also denote the DSMC simulation results from \cite{MSG2005} obtained with another method---by the implementation of an external force which compensates for the collisional cooling. The DSMC simulation data from this method for $e=\{0.6,0.7,0.8,0.9,1\}$ were obtained by \citet{MSG2005} while those for $e=\{0.2,0.3,0.4,0.5\}$ were obtained by \citet{GSM2007}.

In general, the reduced shear viscosity from the first Sonine approximation (dashed green line) agrees with that from the DSMC simulations using GK relations (triangles). Nevertheless, the results from the DSMC simulations with the implementation of external force (shown by circles) differ from those obtained with the GK relations (shown by triangles), especially for $e\lesssim 0.7$. The difference between the results from the two simulations could be due to velocity correlations in the correlation function present in the GK relation for the shear viscosity \citep{GSM2007}. In comparison to the results from the modified version of the first Sonine approximation (dash-dotted magenta line), the present results  for the reduced shear viscosity (solid black line) are in better agreement with the DSMC simulations of \cite{GSM2007} even for small coefficient of restitution. Furthermore, the present results for the reduced shear viscosity (solid black line) coincide with those obtained from the computer-aided method devised by \cite{NBSG2007} (shown by squares) for almost all values of the coefficient of restitution. Nevertheless, in the elastic ($e=1$) case, the theoretical expressions of \citet{NBSG2007} lead to $\eta^\ast = 1.01205$ while the DSMC simulations of \citet{GSM2007} give $\eta^\ast \approx 1.016$. Thus, in the elastic ($e=1$) case, $\eta^\ast \approx 1.014851$ from the expression \eqref{ReducedTransCoeff}$_1$ obtained here is much closer to its true value in comparison to that from  the theoretical expressions of \citet{NBSG2007}. 

The expressions for the reduced transport coefficients corresponding to the heat flux, (i.e., for $\kappa^\ast$ and $\lambda^\ast$) derived here are the same as those obtained via the first Sonine approximation \citep{BDKS1998} and via Grad moment method of \cite{Garzo2013} because in the considered linearized form, the RHS of heat flux balance equation \eqref{eqn:HFSimp} does not couple with any other vectorial or tensorial moments. Therefore, the curves of the reduced thermal conductivity $\kappa^\ast$ and the reduced coefficient $\lambda^\ast$ in figure~\ref{fig:kappa_lambda} from \eqref{ReducedTransCoeff}$_{2,3}$ (solid black lines) coincide with those from the first Sonine approximation \citep{BDKS1998} or from Grad moment method of \cite{Garzo2013} (dashed green lines). It is clear from figure~\ref{fig:kappa_lambda} that the results for $\kappa^\ast$ and $\lambda^\ast$ from the present work, from first Sonine approximation and from \cite{Garzo2013} agree with the DSMC simulations as well as with those from the computer-aided method of \cite{NBSG2007} only for large coefficient of restitution ($e \gtrsim 0.7$). For $e \lesssim 0.7$, similar to the first Sonine approximation
and Grad moment method of \cite{Garzo2013}, the present work also overestimates both the transport coefficients.

Motivated from the satisfactory result for the reduced shear viscosity through the G26 equations, it is expected that a suitable coupling on the RHS of the heat flux balance equation \eqref{eqn:HFSimp} would improve the reduced transport coefficients $\kappa^\ast$ and $\lambda^\ast$ significantly. The first such coupling of RHS of the heat flux balance equation in the semi-linearized setting as considered here will be introduced on including the full trace of fifth order moment into the moment system, i.e., one would need to consider the system of 29 moment equations. This is beyond the scope of the present paper and will be considered elsewhere in future.

Although, the modified Sonine approach  \citep{GSM2007} is an excellent theoretical approach whose predictions for all the transport coefficients ($\eta^\ast$, $\kappa^\ast$ and $\lambda^\ast$) are in very good agreement with the simulation results, the G26 theory slightly improves the results for the reduced shear viscosity coefficient $\eta^\ast$. However, the latter could not yield the other two transport coefficients related to the heat flux ($\kappa^\ast$ and $\lambda^\ast$) correctly for $e \lesssim 0.7$, exactly in the same way as the first Sonine approach \citep{BDKS1998}. Here, it is worthwhile to note that while it is an important validity test for moment equations to compute the precise form of the transport coefficients, it is well-known that these coefficients have only a little relevance in non-equilibrium situations when a linear relation between fluxes and gradients does not hold anymore. Instead, it has been shown (for elastic gases) that moment equations allow for cross effects like thermal stresses and non-gradient heat fluxes \citep{TorrilhonARFM} that are expected to also have a systematic influence in granular flows. The actual goal of moment equations is  not only to match the transport coefficients but also to go beyond these classical theories and provide an enhanced fluid dynamic theory for granular gases.
Moreover, our long-term perspective is to establish a complete set of predictive moment equations---somewhat similar to the R13 equations for monatomic gases---for granular flows, which certainly requires the G26 equations or beyond. Thus this work is the first step towards this long-term goal.

\section{Linear stability analysis}
\label{Sec:Stability}
In this section, we investigate the stability of the HCS due to small perturbations through various moment theories---particularly, with the G26 theory (equations~\eqref{massBalSimp}--\eqref{eqn:DeltaSimp} without the underlined term in  \eqref{eqn:DeltaSimp})---developed in the earlier sections. We assume that the amplitudes of these perturbations are sufficiently small so that the linear analysis remains valid.

For the linear stability analysis, we decompose all the field variables into their reference values---i.e., their respective solutions in the HCS---and into perturbations from their respective solutions in the HCS, i.e., we define
%
\begin{align}
\label{perturbations}
\left.
\begin{aligned}
n(t,\bm{x}) &= n_0 \big[1+\tilde{n}(t,\bm{x})\big],
\\
T(t,\bm{x}) &= T_H(t) \big[1+\tilde{T}(t,\bm{x})\big],
\\
v_i(t,\bm{x}) &= v_H(t) \, \tilde{v}_i(t,\bm{x}),
\\
\sigma_{ij}(t,\bm{x}) &= \sigma_{ij}^{(H)}(t) + n_0 \,T_H(t)\, \tilde{\sigma}_{ij}(t,\bm{x}),
\\
q_i(t,\bm{x}) &= q_i^{(H)}(t) + n_0\, T_H(t)\,v_H(t)\, \tilde{q}_i(t,\bm{x}),
\\
m_{ijk}(t,\bm{x}) &= m_{ijk}^{(H)}(t) + n_0 \,T_H(t)\,v_H(t)\, \tilde{m}_{ijk}(t,\bm{x}),
\\
R_{ij}(t,\bm{x}) &= R_{ij}^{(H)}(t) + n_0 \,T_H(t)\,v_H(t)^2\, \tilde{R}_{ij}(t,\bm{x}),
\\
\Delta(t,\bm{x}) &= \Delta_H + \tilde{\Delta}(t,\bm{x}),
\end{aligned}
\right\}
\end{align}
where $n_0$ is the constant number density and $T_H(t)$ is the granular temperature in the HCS; $\Delta_H$ is the constant solution for $\Delta$ in the HCS; $v_H(t)=\sqrt{T_H(t)/m}$ is a reference speed in the HCS and is proportional to the adiabatic sound speed in the HCS; the reference values (HCS solutions) for all other field variables are denoted by the superscript `{\scriptsize$(H)$}'; the quantities with tilde denote the dimensionless perturbations in the field variables from their respective solutions in the HCS. 
Note that as the underlined term in \eqref{eqn:DeltaSimp} is discarded, the HCS solution for $\Delta$ is $\Delta_H = 15\,a_2$.

Inserting the field variables from \eqref{perturbations} into the G26 equations, and neglecting all the nonlinear terms of the perturbed field variables (denoted with tilde in \eqref{perturbations}), one obtains the system of linear partial differential equations in (dimensionless) perturbed field variables  with time-dependent coefficients, which is given in appendix~\ref{App:PertSys}. These equations are further simplified by exploiting the fact---concluded in \S\,\ref{Relax}---that in the HCS, the non-equilibrium vectorial and tensorial moments---$\sigma_{ij}^{(H)}(t)$, $q_i^{(H)}(t)$, $m_{ijk}^{(H)}(t)$, $R_{ij}^{(H)}(t)$---decay faster than the granular temperature $T_H(t)$; therefore, we drop the terms containing $\sigma_{ij}^{(H)}(t)$, $q_i^{(H)}(t)$, $m_{ijk}^{(H)}(t)$, $R_{ij}^{(H)}(t)$ in \eqref{massBalPert}--\eqref{DeltaBalPert}. Now, it is possible to convert this system of partial differential equations to a new system of partial differential equations having constant coefficients as follows. We introduce a length scale 
\begin{align}
\label{ell}
\ell = \frac{v_H(t)}{\nu_H(t)}, \quad\textrm{where}\quad
\nu_H(t) = \frac{16}{5}\sqrt{\pi}n_0 d^2 \sqrt{\frac{T_H(t)}{m}},
\end{align}
to make the space variables dimensionless (i.e., $\tilde{x}_i = x_i/\ell$, where tilde again denotes the dimensionless space variable), and a dimensionless time $\tilde{t}$ \citep[see][]{McNamara1993} such that
\begin{align}
\frac{1}{\nu_H(t)} \frac{\partial}{\partial t} (\cdot) = \frac{\partial}{\partial \tilde{t}} (\cdot).
\end{align}
This leads to
\begin{align}
\tilde{t} = \nu_H(t) \, \tau_1 \left(1 + \frac{t}{\tau_1}\right) \ln{\left(1 + \frac{t}{\tau_1}\right)} \approx \nu_H(t) \, t + \mathcal{O}(t^2),
\end{align}
where 
\begin{align*}
\tau_1^{-1} = \frac{1}{2} \zeta_0^\ast \,\nu_0 
\qquad\textrm{with}\qquad \nu_0 = \frac{16}{5} \sqrt{\pi}\, n_0 d^2 \sqrt{\frac{T_0}{m}}.
\end{align*}
With these definitions of dimensionless space and time, system \eqref{massBalPert}--\eqref{DeltaBalPert} now simplifies to
\begin{align}
\label{massBalPertDimless}
\frac{\partial \tilde{n}}{\partial \tilde{t}} + \frac{\partial \tilde{v}_i}{\partial \tilde{x}_i} &= 0,
\\
\label{momentBalPertDimless}
\frac{\partial \tilde{v}_i}{\partial \tilde{t}} + \frac{\partial \tilde{\sigma}_{ij}}{\partial \tilde{x}_j}  + \frac{\partial \tilde{n}}{\partial \tilde{x}_i} + \frac{\partial \tilde{T}}{\partial \tilde{x}_i} - \frac{1}{2}\xi_0 \tilde{v}_i &= 0,
\\
\label{energyBalPertDimless}
\frac{\partial \tilde{T}}{\partial \tilde{t}}+ \frac{2}{3} \left(\frac{\partial \tilde{q}_i}{\partial \tilde{x}_i} + \frac{\partial \tilde{v}_i}{\partial \tilde{x}_i} \right) + \xi_0 \left(\tilde{n} + \frac{1}{2} \tilde{T}\right) + \xi_1 \tilde{\Delta} &= 0,
\\
\label{stressBalPertDimless}
\frac{\partial \tilde{\sigma}_{ij}}{\partial \tilde{t}} + \frac{\partial \tilde{m}_{ijk}}{\partial \tilde{x}_k} + \frac{4}{5} \frac{\partial \tilde{q}_{\langle i}}{\partial \tilde{x}_{j \rangle}} + 2 \frac{\partial \tilde{v}_{\langle i}}{\partial \tilde{x}_{j \rangle}}  - \xi_2 \tilde{\sigma}_{ij} + \xi_3 \tilde{R}_{ij} 
&= 0,
\\
\label{HFBalPertDimless}
\frac{\partial \tilde{q}_i}{\partial \tilde{t}} + \frac{1}{2}\frac{\partial \tilde{R}_{ij}}{\partial \tilde{x}_j} + \frac{1}{6}\frac{\partial \tilde{\Delta}}{\partial \tilde{x}_i} + \frac{5}{2} a_2 \frac{\partial \tilde{n}}{\partial \tilde{x}_i} + \xi_4 \frac{\partial \tilde{T}}{\partial \tilde{x}_i} + \frac{\partial \tilde{\sigma}_{ij}}{\partial \tilde{x}_j}  - \xi_5 \tilde{q}_i &= 0,
\\
\label{mijkBalPertDimless}
\frac{\partial \tilde{m}_{ijk}}{\partial \tilde{t}} + \frac{3}{7}\frac{\partial \tilde{R}_{\langle ij}}{\partial \tilde{x}_{k\rangle}} + 3\frac{\partial \tilde{\sigma}_{\langle ij}}{\partial \tilde{x}_{k\rangle}} - \xi_6 \tilde{m}_{ijk} &= 0,
\\
\label{RijBalPertDimless}
\frac{\partial \tilde{R}_{ij}}{\partial \tilde{t}} + \frac{28}{5} \frac{\partial \tilde{q}_{\langle i}}{\partial \tilde{x}_{j\rangle}}
+ 2\frac{\partial \tilde{m}_{ijk}}{\partial \tilde{x}_k} 
+14\,a_2 \frac{\partial \tilde{v}_{\langle i}}{\partial \tilde{x}_{j\rangle}} - \xi_7 \tilde{R}_{ij} - \xi_8 \tilde{\sigma}_{ij} &= 0,
\\
\label{DeltaBalPertDimless}
\frac{\partial \tilde{\Delta}}{\partial \tilde{t}} + \xi_9 \frac{\partial \tilde{q}_i}{\partial \tilde{x}_i}
+\xi_{10} \tilde{\Delta} &= 0,
\end{align}
where the coefficients
\begin{align}
\left.
\begin{alignedat}{4}
\xi_0 &= \zeta_0^\ast,
\qquad &
\xi_1 &= \frac{1-e^2}{192},
\qquad &
\xi_2 &= \xi_0 - \nu_\sigma^\ast,
\qquad &
\xi_3 &= \nu_{\sigma R}^\ast,
\\ 
\xi_4 &= \frac{5}{2}(1 + 2 \, a_2),
\qquad &
\xi_5 &= \frac{3}{2} \xi_0 - \nu_q^\ast,
\qquad &
\xi_6 &= \frac{3}{2} \xi_0 - \nu_m^\ast,
\qquad &
\xi_7 &= 2 \,\xi_0 - \nu_R^\ast,
\\
\xi_8 &= \nu_{R\sigma}^\ast,
\qquad &
\xi_9 &= 8-20\,a_2,
\qquad &
\xi_{10} &= \nu_\Delta^\ast 
\end{alignedat}
\right\}
\end{align}
with
\begin{align}
\left.
\begin{aligned}
\nu_m^\ast&=\frac{3}{8}(1+e)(3-e) \left(1-\frac{3\,a_2}{224}\right),
\\
\nu_\Delta^\ast &= \frac{1}{192}(1+e) (81-17 e+30 e^2-30 e^3)
\end{aligned}
\right\}
\end{align}
depend only on the parameter $e$, the coefficient of restitution. 

Now, we assume a normal mode solution of the form
\begin{align}
\label{normalmodesol}
(\tilde{n}, \tilde{v}_i, \tilde{T}, \tilde{\sigma}_{ij}, \tilde{q}_i, \tilde{m}_{ijk}, \tilde{R}_{ij}, \tilde{\Delta} )^{\top} 
= (\hat{n}, \hat{v}_i, \hat{T}, \hat{\sigma}_{ij}, \hat{q}_i, \hat{m}_{ijk}, \hat{R}_{ij}, \hat{\Delta})^{\top} \, \exp {\big[ \mathbbm{i} ( \bm{k} \cdot \tilde{\bm{x}} - \omega \, \tilde{t}) \big]}
\end{align}
for system \eqref{massBalPertDimless}--\eqref{DeltaBalPertDimless}.
Here, quantities with hats denote the complex amplitudes of the perturbed field variables; $\mathbbm{i}$ is the imaginary unit; $\bm{k}$ and $\omega$ are the (dimensionless) wavevector and (dimensionless) frequency, respectively, of the disturbance. For the temporal stability analysis---to be analyzed here---the wavevector $\bm{k}$ is assumed to be real and the frequency $\omega$ is assumed to be complex. The real part of the complex frequency, $\mathrm{Re}(\omega)$, determines the phase velocity $\bm{v}_{\mathrm{ph}} = \mathrm{Re}(\omega)/\bm{k}$ of the corresponding wave whereas the imaginary part of the complex frequency, $\mathrm{Im}(\omega)$, determines whether the amplitude of the disturbance grows or decays in time. 
The imaginary part of the complex frequency, $\mathrm{Im}(\omega)$, is referred to as the \emph{growth rate}. From the normal mode solution \eqref{normalmodesol}, it is clear that the solution will decay (or grow) in time if the growth rate is negative (or positive). Consequently, stability of the system requires the growth rate to be non-positive, i.e., $\mathrm{Im}(\omega) \leq 0$. 


If we assume that the wavevector of the disturbance is parallel to the $x$-axis, i.e., $\bm{k} = k \, \hat{\bm{x}}$ where the wavenumber $k$ is the magnitude of the wavevector $\bm{k}$ and $\hat{\bm{x}}$ is the unit vector in $x$-direction, we get two independent eigenvalue problems---namely, the longitudinal problem and the transverse problem---for the amplitude of the disturbance in two dimensions. These problems read
\begin{align}
\label{eigvalProbs}
\mathcal{A} \begin{bmatrix}
                \hat{n} \\
                \hat{v}_x \\
                \hat{T} \\
                \hat{\sigma}_{xx} \\
                \hat{q}_x \\
                \hat{m}_{xxx} \\
                \hat{R}_{xx} \\
                \hat{\Delta}
               \end{bmatrix}
= 
\begin{bmatrix}
0 \\
0 \\
0 \\
0 \\
0 \\
0 \\
0 \\
0 
\end{bmatrix}
\qquad\textrm{and}\qquad
\mathcal{B} 
\begin{bmatrix}
\hat{v}_y \\
\hat{\sigma}_{xy} \\
\hat{q}_y \\
\hat{m}_{xxy} \\
\hat{R}_{xy} \\
\end{bmatrix}
= 
\begin{bmatrix}
0 \\
0 \\
0 \\
0 \\
0 
\end{bmatrix}
\end{align}
respectively, where
\begin{align}
\label{matrixA}
\mathcal{A} =
\begin{bmatrix}
\omega & -k & 0 & 0 & 0 & 0 & 0 & 0
\\
-k & \omega - \mathbbm{i} \frac{\xi_0}{2}  & -k & -k & 0 & 0 & 0 & 0
\\
\mathbbm{i} \xi_0 & - \frac{2k}{3} & \omega + \mathbbm{i} \frac{\xi_0}{2} & 0 & - \frac{2k}{3} & 0 & 0 & \mathbbm{i} \xi_1 
\\
0 & -\frac{4k}{3} & 0 & \omega - \mathbbm{i} \xi_2 & -\frac{8k}{15} & -k & \mathbbm{i} \xi_3 & 0 
\\
-\frac{5k a_2}{2} & 0 & -k \xi_4 & -k & \omega - \mathbbm{i} \xi_5 & 0 & -\frac{k}{2} & -\frac{k}{6}  
\\
0 & 0 & 0 & -\frac{9k}{5} & 0 & \omega - \mathbbm{i} \xi_6 & -\frac{9k}{35} & 0  
\\
0 & -\frac{28 k a_2}{3} & 0 & - \mathbbm{i} \xi_8 & -\frac{56 k}{15} & -2k & \omega - \mathbbm{i} \xi_7 & 0  
\\
0 & 0 & 0 & 0 & -k \xi_9 & 0 & 0 & \omega + \mathbbm{i} \xi_{10} 
\end{bmatrix}
,
\end{align}
and
\begin{align}
\label{matrixB}
\mathcal{B} =
\begin{bmatrix}
\omega - \mathbbm{i} \frac{\xi_0}{2} & -k & 0 & 0 & 0
\\
-k & \omega - \mathbbm{i}\xi_2  & -\frac{2 k}{5} & -k & \mathbbm{i}\xi_3 
\\
0 & -k & \omega - \mathbbm{i}\xi_5  & 0 & -\frac{k}{2} 
\\
0 & -\frac{3 k}{5} & 0 & \omega - \mathbbm{i}\xi_6 & -\frac{3 k}{35} 
\\
-7 k a_2 & - \mathbbm{i}\xi_8 & -\frac{14 k}{5} & -2 k & \omega - \mathbbm{i}\xi_7  
\end{bmatrix}
.
\end{align}
%
For nontrivial solution of each eigenvalue problem in \eqref{eigvalProbs}, the determinant of each of the two matrices $\mathcal{A}$ and $\mathcal{B}$ must vanish, i.e., $\mathrm{det}(\mathcal{A})=0$ and $\mathrm{det}(\mathcal{B})=0$. These two conditions on the determinant of matrices $\mathcal{A}$ and $\mathcal{B}$ lead to the following dispersion relations
\begin{align}
\label{dispRelLong}
\omega^8 + \mathbbm{a}_1\, \omega^7 + \mathbbm{a}_2 \, \omega^6 + \mathbbm{a}_3 \, \omega^5 + \mathbbm{a}_4 \, \omega^4 + \mathbbm{a}_5 \, \omega^3 + \mathbbm{a}_6 \, \omega^2 + \mathbbm{a}_7 \, \omega + \mathbbm{a}_8 &= 0, 
\\
\label{dispRelTran}
\omega^5 + \mathbbm{b}_1 \, \omega^4 + \mathbbm{b}_2 \, \omega^3 + \mathbbm{b}_3 \, \omega^2 + \mathbbm{b}_4 \, \omega + \mathbbm{b}_5 &= 0
\end{align}
for the longitudinal and transverse problems \eqref{eigvalProbs}, respectively.
The coefficients $\mathbbm{a}_1,\mathbbm{a}_2,\dots,\mathbbm{a}_8$ and $\mathbbm{b}_1,\mathbbm{b}_2,\dots,\mathbbm{b}_5$ in \eqref{dispRelLong} and \eqref{dispRelTran} are functions of the wavenumber $k$ and the coefficient of restitution $e$; although the explicit values of these coefficients are not given here for brevity. 
\subsection{Eigenmodes}\label{Eigmodes}
\begin{figure}
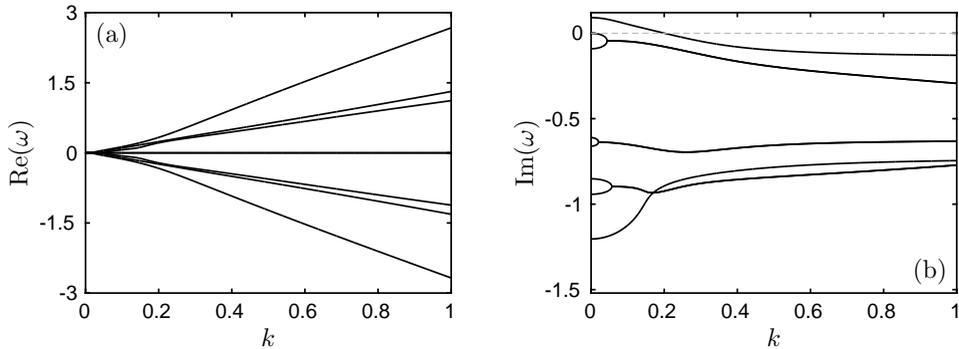

\begin{center}
\includegraphics[scale=0.32]{plots/longRe_e0p75.eps}
\qquad
\includegraphics[scale=0.32]{plots/longIm_e0p75.eps}
\caption{Eigenmodes from the longitudinal system \eqref{eigvalProbs}$_1$ for the coefficient of restitution $e=0.75$: (a) the real part of frequency $\omega$ and (b) the imaginary part of frequency $\omega$ representing the growth rate.
}
\label{fig:long_e0p75}
\end{center}
\end{figure}

\begin{figure}
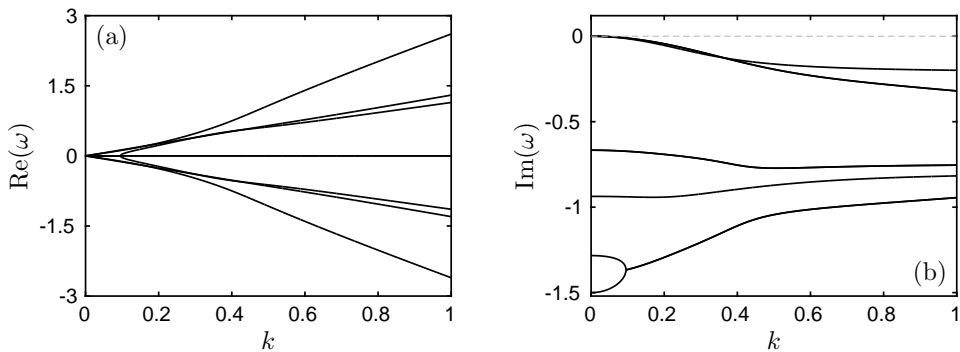

\begin{center}
\includegraphics[scale=0.32]{plots/longRe_e1.eps}
\qquad
\includegraphics[scale=0.32]{plots/longIm_e1.eps}
\caption{Same as figure~\ref{fig:long_e0p75} but in the elastic case, i.e., for $e=1$.
}
\label{fig:long_e1}
\end{center}
\end{figure}

Condition \eqref{dispRelLong} results into eight eigenmodes for the longitudinal system \eqref{eigvalProbs}$_1$. These eigenmodes for the coefficient of restitution $e=0.75$ and $e=1$ are shown in figures~\ref{fig:long_e0p75} and~\ref{fig:long_e1}, respectively. The left panels in each of figures~\ref{fig:long_e0p75} and~\ref{fig:long_e1} depict the real part of frequency, $\mathrm{Re}(\omega)$, while the right panels display the growth rate, 
$\mathrm{Im}(\omega)$, all plotted over the wavenumber $k$. For very small wavenumbers ($k \lesssim 0.02$), all the eigenmodes of the longitudinal system \eqref{eigvalProbs}$_1$ are stationary in the case of $e=0.75$ (see the left panel of figure~\ref{fig:long_e0p75}); however, as the wavenumber starts increasing, a pair of traveling modes commences at $k \approx 0.02$, another pair of slow traveling modes commences at $k \approx 0.043$ and a third pair of even slower traveling modes starts at $k \approx 0.057$. These six eigenmodes (which are traveling for large $k$) are referred to as the \emph{sound modes} \citep{BP2004,Garzo2005}. Each pair of sound modes propagates in the opposite directions (for large $k$) since the corresponding eigenvalues are in complex conjugate pairs. 
The remaining two eigenmodes continue to be stationary for all wavenumbers since the frequencies associated with them are purely imaginary. 
These eigenmodes (which remain stationary for all wavenumbers) are referred to as the \emph{heat modes} \citep{BP2004,Garzo2005}. 
The left panel of figure~\ref{fig:long_e1} illustrates that there are two (heat) eigenmodes which remain stationary for all wavenumbers and three pairs of traveling (sound) eigenmodes in the elastic case (i.e., for $e=1$) as well. Nevertheless, two pairs of  sound modes commence propagating in the opposite directions at $k=0$ itself in this case. These two pairs of sound modes travel almost with the same speeds for $0 \leq k \lesssim 0.18$ whereas for $k \gtrsim 0.18$ one pair of sound modes travels faster than the other. It may also be interesting to note that out of these two pairs of sound modes, the speed of the one pair traveling with slower speed coincides with that of a third pair of sound modes---which starts propagating at $k \approx 0.095$ in the opposite directions with even slower speed---for $0.28 \lesssim k \lesssim 0.45$. It can also be noticed by comparing the left panels of figures~\ref{fig:long_e0p75} and~\ref{fig:long_e1} that the speeds of sound modes in the case of $e=0.75$ (figure~\ref{fig:long_e0p75}) are almost same as those of corresponding sound modes in the elastic case (figure~\ref{fig:long_e1}) for wavenumbers $k \gtrsim 0.7$. 
%

Since the complex eigenvalues of a matrix occur in pairs, the growth rates, $\mathrm{Im}(\omega)$, of a pair of sound modes coincide beyond the wavenumber at which they start propagating.   
This is clearly reflected in right panels of figures~\ref{fig:long_e0p75} and~\ref{fig:long_e1}: the imaginary parts of frequencies, $\mathrm{Im}(\omega)$, of the three pairs of traveling waves---in the left panels $(i)$ starting at $k \approx 0.02$, $k \approx 0.043$ and $k \approx 0.057$ in the case of $e=0.75$ (figure~\ref{fig:long_e0p75}) and $(ii)$ two pairs starting at $k=0$ and one starting at $k \approx 0.095$  in the elastic case (figure~\ref{fig:long_e1})---merge together beyond these wavenumber values
(compare the corresponding left and right panels of figures~\ref{fig:long_e0p75} and~\ref{fig:long_e1}). 
As we have discussed above that non-positive growth rate ($\mathrm{Im}(\omega) \leq 0$) implies stability and vice versa, it is clear from the right panel of figure~\ref{fig:long_e0p75} that for $e=0.75$, one heat mode and all the sound modes are stable for all wavenumbers whereas the other heat mode (which has $\mathrm{Im}(\omega) > 0$ for some wavenumbers) is unstable for small wavenumbers. 
On the other hand, it is evident from the right panel of figure~\ref{fig:long_e1} that in the elastic case (i.e., for $e=1$), all the heat and sound modes of the longitudinal system \eqref{eigvalProbs}$_1$ are stable for all wavenumbers. The unstable heat mode in the case of  $e=0.75$ turns stable beyond $k \approx 0.2005$ (see the right panel of figure~\ref{fig:long_e0p75}) since beyond this value of the wavenumber, $\mathrm{Im}(\omega)$ becomes negative. A value of wavenumber $k$ at which the growth rate, $\mathrm{Im}(\omega)$, switches its sign is referred to as the \emph{critical} wavenumber. In other words, the corresponding eigenmode is unstable (or stable) for the wavenumbers below the critical wavenumber since $\mathrm{Im}(\omega)>0$ (or $\mathrm{Im}(\omega) \leq 0$) for them while it is stable (or unstable) for those above the critical wavenumber since $\mathrm{Im}(\omega) \leq 0$ (or $\mathrm{Im}(\omega)>0$) for them. Thus, $k \approx 0.2005$ is the critical wavenumber for the longitudinal system \eqref{eigvalProbs}$_1$ in the case of $e=0.75$.

\begin{figure}
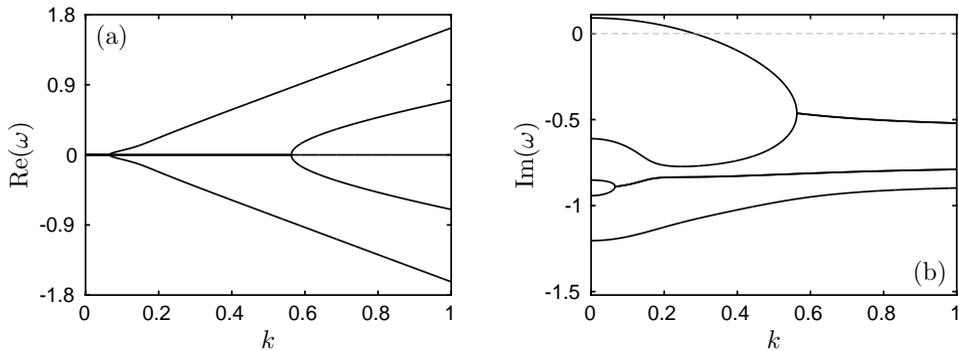

\begin{center}
\includegraphics[scale=0.32]{plots/tranRe_e0p75.eps}
\qquad
\includegraphics[scale=0.32]{plots/tranIm_e0p75.eps}
\caption{Eigenmodes from the transverse system \eqref{eigvalProbs}$_2$ for the coefficient of restitution $e=0.75$: (a) the real part of frequency $\omega$ and (b) the imaginary part of frequency $\omega$ representing the growth rate.
}
\label{fig:tran_e0p75}
\end{center}
\end{figure}

\begin{figure}
\begin{center}
\includegraphics[scale=0.32]{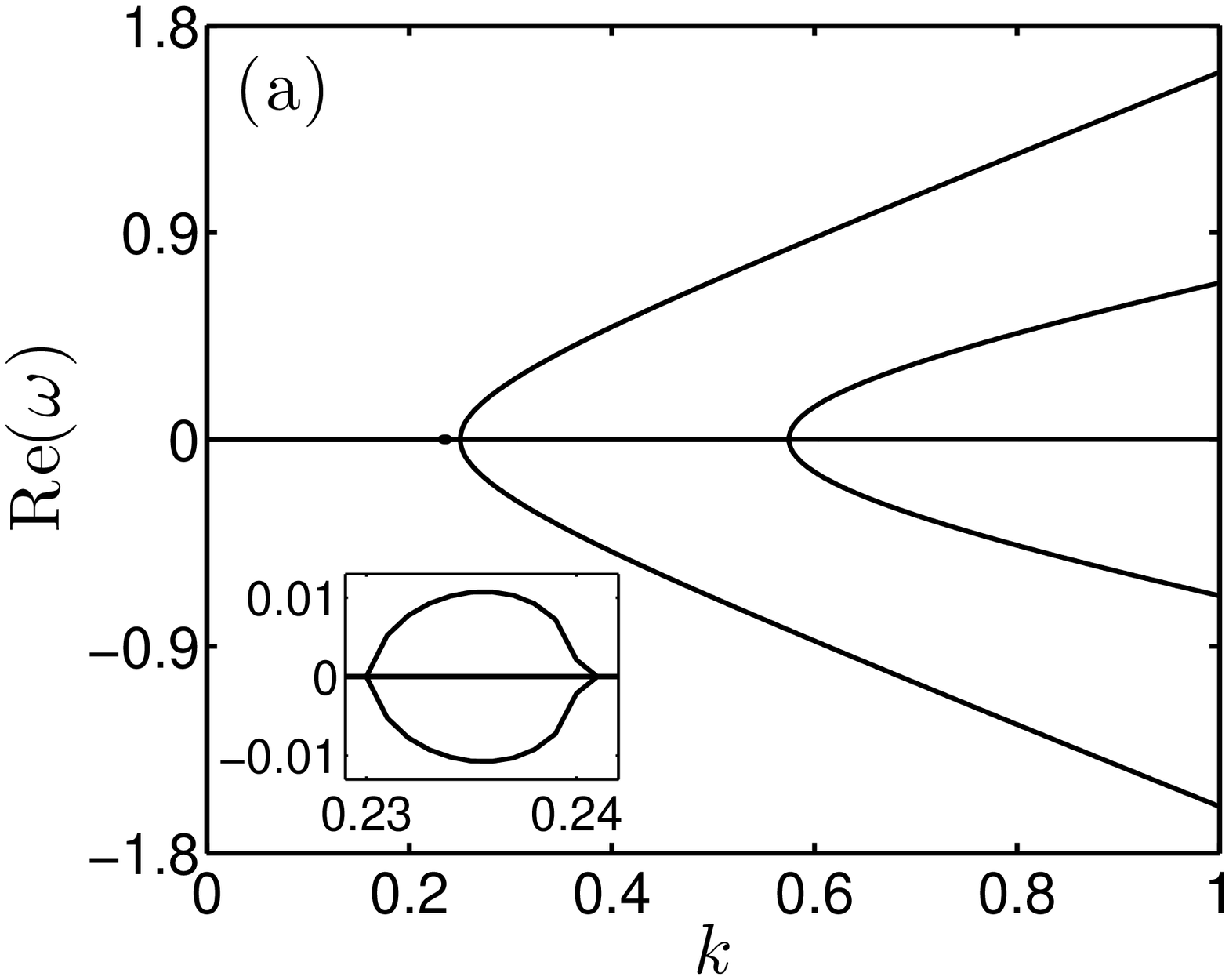}
\qquad
\includegraphics[scale=0.32]{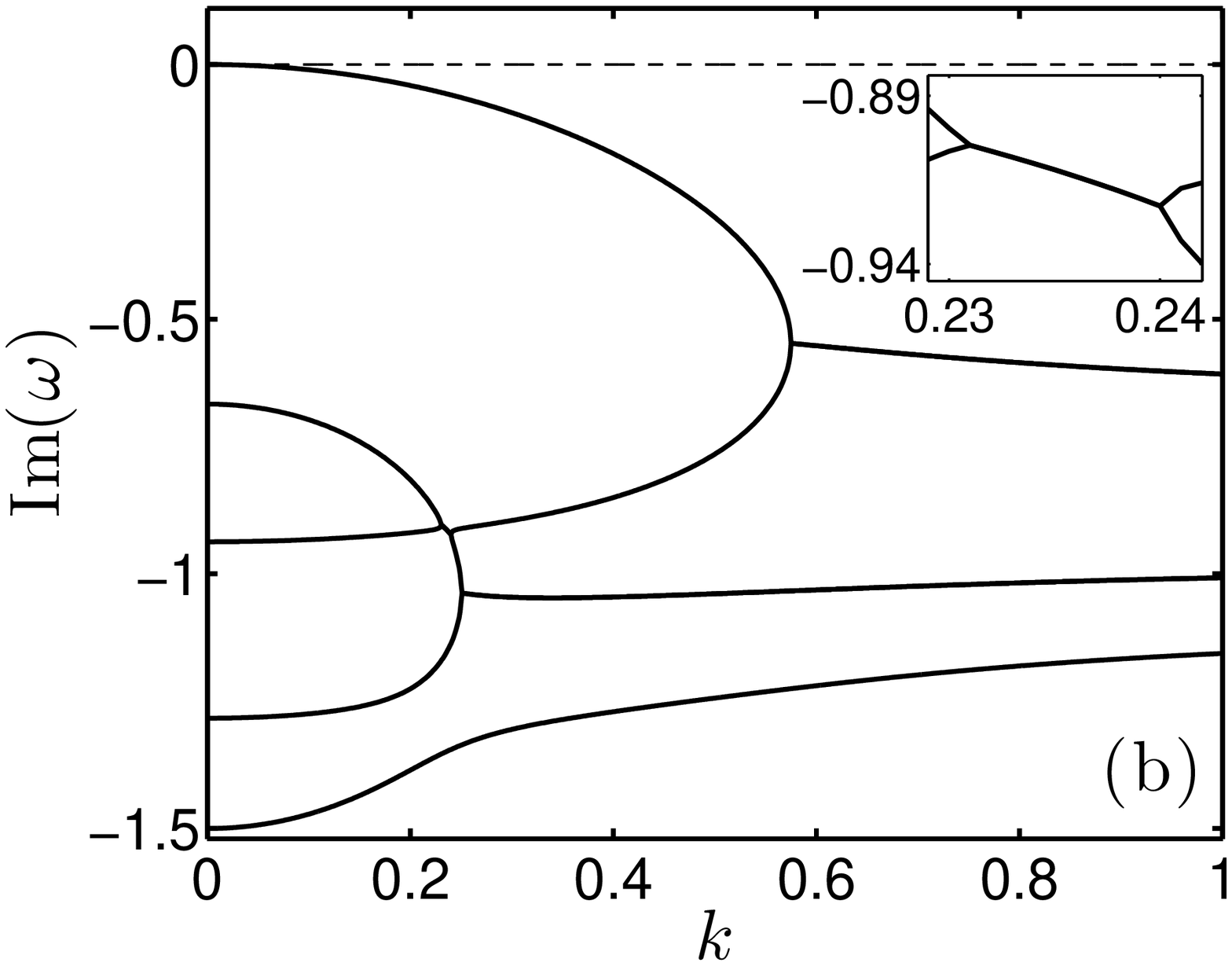}
\caption{Same as figure~\ref{fig:tran_e0p75} but in the elastic case, i.e., for $e=1$. The insets exhibit the zoomed region where a pair of stationary eigenmodes changes to a pair of traveling eigenmodes with increasing wavenumber and turns back to become stationary with further increase in the wavenumber.
}
\label{fig:tran_e1}
\end{center}
\end{figure}

Condition \eqref{dispRelTran} leads to five eigenmodes for the transverse system \eqref{eigvalProbs}$_2$. The eigenmodes of the transverse system \eqref{eigvalProbs}$_2$ are referred to as the \emph{shear modes} \citep{BP2004,Garzo2005}. These shear modes for the coefficient of restitution $e=0.75$ and $e=1$ are illustrated in figures~\ref{fig:tran_e0p75} and~\ref{fig:tran_e1}, respectively. 
The left panels in each of figures~\ref{fig:tran_e0p75} and~\ref{fig:tran_e1} again delineate the real part of frequency, $\mathrm{Re}(\omega)$, while the right panels portray the growth rate, $\mathrm{Im}(\omega)$, all plotted over the wavenumber $k$. For very small wavenumbers ($k \lesssim 0.065$), all the shear modes are stationary in the case of $e=0.75$ (see the left panel of figure~\ref{fig:tran_e0p75}). A pair of traveling shear modes commences at $k \approx 0.065$ and another pair of slow traveling shear modes commences at $k \approx 0.562$. Each pair of traveling shear modes propagates in the opposite directions, and one remaining shear mode continues to be stationary for all wavenumbers since the frequency associated with it is purely imaginary. For the same reason as discussed above, the imaginary parts of frequencies of each pair of traveling shear modes coincide beyond the wavenumbers at which they start propagating (one pair coincides for $k \gtrsim 0.066$ and another for $k \gtrsim 0.563$ in the right panel of figure~\ref{fig:tran_e0p75}). Furthermore, it is noted from the right panel of figure~\ref{fig:tran_e0p75} that one shear mode in the case of $e=0.75$ is also unstable for wavenumber values below the critical wavenumber which is $k \approx 0.2827$ while it is stable for wavenumber values above the critical wavenumber; the remaining four shear modes are always stable for all values of wavenumber $k$.

In the elastic case, for $e=1$ (figure~\ref{fig:tran_e1}), all the shear modes are stationary for small wavenumbers ($k \lesssim 0.23$). A pair of traveling shear modes commences at $k \approx 0.23$ but turns back to become stationary on slight increase in the wavenumber (at $k \approx 0.241$), refer to the zoomed region shown in the insets of figure~\ref{fig:tran_e1}. On further increase in the wavenumber, a pair of traveling shear modes starts at $k \approx 0.25$ and another pair of slow traveling shear modes commences at $k \approx 0.574$. Each pair of traveling shear modes propagates in the opposite directions, and one shear mode remains stationary for all wavenumbers since the frequency associated with it is purely imaginary. For the same reason as discussed above, the imaginary parts of frequencies of each pair of traveling shear modes coincide for the wavenumbers for which they are propagating---one pair coincides for $0.231 \lesssim k \lesssim 0.24$ (see the inset in the right panel of figure~\ref{fig:tran_e1}), one pair for $k \gtrsim 0.251$ and another for $k \gtrsim 0.575$ in the right panel of figure~\ref{fig:tran_e1}). Furthermore, it is noted from the right panel of figure~\ref{fig:tran_e1} that all five shear modes are always stable in the elastic case. It can again be noticed from the left panels of figures~\ref{fig:tran_e0p75} and~\ref{fig:tran_e1} that the speeds of traveling shear modes in the case of $e=0.75$ (figure~\ref{fig:tran_e0p75}) are almost same as those of corresponding traveling shear modes in elastic case (figure~\ref{fig:tran_e1}) for wavenumbers $k \gtrsim 0.6$.%
\subsection{Eigenmodes in small wavenumber limit}
There is another interesting classification of eigenmodes in the small wavenumber limit (or long wavelength limit), i.e., in the limit $k \to 0$: an eigenmode is referred to as a \emph{hydrodynamic} mode if the frequency $\omega(k)$ of this eigenmode vanish in the limit $k \to 0$ while it is referred to as a \emph{kinetic} mode if its frequency $\omega(k)$ attains a nonzero constant value in the limit $k \to 0$ \citep{KM2011}. In order to explore the behavior of the eigenmodes in the small wavenumber limit (i.e., in the limit $k \to 0$), the frequency $\omega$ is expressed in powers of $k$ \cite[see][]{KM2011}:
\begin{align}
\label{OmegaExpansion}
\omega = \gamma_0 + \gamma_1 k + \gamma_2 k^2 + \dots
\end{align}
The unknown coefficients $\gamma_0,\gamma_1,\gamma_2,\dots$ in the above expansion are determined by inserting expansion \eqref{OmegaExpansion} for $\omega$ into \eqref{dispRelLong} and \eqref{dispRelTran},
and solving the algebraic equations resulting from the comparison of coefficients of each power of $k$ on both sides of each equation. 
With this technique, it turns out that in the limit $k \to 0$, the frequencies of eight eigenmodes of the longitudinal system \eqref{eigvalProbs}$_1$ in the inelastic (i.e., $e \neq 1$) case are related to the wavenumber $k$ via 
\begin{align}
\label{analOmegaLong}
\left.
\begin{aligned}
\omega^{(1)} &= - \frac{2 \, \mathbbm{i}}{\xi_0} k^2 + \dots
\\
\omega^{(2)} &= - \frac{\mathbbm{i}\,\xi_0}{2} + \frac{4 \, \mathbbm{i}}{3} \left[\frac{2}{\xi_0} + \frac{\xi_4}{(\xi_0+2\,\xi_5)} \left\{1 + \frac{3\,\xi_1\xi_9}{(\xi_0-2\,\xi_{10})}\right\} \right]
k^2 + \dots
\\
\omega^{(3)} &= \frac{\mathbbm{i}\,\xi_0}{2} - \frac{2 \, \mathbbm{i}}{3} \left[\frac{1}{\xi_0} - \frac{4\,(14 \, a_2 \, \xi_3 - \xi_0 + 2\,\xi_7)}{(\xi_0-2\,\xi_2)(\xi_0-2\,\xi_7) + 4\xi_3\xi_8} \right] k^2 + \dots
\\
\omega^{(4)} &= - \mathbbm{i}\,\xi_{10} + \frac{\mathbbm{i}}{6} \frac{\xi_9}{(\xi_5+\xi_{10})} \left(1 - \frac{12\,\xi_1\xi_4}{\xi_0-2\,\xi_{10}}\right) k^2 + \dots
\\
\omega^{(5)} &= \mathbbm{i}\,\xi_5 - \frac{4 \, \mathbbm{i}}{3} \vartheta\, k^2 + \dots
\\
\omega^{(6)} &= \mathbbm{i}\,\xi_6 - \frac{9 \, \mathbbm{i}}{35} \left[\frac{2\xi_2+14\xi_3-9\xi_6+7\xi_7-\xi_8}{(\xi_2-\xi_6)(\xi_6-\xi_7) - \xi_3\xi_8}\right]k^2 + \dots
\\
\omega^{(7)} &= \frac{\mathbbm{i}\,\xi_{-}}{2} + \frac{\mathbbm{i}}{210}  \left( \frac{\vartheta_1 + \vartheta_2 \xi_{-} }{\vartheta_0} \right) k^2  + \dots
\\
\omega^{(8)} &= \frac{\mathbbm{i}\,\xi_{+}}{2} + \frac{\mathbbm{i}}{210}  \left( \frac{\vartheta_1 + \vartheta_2 \xi_{+} }{\vartheta_0} \right) k^2 + \dots,
\end{aligned}
\right\}
\end{align}
and those in the elastic (i.e., $e = 1$) case are related to the wavenumber $k$ via
\begin{align}
\label{analOmegaLongElastic}
\left.
\begin{aligned}
\omega^{(1)} &= - \frac{3\,\mathbbm{i}}{2} k^2 + \dots
\\
\omega^{(2)} &= - \sqrt{\frac{5}{3}} \, k - \frac{713\,\mathbbm{i}}{606}  k^2 + \dots
\\
\omega^{(3)} &=  \sqrt{\frac{5}{3}} \, k - \frac{713\,\mathbbm{i}}{606}  k^2 + \dots
\\
\omega^{(4)} &= - \frac{2\,\mathbbm{i}}{3} - \frac{2}{\sqrt{3}} k - \frac{59\,\mathbbm{i}}{84}  k^2 + \dots
\\
\omega^{(5)} &= - \frac{2\,\mathbbm{i}}{3} + \frac{2}{\sqrt{3}} k - \frac{59\,\mathbbm{i}}{84}  k^2 + \dots
\\
\omega^{(6)} &=  - \frac{3\,\mathbbm{i}}{2} + \frac{342\,\mathbbm{i}}{41}  k^2 + \dots
\\
\omega^{(7)} &= - \frac{\big(373 - \sqrt{3385}\big)\,\mathbbm{i}}{336}  - \frac{\big(453843875 - 6402969 \sqrt{3385}\big)\,\mathbbm{i}}{294362985} k^2  + \dots
\\
\omega^{(8)} &=- \frac{\big(373 + \sqrt{3385}\big)\,\mathbbm{i}}{336} - \frac{\big(453843875 + 6402969 \sqrt{3385}\big)\,\mathbbm{i}}{294362985} k^2  + \dots
\end{aligned}
\right\}
\end{align}
The unknown constants $\xi_{-}$, $\xi_{+}$, $\vartheta$ and $\vartheta_0$ in \eqref{analOmegaLong} are given by
\begingroup
\allowdisplaybreaks
\begin{align*}
\xi_{-} &= \xi_2 + \xi_7 - \mathbbm{i} \sqrt{4\xi_3\xi_8-(\xi_2 - \xi_7)^2}\,,
\\
\xi_{+} &= \xi_2 + \xi_7 + \mathbbm{i} \sqrt{4\xi_3\xi_8-(\xi_2 - \xi_7)^2}\,,
\\
\vartheta &= \frac{1}{8} \frac{\xi_9}{(\xi_5+\xi_{10})} + \frac{\xi_4}{(\xi_0+2\,\xi_5)} \left[1-\frac{3}{2}\frac{\xi_1\xi_9}{(\xi_5+\xi_{10})}\right] + \frac{1}{5} \left[\frac{7\xi_2+14\xi_3-9\xi_5+2\xi_7-\xi_8}{(\xi_2-\xi_5)(\xi_5-\xi_7) - \xi_3\xi_8}\right],
\\
\vartheta_0 &= \big[(\xi_2 - \xi_7)^2 - 4\xi_3\xi_8\big] 
\big[(\xi_2-\xi_5)(\xi_5-\xi_7) - \xi_3\xi_8\big]
\big[(\xi_2-\xi_6)(\xi_6-\xi_7) - \xi_3\xi_8\big]
\nonumber\\
&\quad \times
\big[(\xi_0-2\xi_2)(\xi_0-2\xi_7) + 4 \xi_3\xi_8\big];
\end{align*}
\endgroup
while the constants $\vartheta_1$ and $\vartheta_2$ in \eqref{analOmegaLong} are too cumbersome to write here.

It is clear from \eqref{analOmegaLong} that, in the limit $k \to 0$, out of eight eigenmodes of the longitudinal system \eqref{eigvalProbs}$_1$, one mode is hydrodynamic while the rest seven modes are kinetic in the inelastic case (i.e., $e \neq 1$); however, from \eqref{analOmegaLongElastic}, it is evident that three modes out of eight eigenmodes of the longitudinal system \eqref{eigvalProbs}$_1$ are hydrodynamic in the elastic case while only the remaining five modes are kinetic. Figure~\ref{fig:long_vs_e} illustrates the real (in the left panel) and imaginary (in the right panel) parts of frequencies $\omega$---associated with the eigenmodes of the longitudinal system \eqref{eigvalProbs}$_1$---plotted over the coefficient of restitution $e$ for $k=0$. The solid lines in figure~\ref{fig:long_vs_e} denote the frequencies obtained directly from the condition $\mathrm{det}(\mathcal{A})=0$ on substituting $k=0$ in the matrix $\mathcal{A}$ while the circles denote the frequencies obtained from the analytic expressions \eqref{analOmegaLong} and \eqref{analOmegaLongElastic} in the limit $k \to 0$; moreover, the numbers in figure~\ref{fig:long_vs_e} represent the numbering of eigenmodes as given in \eqref{analOmegaLong} and \eqref{analOmegaLongElastic}. 
Clearly, the frequencies associated with the eigenmodes of the longitudinal system \eqref{eigvalProbs}$_1$ in the limit $k \to 0$ from both the methods match perfectly well.
From the right panel of figure~\ref{fig:long_vs_e}, one may notice that the magnitude of the imaginary part of frequency, $\mathrm{Im}(\omega)$, for eigenmodes $2$ and $3$ decrease as the coefficient of restitution increases and they vanish for $e=1$; thus, it is concluded that the longitudinal system \eqref{eigvalProbs}$_1$ has three hydrodynamic modes exclusively for the elastic case ($e=1$); in the inelastic case ($e \neq 1$), the longitudinal system \eqref{eigvalProbs}$_1$ always has only a single hydrodynamic mode. 
The left panel of figure~\ref{fig:long_vs_e} shows that for the wavenumber $k=0$ only two eigenmodes ($7$ and $8$) have nonzero real part of frequency for the coefficient of restitution $0 \leq e \lesssim 0.73$ while the rest six eigenmodes ($1,2,\dots,6$) are purely diffusive since the frequencies associated with them are purely imaginary. This can also be seen from \eqref{analOmegaLong}: apart from frequency $\omega^{(1)}$ which is zero in the limit $k \to 0$ for all $e$, the frequencies $\omega^{(2)},\omega^{(3)},\dots,\omega^{(6)}$ are purely imaginary in the limit $k \to 0$; the frequencies $\omega^{(7)}$ and $\omega^{(8)}$ are, in general, complex with nonzero real part---even in the limit $k \to 0$---since $\xi_{-}$ and $\xi_{+}$ may be complex depending on the value of the coefficient of restitution; nonetheless, in the limit $k \to 0$, $\omega^{(7)}$ and $\omega^{(8)}$ also become purely imaginary for $0.73 \lesssim e \leq 1$---it can be seen form \eqref{analOmegaLongElastic} for $e=1$ as well as in the left panel of figure~\ref{fig:long_vs_e}. One can further perceive from the right panel of figure~\ref{fig:long_vs_e} that the imaginary part of frequency of one eigenmode (denoted by number `$3$') is always positive for $e \neq 1$; consequently, one eigenmode of longitudinal system \eqref{eigvalProbs}$_1$ in the inelastic case ($e \neq 1$) is unstable in the limit $k \to 0$ whereas all the other eigenmodes are stable in this limit. Nevertheless, all the  eigenmodes of longitudinal system \eqref{eigvalProbs}$_1$ in the elastic case ($e = 1$) are always stable for any wavenumber (see also the right panel of figure~\ref{fig:long_e1}).

\begingroup
\begin{figure}
\begin{center}
\includegraphics[scale=0.32]{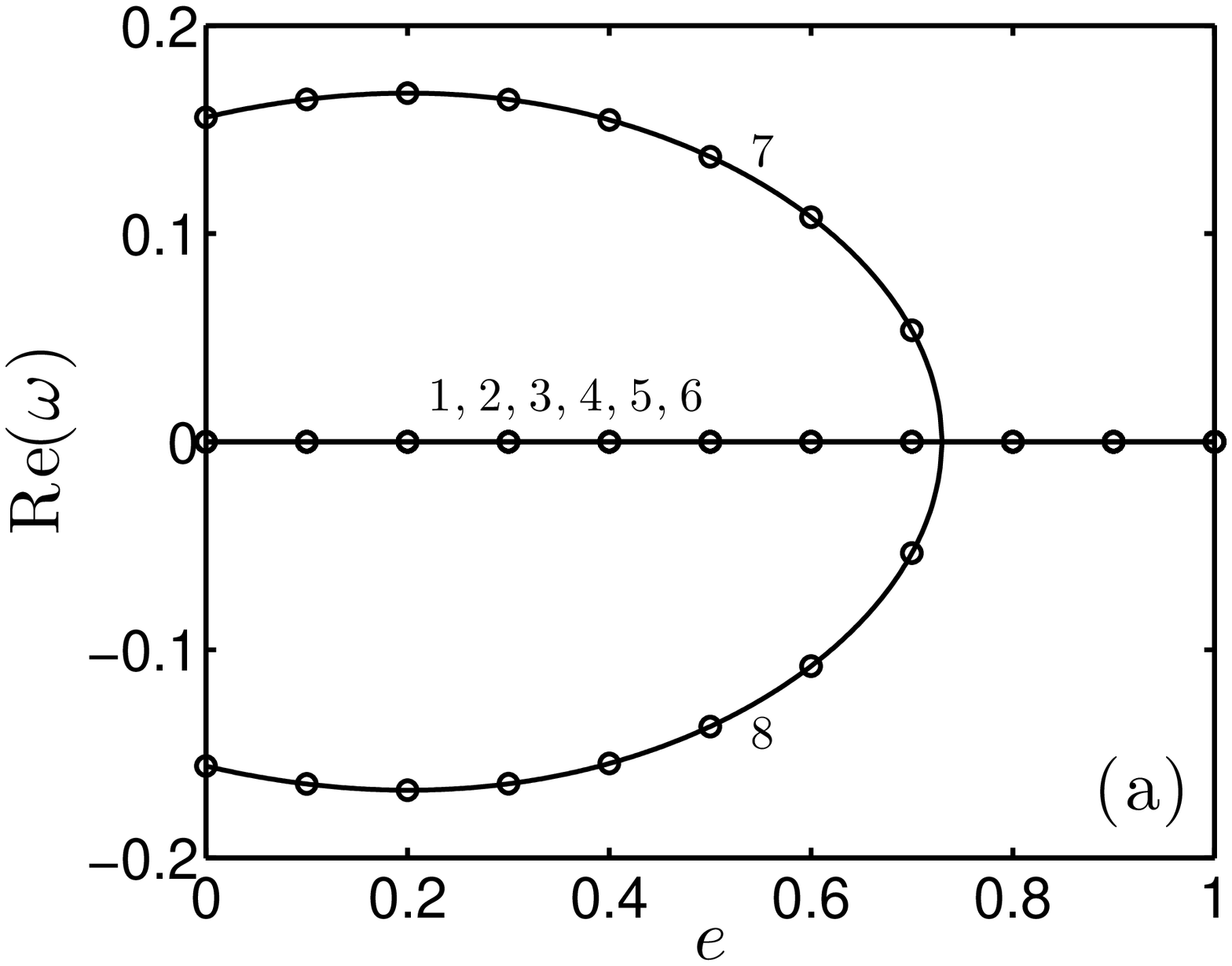}
\qquad
\includegraphics[scale=0.32]{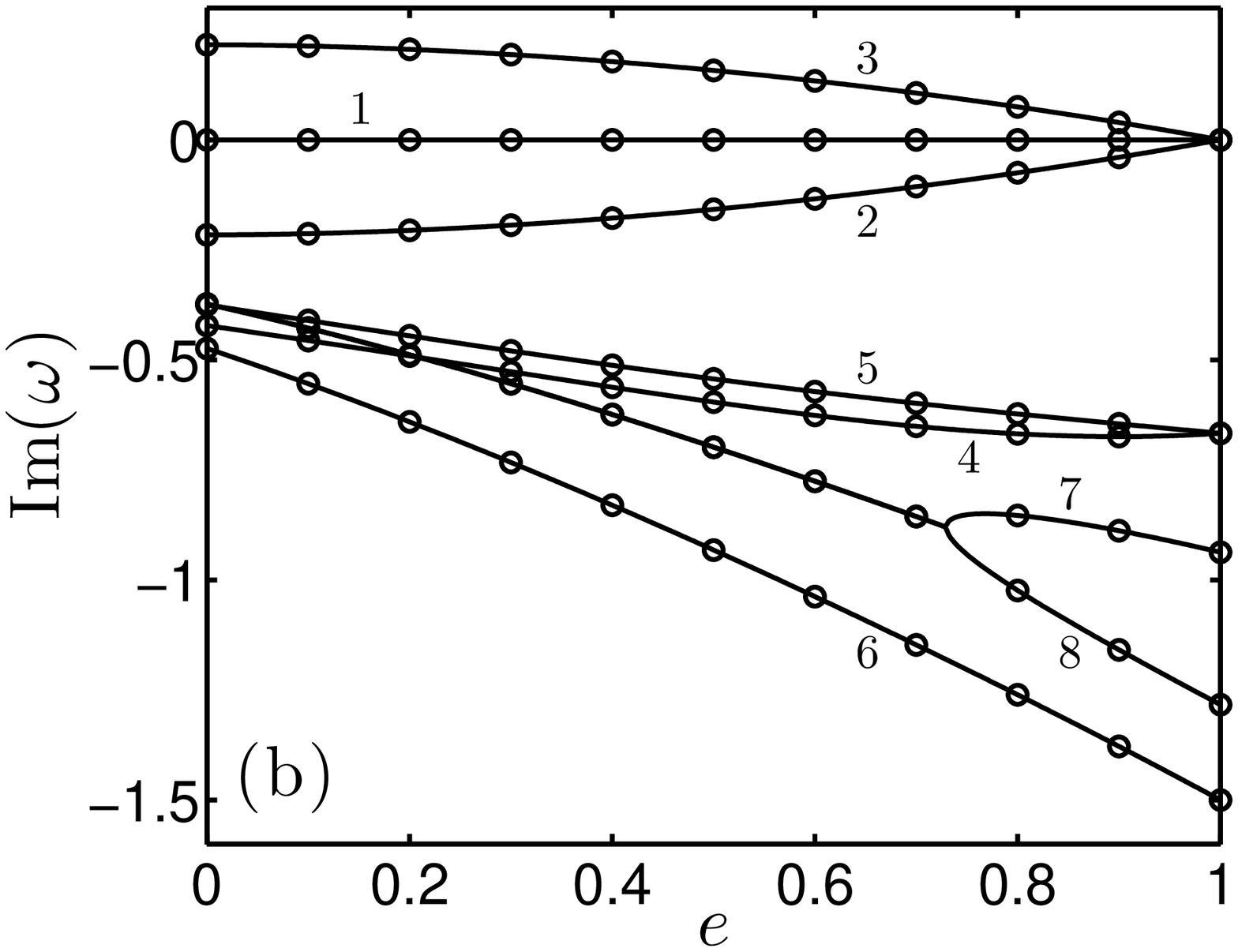}
\caption{Real and imaginary parts of frequencies associated with the eigenmodes of the longitudinal system \eqref{eigvalProbs}$_1$ plotted over the coefficient of restitution $e$ for wavenumber $k=0$. The lines denote the frequencies obtained directly from the condition $\mathrm{det}(\mathcal{A})=0$ on substituting $k=0$ in the matrix $\mathcal{A}$ while the circles denote the frequencies obtained from the analytic expressions \eqref{analOmegaLong} and \eqref{analOmegaLongElastic} in the limit $k \to 0$. The numbers represent the numbering of eigenmodes as given in \eqref{analOmegaLong} and \eqref{analOmegaLongElastic}.
}
\label{fig:long_vs_e}
\end{center}
\end{figure}
\begin{figure}
\begin{center}
\includegraphics[scale=0.32]{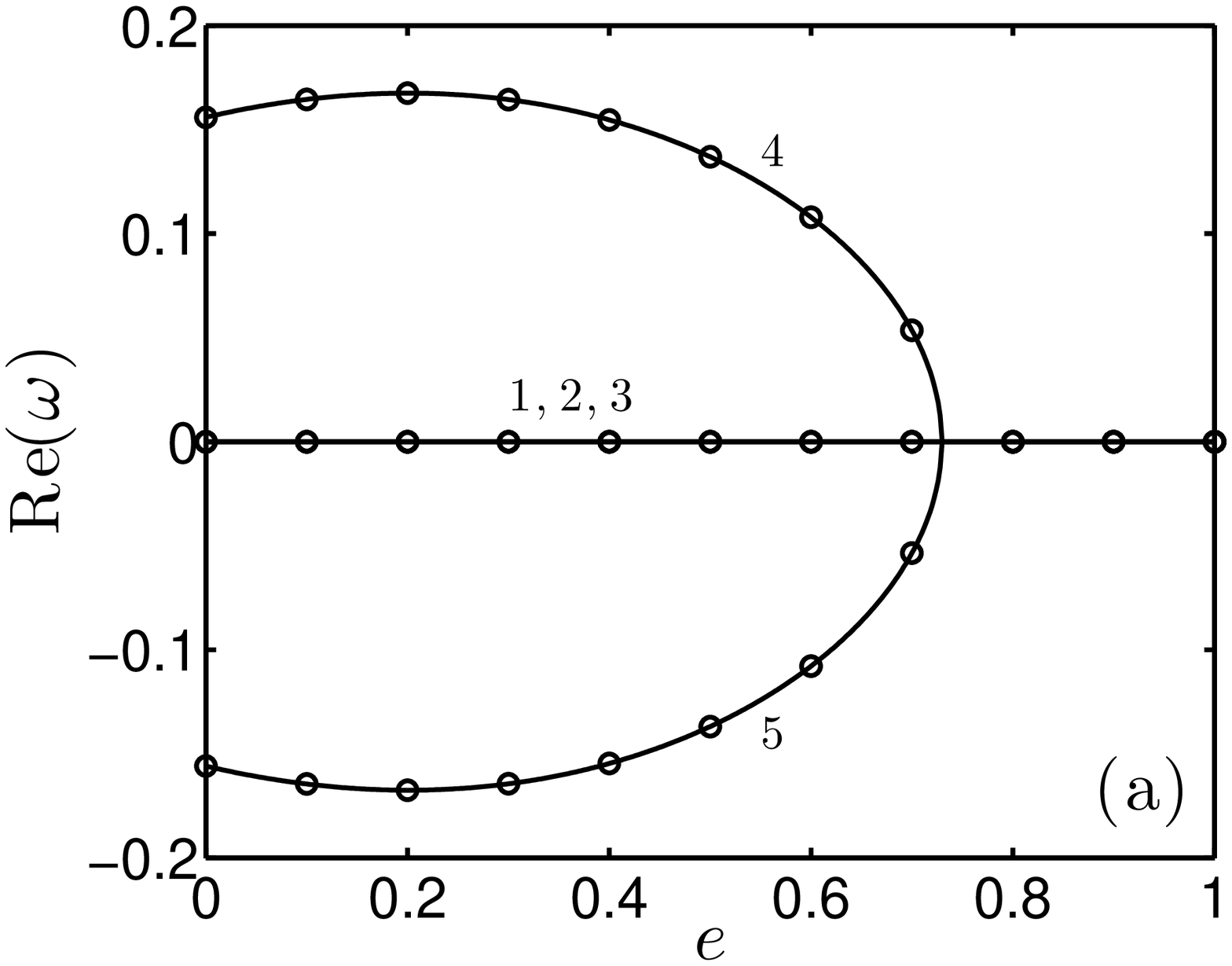}
\qquad
\includegraphics[scale=0.32]{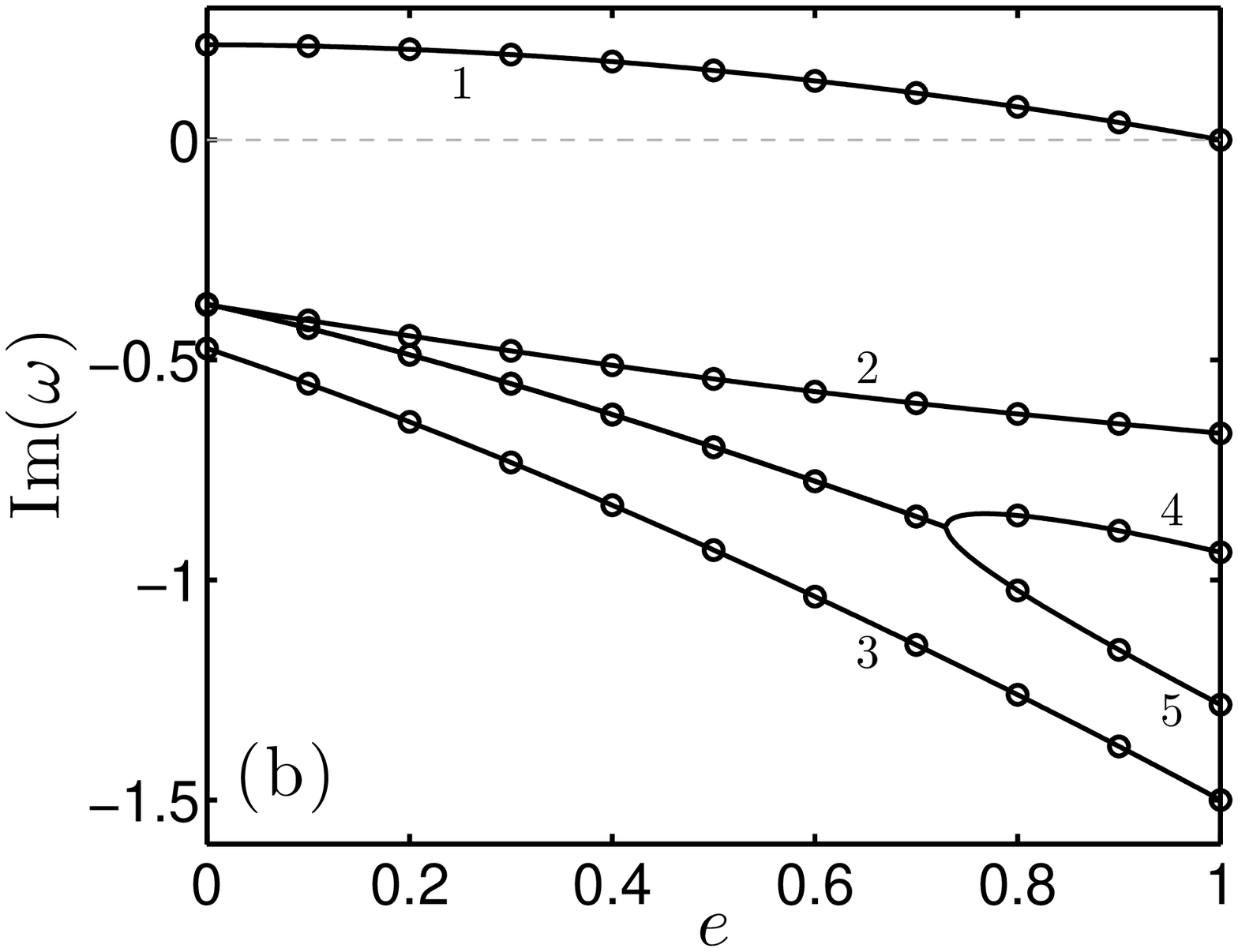}
\caption{Same as figure~\ref{fig:long_vs_e} but for the transverse system \eqref{eigvalProbs}$_2$. The lines denote the frequencies obtained directly from the condition $\mathrm{det}(\mathcal{B})=0$ on substituting $k=0$ in the matrix $\mathcal{B}$ while the circles denote the frequencies obtained from the analytic expressions \eqref{analOmegaTran} and \eqref{analOmegaTranElastic} in the limit $k \to 0$. The numbers represent the numbering of eigenmodes as given in \eqref{analOmegaTran} and \eqref{analOmegaTranElastic}.
}
\label{fig:tran_vs_e}
\end{center}
\end{figure}
\endgroup

By the similar frequency expansion technique employed above, the frequencies of five eigenmodes of the transverse system \eqref{eigvalProbs}$_2$ in the limit $k \to 0$ are related to the wavenumber $k$ in the inelastic ($e\neq 1$) case via 
\begin{align}
\label{analOmegaTran}
\left.
\begin{aligned}
\omega^{(1)} &= \frac{\mathbbm{i}\,\xi_0}{2} + 2 \, \mathbbm{i} \left[\frac{14\,a_2\, \xi_3 - \xi_0 + 2\xi_7}{(\xi_0-2\xi_2)(\xi_0-2\xi_7) + 4\xi_3\xi_8}\right]
k^2 + \dots
\\
\omega^{(2)} &= \mathbbm{i}\,\xi_5 - \frac{\mathbbm{i}}{5} \left[\frac{7\xi_2+14\xi_3-9\xi_5+2\xi_7-\xi_8}{(\xi_2-\xi_5)(\xi_5-\xi_7) - \xi_3\xi_8}\right] k^2 + \dots
\\
\omega^{(3)} &= \mathbbm{i}\,\xi_6 - \frac{3 \, \mathbbm{i}}{35} \left[\frac{2\xi_2+14\xi_3-9\xi_6+7\xi_7-\xi_8}{(\xi_2-\xi_6)(\xi_6-\xi_7) - \xi_3\xi_8}\right]k^2 + \dots
\\
\omega^{(4)} &= \frac{\mathbbm{i}\,\xi_{-}}{2} + \frac{\mathbbm{i}}{210}  \left( \frac{\vartheta_3 + \vartheta_4 \xi_{-} }{\vartheta_0} \right) k^2  + \dots
\\
\omega^{(5)} &= \frac{\mathbbm{i}\,\xi_{+}}{2} + \frac{\mathbbm{i}}{210}  \left( \frac{\vartheta_3 + \vartheta_4 \xi_{+} }{\vartheta_0} \right) k^2 + \dots
\end{aligned}
\right\}
\end{align}
and, in the elastic ($e = 1$) case, these frequencies simplify to
\begin{align}
\label{analOmegaTranElastic}
\left.
\begin{aligned}
\omega^{(1)} &= - \frac{205\,\mathbbm{i}}{202} k^2 + \dots
\\
\omega^{(2)} &= - \frac{2\,\mathbbm{i}}{3} - \frac{41\,\mathbbm{i}}{14}  k^2 + \dots
\\
\omega^{(3)} &=  - \frac{3\,\mathbbm{i}}{2} + \frac{114\,\mathbbm{i}}{41}  k^2 + \dots
\\
\omega^{(4)} &= - \frac{\big(373 - \sqrt{3385}\big)\,\mathbbm{i}}{336}  + \frac{5\big(2282167 - 12093\sqrt{3385}\big)\,\mathbbm{i}}{19624199} k^2  + \dots
\\
\omega^{(5)} &=- \frac{\big(373 + \sqrt{3385}\big)\,\mathbbm{i}}{336}  + \frac{5\big(2282167 + 12093\sqrt{3385}\big)\,\mathbbm{i}}{19624199} k^2  + \dots
\end{aligned}
\right\}
\end{align}
The constants $\vartheta_3$ and $\vartheta_4$ in \eqref{analOmegaTran} are also too cumbersome to write here.

Expressions \eqref{analOmegaTran} and \eqref{analOmegaTranElastic} suggest that, in the limiting case of limit $k \to 0$, all five eigenmodes of the transverse system \eqref{eigvalProbs}$_2$ are kinetic for all values of the coefficient of restitution except in the elastic case (i.e., for $e=1$) for which $\xi_0 = 0$ and consequently, one of the five eigenmodes is hydrodynamic in the elastic limit. Figure~\ref{fig:tran_vs_e} illustrates the real (in the left panel) and imaginary (in the right panel) parts of frequencies $\omega$---associated with the eigenmodes of the transverse system \eqref{eigvalProbs}$_2$---plotted over the coefficient of restitution $e$ for $k=0$. The solid lines in figure~\ref{fig:tran_vs_e} denote the frequencies obtained directly from the condition $\mathrm{det}(\mathcal{B})=0$ on substituting $k=0$ in the matrix $\mathcal{B}$ while the circles denote the frequencies obtained from the analytic expressions \eqref{analOmegaTran} and \eqref{analOmegaTranElastic} in the limit $k \to 0$; moreover, the numbers in figure~\ref{fig:tran_vs_e} represent the numbering of eigenmodes as given in \eqref{analOmegaTran}. 
Clearly, the frequencies associated with the eigenmodes of the transverse system \eqref{eigvalProbs}$_2$ in the limit $k \to 0$ from both the methods match perfectly well.
From the right panel of figure~\ref{fig:tran_vs_e}, one may notice that the imaginary part of frequency, $\mathrm{Im}(\omega)$, for one eigenmode (denoted by `$1$') is non-negative and decreases as the coefficient of restitution increases and, finally, vanishes for $e=1$; therefore, it is concluded that one eigenmode (denoted by `$1$') of the transverse system \eqref{eigvalProbs}$_2$ in the elastic case (i.e., for $e=1$) is hydrodynamic since its real and imaginary parts are zero (cf.~\eqref{analOmegaTranElastic}$_1$) and that this mode of the transverse system \eqref{eigvalProbs}$_2$ in the inelastic case ($e \neq 1$) is unstable in the limit $k \to 0$ whereas all the other eigenmodes are stable in this limit. Nonetheless, all the  eigenmodes of transverse system \eqref{eigvalProbs}$_2$ in the elastic case ($e = 1$) are always stable for any wavenumber (see also the right panel of figure~\ref{fig:tran_e1}).
The left panel of figure~\ref{fig:tran_vs_e} shows that for the wavenumber $k=0$, only two eigenmodes ($4$ and $5$) have nonzero real part of frequency for the coefficient of restitution $0 \leq e \lesssim 0.73$ while the remaining three eigenmodes ($1,2$ and $3$) are purely diffusive since the frequencies associated with them are purely imaginary. This can also be seen from \eqref{analOmegaTran}: the frequencies $\omega^{(1)},\omega^{(2)}$ and $\omega^{(3)}$ are purely imaginary in the limit $k \to 0$; the frequencies $\omega^{(4)}$ and $\omega^{(5)}$ are, in general, complex with nonzero real part---even in the limit $k \to 0$---since $\xi_{-}$ and $\xi_{+}$ may be complex depending on the value of the coefficient of restitution; nonetheless, in the limit $k \to 0$, $\omega^{(4)}$ and $\omega^{(5)}$ also become purely imaginary for $0.73 \lesssim e \leq 1$---it can be seen form \eqref{analOmegaTranElastic} for $e=1$ as well as in the left panel of figure~\ref{fig:tran_vs_e}. %
\subsection{Comparison among various Grad moment theories}\label{CompTheories}
As discussed in \S\,\ref{Subsec:various}, one can obtain a lower-level system of Grad moment equations by dropping the appropriate field variables in the system of G26 equations. In the same way, one can obtain the longitudinal and transverse systems associated with the G13, G14, G20 and G21 equations by dropping the appropriate variables and corresponding rows and columns of the matrices $\mathcal{A}$ and $\mathcal{B}$ in \eqref{eigvalProbs}. Note that the transverse systems for the G13 and G14 equations are same since the field variable $\hat{\Delta}$ does not appear in the list of unknowns of the transverse system (see \eqref{eigvalProbs}$_2$); for the same reason, the transverse systems for the G20 and G21 equations are also same. It is also important to note that the results for the G13 theory presented below are equivalent to those of \cite{KM2011} since they assumed a constant value for the field variable $\Delta$ (given in \eqref{DeltaZerothOrder}) and studied the eigenmodes of the G13 theory essentially.  

\begin{figure}
\begin{center}
\includegraphics[scale=0.32]{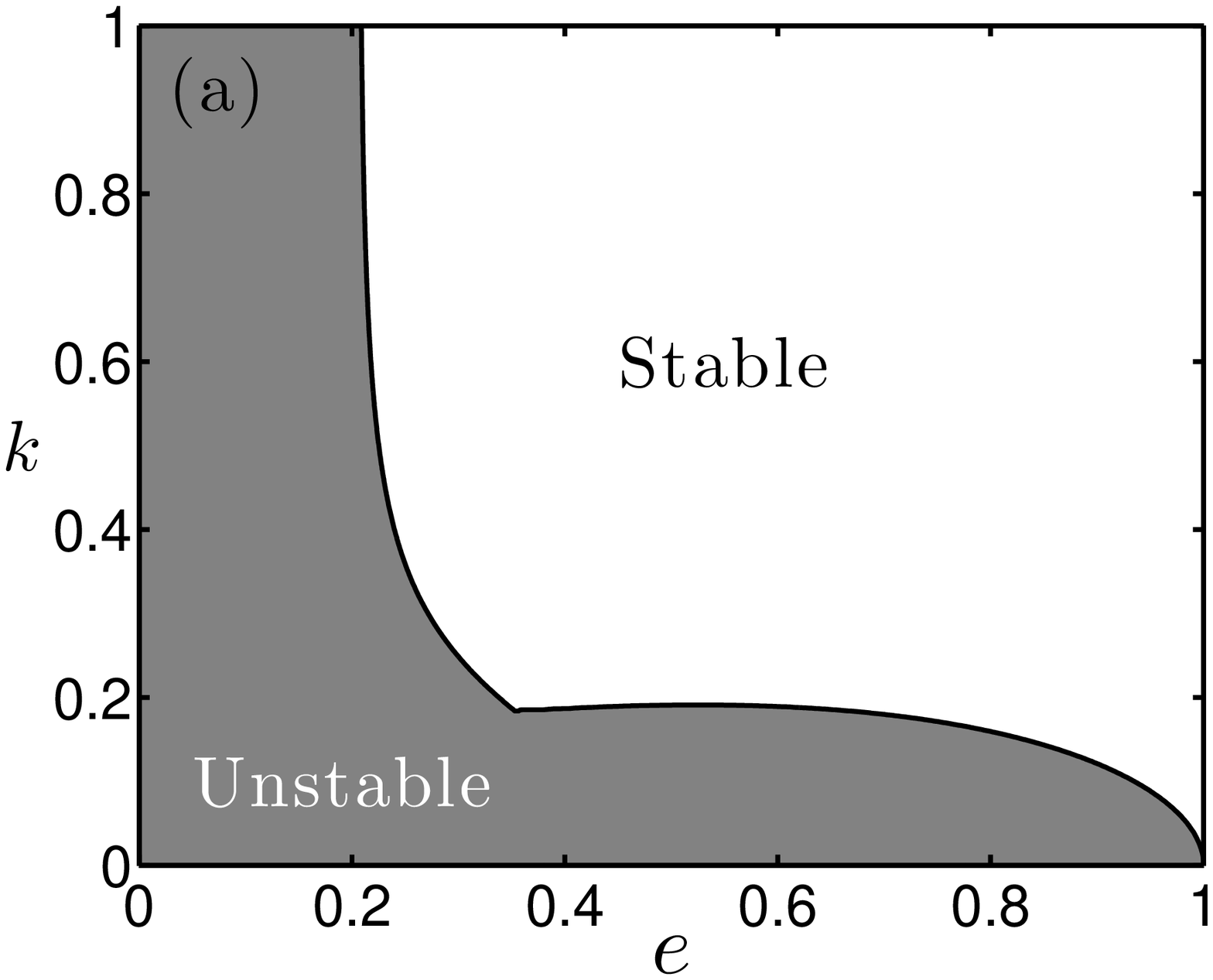}
\qquad
\includegraphics[scale=0.32]{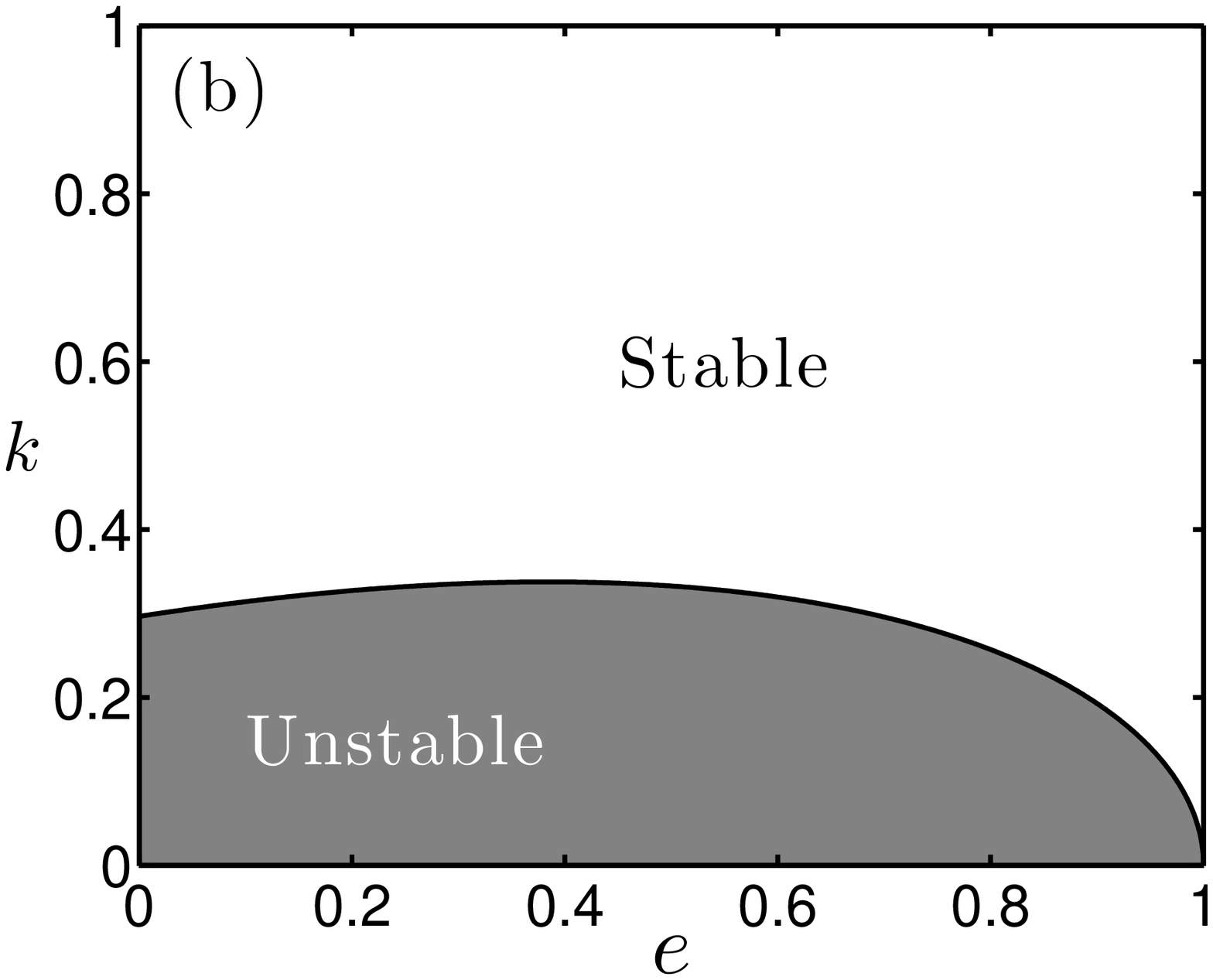}
\caption{Stability diagram in $(e,k)$-plane showing the unstable (in gray color) and stable (in white color) regions for (a) longitudinal and (b) transverse systems associated with G13 theory.
%
}
\label{fig:zero_contours_G13}
\end{center}
\end{figure}
\begin{figure}
\begin{center}
\includegraphics[scale=0.32]{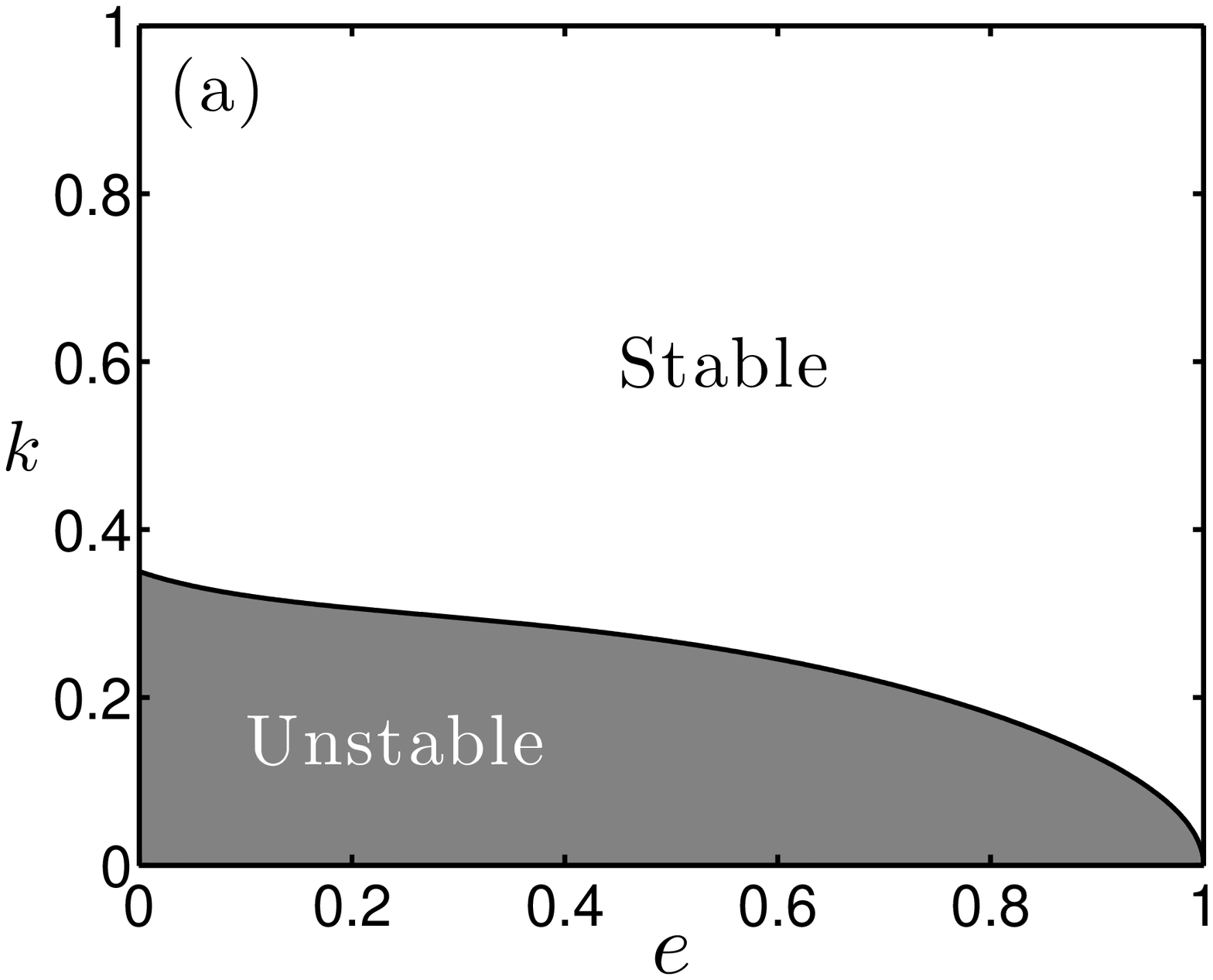}
\qquad
\includegraphics[scale=0.32]{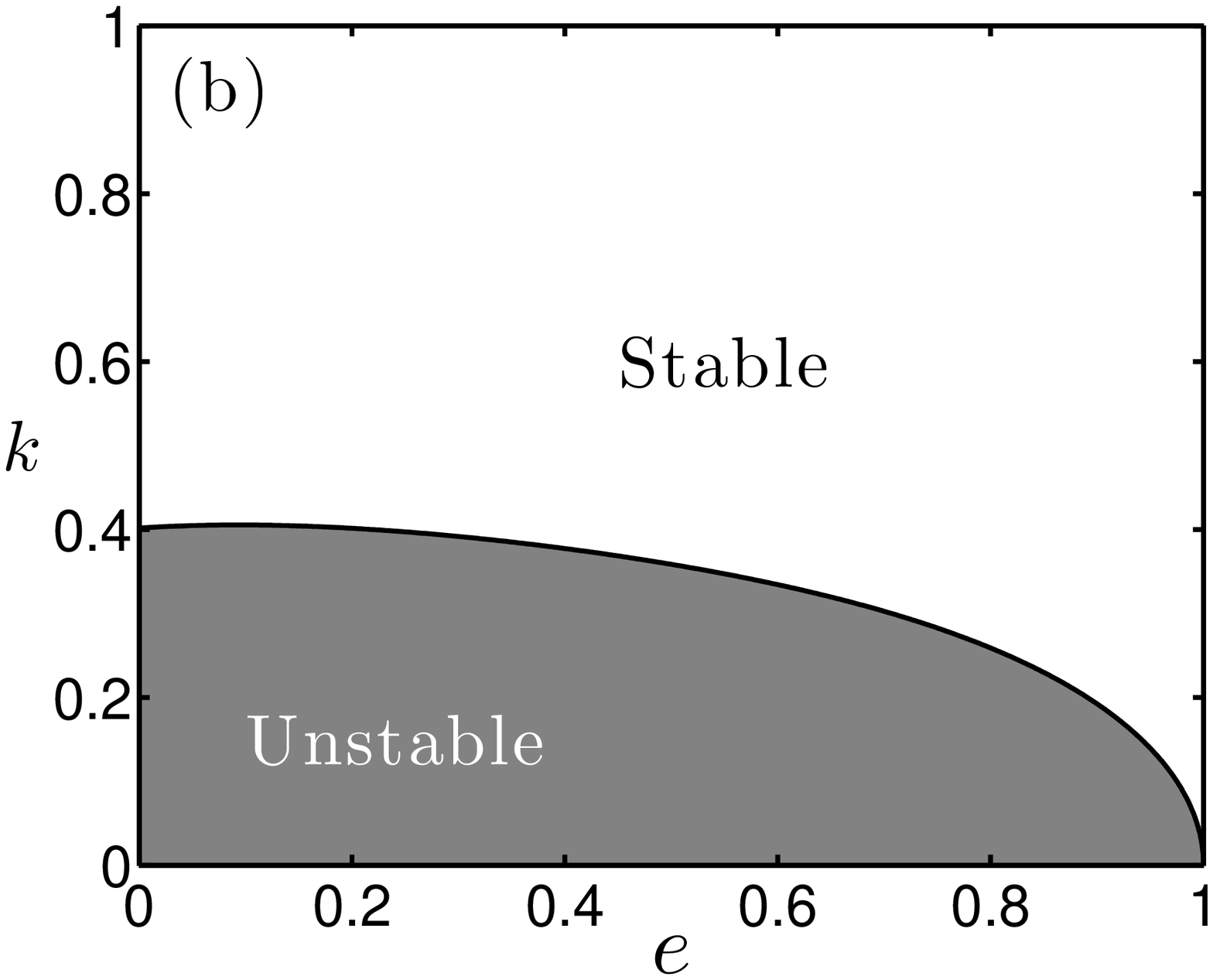}
\caption{
Stability diagram in $(e,k)$-plane showing the unstable (in gray color) and stable (in white color) regions for (a) longitudinal and (b) transverse systems associated with G26 theory.
%
}
\label{fig:zero_contours_G26}
\end{center}
\end{figure}
We have demonstrated in \S\,\ref{Eigmodes} that in the inelastic ($e \neq 1$) case, one eigenmode of both the longitudinal and transverse systems \eqref{eigvalProbs}---a heat mode of the longitudinal system \eqref{eigvalProbs}$_1$ and a shear mode of the transverse system \eqref{eigvalProbs}$_2$---is unstable below some critical wavenumber values (see the right panels of figures~\ref{fig:long_e0p75} and \ref{fig:tran_e0p75}). 
To analyze the stability of the longitudinal and transverse systems \eqref{eigvalProbs}, we investigate the complex frequency of the least stable eigenmode in each system.
We have found (but not shown here) that the real part of the complex frequency of the least stable eigenmode in each system is always zero; therefore, the instabilities of both the longitudinal and transverse systems are stationary. This means that the complex frequency $\omega$ vanishes (i.e., $\omega=0$) for the least stable eigenmode in each of the longitudinal and transverse systems. Consequently, the principle of \emph{exchange of instabilities} \citep{DR1981} is valid for both the systems.%

Figures~\ref{fig:zero_contours_G13} and \ref{fig:zero_contours_G26} illustrate the zero contours of (a) the least stable heat mode of the longitudinal system and (b) the least stable shear mode of the  transverse system associated with the G13 and G26 equations, respectively in $(e,k)$-plane. The black solid line in both the figures represents the critical wavenumber below which the system is unstable (region depicted with gray color) and above which the system is stable (region shown with white color).
%
%
The left panels of figures~\ref{fig:zero_contours_G13} and \ref{fig:zero_contours_G26} unveil that the longitudinal system associated with the G13 equations is unstable for $0 \leq e \lesssim 0.21$ and for $0 \leq k \leq 1$ (in fact, for any wavenumber apparently) whereas that associated with the G26 equations is stable above some critical wavenumber for all values of $e$. Thus, the G13 theory of \citet{KM2011} may not be suitable to the granular flows having coefficient of restitution $e \lesssim 0.35$ (the point where the critical wavenumber from the G13 theory attains a sudden jump). The critical wavenumber in the transverse system associated with the G13 equations (right panel of figure~\ref{fig:zero_contours_G13})---for small values of the coefficient of restitution ($0 \leq e \lesssim 0.4$)---increases with increasing $e$, however, that in the transverse system associated with the G26 equations (right panel of figure~\ref{fig:zero_contours_G26}) decreases with increasing $e$ in the same region. Moreover, the instability region is more in the transverse system associated with the G26 equations than that in the transverse system associated with the G13 equations (cf.~the right panels of figures~\ref{fig:zero_contours_G13} and \ref{fig:zero_contours_G26}). 

\begin{figure}
\begin{center}
\includegraphics[scale=0.32]{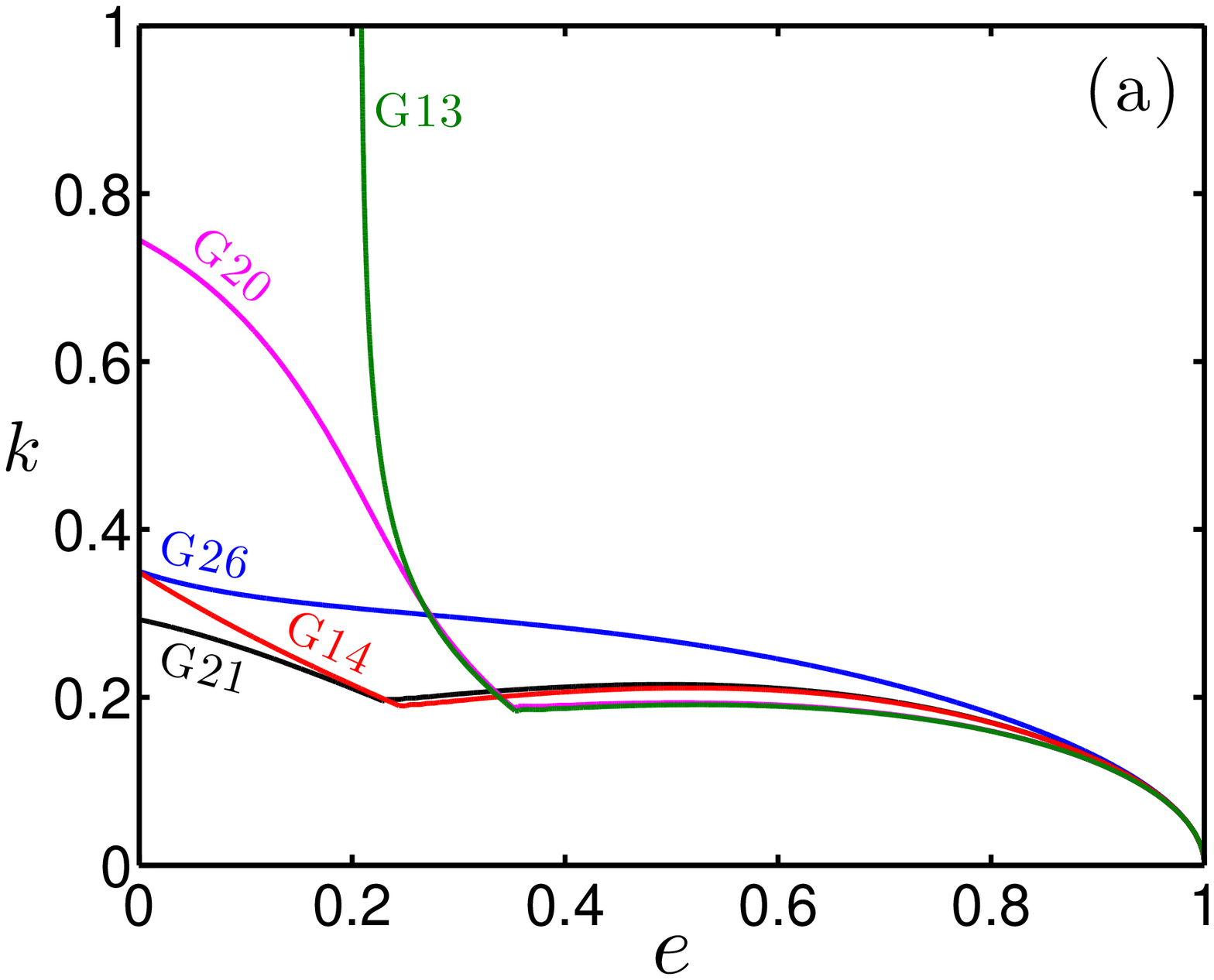}
\qquad
\includegraphics[scale=0.32]{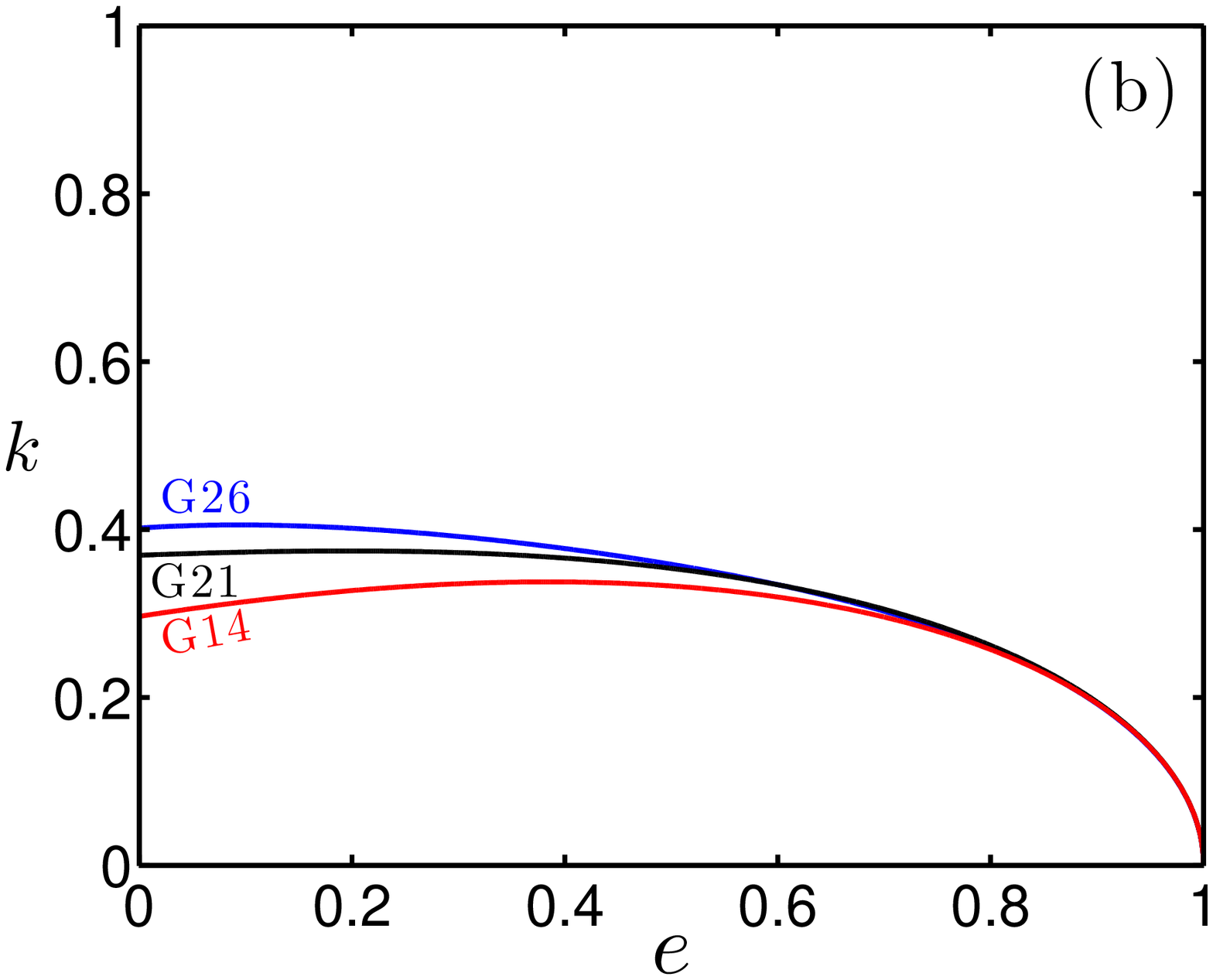}
\caption{Stability diagram in $(e,k)$-plane for (a) longitudinal and (b) transverse 
systems associated with various moment theories (G13, G14, G20, G21, G26).
The system is unstable below or left of the curves while stable above or right of the curves. The curves from G13 theory are equivalent to those from the theory of \citet{KM2011}.
%
}
\label{fig:zero_contours_all}
\end{center}
\end{figure}

In order to compare the critical wavenumbers for the longitudinal and transverse systems associated with various moment theories (G13, G14, G20, G21 and G26), we again plot the zero contours of the least stable eigenmode of the (a) longitudinal and (b) transverse systems associated with these moment theories in $(e,k)$-plane in figure~\ref{fig:zero_contours_all}. Recall that the transverse systems for the G13 and G14 equations  are same and those for the G20 and G21 equations  are also same. Therefore, the curves representing the G14 (in red color) and G21 (in black color) theories in right panel of  figure~\ref{fig:zero_contours_all} also represent the G13 and G20 theories, respectively. Apparently, all moment theories predict same  critical wavenumber for the longitudinal system when $0.94 \lesssim e \leq 1$ (see the left panel of figure~\ref{fig:zero_contours_all}) and same critical wavenumber for the transverse system when $0.75 \lesssim e \leq 1$ (see the right panel of figure~\ref{fig:zero_contours_all}). From the right panel of figure~\ref{fig:zero_contours_all}, we see that for the transverse system, the critical wavenumber profiles predicted by all Grad moment theories are qualitatively similar except for that in the region where $0 \leq e \lesssim 0.4$; in this region, the critical wavenumber predicted by the G14 equations increases with increasing $e$ while that predicted by the G21 equations remains more or less constant and that predicted by the G26 equations decreases with increasing $e$. Also, the stability region for the transverse system is decreasing as the number of moments in the system are increasing. On the other hand, the critical wavenumber profiles for the longitudinal system predicted by various moment theories are quite different from each other, especially for small values of the coefficient of restitution ($0 \leq e \lesssim 0.4$). From the left panel of figure~\ref{fig:zero_contours_all}, it seems that based on the critical wavenumber profiles, one can classify the Grad moment theories into two groups: one containing the G13 and G20 theories and the other containing the G14, G21 and G26 theories. Notice that the first group of theories does not contain the perturbed part of the scalar moment $\tilde{\Delta}$ in contrast to the other group of theories. One can see that the critical wavenumber profiles predicted by the G13 and G20 theories, which do not contain $\tilde{\Delta}$, are very different from those predicted by the G14, G21 and G26 theories,  which contain $\tilde{\Delta}$. The  critical wavenumber from the G20 theory closely follows that from the G13 theory for $0.26 \lesssim e \leq 1$ and they both have a sudden jump at around $e \approx 0.35$; nevertheless, for $0 \leq e \lesssim 0.21$, the G13 theory does not give any critical wavenumber whereas the G20 theory does give critical wavenumbers, which means that the longitudinal system associated with G13 equations is always unstable for $0 \leq e \lesssim 0.21$ while that associated with G20 equations is stable above a critical wavenumber for all value of $e$. This also means that the addition of more moments into the system is stabilizing the longitudinal system associated with the G13 equations. In the other group of theories (G14, G21 and G26), the critical wavenumber profiles from the longitudinal system associated with the G14 and G21 are qualitatively similar---including kinks at $e \approx 0.24$ and $e \approx 0.23$ in critical wavenumber profiles for G14 and G21 theories, respectively---since they both contain $\tilde{\Delta}$ and do not contain $\hat{R}_{ij}$. The instability region for the longitudinal system associated with the G26 equations is also more than that associated with G14 and G21 equations. It may also be stated from the left panel of figure~\ref{fig:zero_contours_all} that the number of moments ought to be increased as the inelasticity increases; while stating this, we have ignored the G20 theory (magenta line in the figure) which is, any way, not meaningful for granular gases since it does not contain the scalar fourth order moment.


\subsection{Critical length}
It is well-known (and have also been discussed above) that the HCS of a freely cooling granular gas is unstable in general, and leads to the formation of velocity vortices (due to the instability of shear mode) and, subsequently, to the formation of density clusters (due to the instability of heat mode). Nevertheless, it is also known that these instabilities are confined to long wavelengths (or to small wavenumbers), and thus may not be observed in systems having small enough system size, \cite[see e.g.,][]{BRM1998, BP2004, Garzo2005}. Here, we want to find the critical system size (length), above which the system will become unstable, through the G26 theory presented in this work.

Recall from \S\,\ref{CompTheories} that the complex frequency $\omega$ is zero for the least stable modes of both the longitudinal and transverse systems. Therefore the critical wavenumber for the least stable shear mode of the longitudinal system and that for  the least stable heat mode of the transverse system can be obtained directly by inserting $\omega=0$ in \eqref{dispRelLong} and \eqref{dispRelTran}, respectively. Inserting $\omega=0$ in \eqref{dispRelLong} and \eqref{dispRelTran}, they yield $\mathbbm{a}_8 = 0$ and $\mathbbm{b}_5 = 0$, respectively. Note that $\mathbbm{a}_8$ is a six degree polynomial in $k$ while $\mathbbm{b}_5$ is a four degree polynomial in $k$. Nevertheless, the solution of $\mathbbm{a}_8 = 0$ leads to only one meaningful value of $k$, which is the critical wavenumber $k_h$ for the least stable heat mode of the longitudinal system \eqref{eigvalProbs}$_1$; similarly, the solution of $\mathbbm{b}_5 = 0$ also leads to only one meaningful value of $k$, which is the critical wavenumber $k_s$ for the least stable shear mode of the transverse system \eqref{eigvalProbs}$_2$. The explicit expressions for $k_h$ and $k_s$, in a compact form, can be written as
\begin{align}
\label{criticalkh} 
k_h &= \frac{1}{3} \sqrt{\frac{\xi_{13} - \sqrt{\xi_{13}^2 - 216 \, \xi_0 \, \xi_5 \, \xi_{10} \, \xi_{11} \, \xi_{12}}}{2 \,\xi_{12}}},
\\
\label{criticalks} 
k_s &= \sqrt{\frac{\xi_{18} -  \sqrt{\xi_{18}^2 - 4 \, \xi_0 \, \xi_5 \, \xi_{11} \, \xi_{17}}}{2 \,\xi_{17}}},
\end{align}
where
\begin{align*}
\xi_{11} &= 35 \, \xi_6 \,(\xi_2 \, \xi_7 + \xi_3 \, \xi_8),
\\
\xi_{12} &= 120 \, (1+a_2) \, \xi_0 \, \xi_{10} - \xi_{14} \, \xi_{15},
\\
\xi_{13} &= \xi_{11} \, \xi_{15} + 2 \, \xi_{16},
\\
\xi_{14} &= 2 \, \xi_2 + 14 \, \xi_3 + 7 \, \xi_7 - \xi_8,
\\
\xi_{15} &= \xi_9 \, \big[\xi_0 + 30 \, (1+a_2) \, \xi_1\big] - 20 \, (1+a_2) \, \xi_{10},
\\
\xi_{16} &= \xi_0 \, \xi_{10} \big[28 \, \xi_6 \, \big\{(8+15 \, a_2) \, (7 \,\xi_3 + \xi_7) - 7 \, \xi_2 + \xi_8\big\} - 27 \, \xi_5 \, \xi_{14}\big],
\\
\xi_{17} &= 15 \, \xi_0 + 6 \, (2-7 \, a_2) \, \xi_5 + 98 \, (1-a_2) \, \xi_6,
\\
\xi_{18} &= 35 \, \xi_6 \, (\xi_0 \, \xi_7 - \xi_0 \, \xi_2 - 2 \, \xi_5 \, \xi_7-14 \, a_2 \, \xi_3 \, \xi_5) - (3\, \xi_5 + 7 \,\xi_6) \, \xi_0 \, \xi_{14}.
\end{align*}
The plots of the  critical wavenumbers $k_h$ and $k_s$ over the coefficient of restitution $e$ are exactly same as the solid line in figure~\ref{fig:zero_contours_G26}~(a) and (b), respectively. Thus, a heat mode of the longitudinal system \eqref{eigvalProbs}$_1$ in the regime $k>k_h$ will always decay while that in the regime $k<k_h$ will grow exponentially; similarly a shear mode of the transverse system \eqref{eigvalProbs}$_2$ in the regime $k>k_s$ will always decay while that in the regime $k<k_s$ will grow exponentially. Notice from figure~\ref{fig:zero_contours_G26} that $k_s>k_h$ for all values of the coefficient of restitution.

In the normal mode analysis considered above, the wavevector is assumed to be in the $x$-direction; thus the $x$-direction is periodic with period $2\pi/k$. Consequently, the smallest admissible wavenumber in a system imposed with periodic boundary conditions is $2\pi/L$, where $L$ is the largest system size (length). Hence, corresponding to the highest critical wavenumber, one can determine a critical system size (length) $L_c$ such that the system is stable for $L<L_c$ while unstable for $L>L_c$. In other words, the critical length $L_c$ for the system considered in this work is determined by
\begin{align}
\frac{2\pi}{\tilde{L}_c}=\max{\{k_h,k_s\}}, \quad\textrm{where}\quad \tilde{L}_c= \frac{L_c}{\ell}
\end{align}
is the dimensionless critical length and $\ell$ is the length scale defined in \eqref{ell} (recall that the wavenumber---and hence $k_h$ and $k_s$---are dimensionless and the length scale for making them dimensionless was $\ell$). Since we have concluded above that $k_s>k_h$, the critical length $L_c$ is given by
\begin{align}
\label{criticalLength}
L_c = \frac{2\pi}{k_s} \times \ell = \frac{2\pi}{k_s} \times \frac{5\sqrt{\pi}}{8\sqrt{2}} \ell_0,
\end{align}
where $\ell_0 = 1/(\sqrt{2}\,\pi n_0 d^2)$ is the mean free path of a hard-sphere dilute gas.

\begin{figure}
\centering
\includegraphics[scale = 0.4]{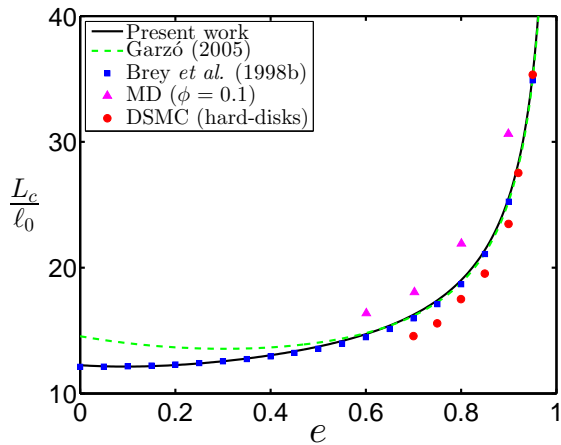}
\caption{The critical length in units of the mean free path $\ell_0$ plotted over the coefficient of restitution $e$. The solid (black) line denotes the critical length computed from \eqref{criticalLength} and \eqref{criticalks} while the dashed (green) line and the squares denote the critical lengths computed from the theoretical expressions obtained in \cite{Garzo2005} and \cite{BRM1998}, respectively. The circles depict the two-dimensional DSMC simulation results of \cite{BRM1998} while the triangles represent the molecular dynamic simulation results of \cite{MDCPH2011} at solid fraction $\phi=0.1$.
}
\label{fig:Critical_length}
\end{figure}

Figure~\ref{fig:Critical_length} illustrates the critical length of the system in units of mean free path ($L_c/\ell_0$) as a function of the coefficient of restitution $e$. The solid (black) line denotes the critical length (in units of mean free path) computed from \eqref{criticalLength} with expression \eqref{criticalks} for $k_s$, which has been obtained through the G26 theory presented above. The dashed (green) line denotes the critical length (in units of mean free path) computed from theoretical expression obtained by the linear stability analysis of Navier--Stokes and Fourier equations in \cite{Garzo2005}. It is evident from the figure that the results on critical length from the G26 (black solid line) and Navier--Stokes--Fourier (green dashed line) theories agree for $0.55\lesssim e \leq 1$ while differ significantly for more inelastic systems as the latter is not reliable for the systems with large inelasticity. The squares delineate the critical length (in units of mean free path) computed from the theoretical expression given in \cite{BRM1998}, which is obtained by the linear stability analysis of a kinetic model due to \citet{BDS1997}. Clearly, the results from the G26 theory and those from the kinetic model by \citet{BDS1997} agree perfectly for all values of $e$. 

The simulation results (depicted by circles and triangles) in figure~\ref{fig:Critical_length} are included only to assess the results obtained from the G26 theory qualitatively.
The circles depict the results of \emph{two-dimensional} DSMC simulations carried out by \citet{BRM1998}. One can see in \cite{BRM1998} that these simulation results are in excellent agreement with the theoretical results from the linear stability of the kinetic model by \citet{BDS1997}, and the latter (shown by squares) are in perfect agreement with those from the G26 theory (black solid line). Therefore it can be expected that the results from full three-dimensional DSMC simulation of a dilute granular gas on critical length would agree with those predicted by the G26 theory (black solid line). The triangles represent the results of   molecular dynamic simulations carried out by \cite{MDCPH2011} at solid fraction $\phi=0.1$. Note that the results from the G26 theory presented above are in the dilute limit ($\phi\to 0$). One may notice from \cite{MDCPH2011} that their results are also in excellent agreement with those obtained from the theoretical expression of \cite{Garzo2005} at $\phi=0.1$ for $0.5\leq e\leq 1$, and the latter (green dashed line)---for $\phi=0$---again agree perfectly with those obtained from the G26 theory for $0.5\leq e\leq 1$. Therefore, it is stated that the G26 theory also provides the correct critical length for all values of the coefficient of restitution. 
\section{Conclusion}\label{Sec:Conclusion}
Higher-order Grad moment equations---up to first 26-moments---for granular gases have been derived by employing the Grad's method of moments to the inelastic Boltzmann equation, and the production terms associated with the derived moment equations have been presented. The production terms have, in fact, been computed and presented for the G27 system (which consists of the first 26-moments and the fully contracted sixth moment). The production terms associated with the G26 equations can easily be computed by dropping the terms containing the fully contracted sixth moment in the given production terms associated with the G27 equations. The HCS of a freely cooling granular gas has been explored with the G26 equations in a semi-linearized setting and it has been found that the temperature decay in the HCS closely follows Haff's law while the other higher-order moments decay on a faster time scale. Further, the nonlinear terms of fully contracted fourth moment have also been included and, by exploiting the stability analysis of fixed points in a dynamical system, it has been shown that some of the fixed points of the system are unstable and with the only stable fixed point it has been concluded that the nonlinear terms, indeed, have only negligible effect on Haff's law. The G27 equations has also been scrutinized and the stability analysis of fixed points in a dynamical system has again been exploited to deduce that even the inclusion of scalar sixth order moment into the G26 system has negligible effect on Haff's law. 
By following the approach of \cite{Garzo2013}, the transport coefficients in the Navier--Stokes and Fourier laws for dilute granular gases have been determined through the G26 equations and compared with those obtained from various theoretical methods and computer simulations. 
The reduced shear viscosity obtained in the present work has been found to be in good agreement with that from the computer-aided method of \cite{NBSG2007} and with the DSMC simulations of \cite{GSM2007} for almost all values of the coefficient of restitution. In fact, the present work yields slightly better results for the reduced shear viscosity in comparison to the modified Sonine approach \citep{GSM2007}. However, in contrast to the modified Sonine approach \citep{GSM2007}, which yields good results for all the transport coefficients, the other two transport coefficients from the present work agree only with the first Sonine approximation, hence overestimate the corresponding coefficients for $e\lesssim 0.7$. 
This suggests to include the full trace of fifth moment into the G26 system in order to get coupling in the semi-linearized RHS of the heat flux balance equation. It is expected that the inclusion of the full trace of fifth moment---i.e., Grad 29-moment system---would improve the reduced transport coefficients $\kappa^\ast$ and $\lambda^\ast$, although this problem will be considered somewhere else in future.
%

The linear stability of the HCS has been analyzed through the G26 system and various sub-systems by decomposing them into the longitudinal and transverse systems. It has been found that a heat mode of the longitudinal system and a shear mode of the transverse system in the case of an inelastic gas are unstable whereas all the eigenmodes for both the systems are stable in the case of an elastic gas. By comparing the least stable eigenmodes from various theories, it has been established that the unstable heat mode of the longitudinal system obtained with the 13-field eigenmode theory of \citet{KM2011} remains unstable for all wavenumbers below a certain coefficient of restitution whereas that obtained with any other higher-order moment theory becomes stable above some critical wavenumber for all values of the coefficient of restitution.
This is apparently an artefact of assuming the dimensionless fourth moment as a constant in \cite{KM2011} rather than considering it as a field variable while analyzing the eigenmodes.
Out of all the theories considered, the G26 theory have produced the smoothest critical wavenumber profile. The critical wavenumber profiles from various Grad moment theories also suggested that the number of moments ought to be increased with increasing inelasticity. 
Further investigation of the critical wavenumbers from the G26 theory has unveiled that the value of the critical length---beyond which the system becomes unstable---is essentially driven by the unstable shear mode of the transverse system. The critical length profile obtained from the G26 theory is in excellent agreement with those obtained by the existing theories and also in qualitatively good agreement with those from the simulations.
%
\section*{Acknowledgments}
The authors acknowledge anonymous reviewers for their valuable suggestions, which significantly improved the paper. VKG gratefully acknowledges financial support from the SRM University in the form of seed grant through the project ``Grain". 
PS acknowledges financial support from IIT Madras in the form of New Faculty Initiation Grant (MAT/15-16/833/NFIG/PRIY) and New Faculty Seed Grant (MAT/16-17/671/NFSC/PRIY). 
\appendix
\section{Production terms}\label{App:ProdTerms}
The fully nonlinear production terms associated with the G27 equations using definition $\Delta=w/(\rho\,\theta^2)$ are given below. The corresponding production terms associated with the G26 equations (eqs.~\eqref{massBal}--\eqref{energyBal} and \eqref{eqn:stress}--\eqref{eqn:Delta}) are obtained from \eqref{P1}--\eqref{P2} on taking $\Xi=0$ and ignoring \eqref{P3}.
\begin{flalign}
\label{P1}
\mathcal{P}^1=&-\frac{5(1-e^2)}{4}\nu\,\rho\,\theta \bigg(1+\frac{1}{80}\Delta - \frac{1}{6720} \Xi +\frac{1}{25600} \Delta^2 + \frac{1}{5160960} \Xi^2 - \frac{1}{215040} \Delta\,\Xi
\nonumber\\
& + \frac{1}{40}\frac{\sigma_{ij}\sigma_{ij}}{\rho^2\theta^2} +\frac{1}{200}\frac{q_i q_i}{\rho^2\theta^3}+\frac{1}{1680}\frac{m_{ijk} m_{ijk}}{\rho^2\theta^3} +\frac{3}{31360}\frac{R_{ij} R_{ij}}{\rho^2\theta^4} -\frac{1}{560}\frac{\sigma_{ij} R_{ij}}{\rho^2\theta^3} \bigg),&
\end{flalign}
\begin{flalign}
\label{Pij0}
\mathcal{P}_{ij}^0=&-\frac{(1+e)(3-e)}{4}\nu \bigg[\bigg(1-\frac{1}{480}\Delta + \frac{1}{13440} \Xi \bigg)\sigma_{ij}
\nonumber\\
&+\frac{1}{28}\bigg(1+\frac{1}{160}\Delta - \frac{1}{2688} \Xi\bigg)\frac{R_{ij}}{\theta}
+\frac{1}{14}\frac{\sigma_{k\langle i}\sigma_{j\rangle k}}{\rho\,\theta}+\frac{1}{100}\frac{q_{\langle i} q_{j\rangle}}{\rho\,\theta^2}&
\nonumber\\
&+\frac{1}{504}\frac{m_{kl\langle i} m_{j\rangle kl}}{\rho\,\theta^2}+\frac{3}{10976}\frac{R_{k\langle i} R_{j\rangle k}}{\rho\,\theta^3}-\frac{1}{196}\frac{\sigma_{k\langle i} R_{j\rangle k}}{\rho\,\theta^2}+\frac{1}{140}\frac{m_{ijk} q_k}{\rho\,\theta^2}\bigg],&
\end{flalign}
\begin{flalign}
\label{Pi1}
\frac{1}{2}\mathcal{P}_i^1=&-\frac{(1+e)}{48}\nu \bigg[\bigg\{(49-33e)+\frac{(19-3e)}{480}\Delta - \frac{(53-21 e)}{13440} \Xi \bigg\}q_i+\frac{3(7+e)}{10}\frac{\sigma_{ij} q_j}{\rho\,\theta}
&\nonumber\\
&+\frac{(13-21e)}{280}\frac{R_{ij} q_j}{\rho\,\theta^2}
+\frac{(11-27e)}{28}\frac{m_{ijk}\sigma_{jk}}{\rho\,\theta}+\frac{(23+9e)}{784}\frac{m_{ijk} R_{jk}}{\rho\,\theta^2}\bigg],&
\end{flalign}
\begin{flalign}
\label{Pijk0}
\mathcal{P}_{ijk}^0=&-\frac{3(1+e)(3-e)}{8}\nu \bigg[\bigg(1-\frac{1}{1120}\Delta + \frac{1}{94080} \Xi\bigg) m_{ijk} 
&\nonumber\\
&- \frac{1}{70}\frac{q_{\langle i}\sigma_{jk\rangle}}{\rho\,\theta} + \frac{1}{280}\frac{q_{\langle i} R_{jk\rangle}}{\rho\,\theta^2}+\frac{1}{14}\frac{\sigma_{l\langle i} m_{jk\rangle l}}{\rho\,\theta}-\frac{1}{1176}\frac{R_{l\langle i} m_{jk\rangle l}}{\rho\,\theta^2}\bigg],&
\end{flalign}
\begin{flalign}
\label{Pij1}
\mathcal{P}_{ij}^1=&-\frac{(1+e)}{336}\nu\bigg[\bigg\{(499 - 288 e + 66 e^2 - 30 e^3)+\frac{(137 - 36 e - 66 e^2 + 30 e^3)}{480}\Delta 
&\nonumber\\
&- \frac{(215 - 90 e - 66 e^2 + 30 e^3)}{13440} \Xi\bigg\} R_{ij}
+28\bigg\{(87-54 e+22 e^2-10 e^3)
&\nonumber\\
&-\frac{(55-18 e-66 e^2+30e^3)}{480}\Delta - \frac{(9 + 22 e^2 - 10 e^3)}{13440} \Xi\bigg\} \,\theta\,\sigma_{ij}
&\nonumber\\
&+2\,(44+27 e+66 e^2-30 e^3)\frac{\sigma_{k\langle i}\sigma_{j\rangle k}}{\rho}+\frac{7(4 - 15 e - 66 e^2 + 30 e^3)}{25}\frac{q_{\langle i} q_{j\rangle}}{\rho\,\theta}
&\nonumber\\
&+\frac{(28 - 45 e - 22 e^2 + 10 e^3)}{6}\frac{m_{kl\langle i} m_{j\rangle kl}}{\rho\,\theta} +\frac{(116 - 99 e - 66 e^2 + 30 e^3)}{392}\frac{R_{k\langle i} R_{j\rangle k}}{\rho\,\theta^2}
&\nonumber\\
&+\frac{(44+27 e+66 e^2-30 e^3)}{7}\frac{\sigma_{k\langle i} R_{j\rangle k}}{\rho\,\theta}+\frac{(169 - 66 e^2 + 30 e^3)}{5}\frac{m_{ijk} q_k}{\rho\,\theta}\bigg],&
\end{flalign}
\begin{flalign}
\label{P2}
\mathcal{P}^2=&-\frac{5(1+e)}{4}\nu\,\rho\,\theta^2\bigg[(1-e)(9+2 e^2)+\frac{(271-207 e+30 e^2-30 e^3)}{240}\Delta 
&\nonumber\\
&+ \frac{(181 - 117 e + 10 e^2 - 10 e^3)}{6720} \Xi + \frac{(137 - 9 e - 30 e^2 + 30 e^3)}{230400}\left(\Delta^2 - \frac{1}{14} \Delta\,\Xi \right)
&\nonumber\\
&+ \frac{(91 - 27 e - 10 e^2 + 10 e^3)}{36126720} \Xi^2 
+\frac{(23+9 e+30 e^2-30 e^3)}{120} \left(\frac{\sigma_{ij}\sigma_{ij}}{\rho^2\theta^2} + \frac{1}{14} \frac{\sigma_{ij} R_{ij}}{\rho^2\theta^3}\right)
&\nonumber\\
& + \frac{(61 + 3 e - 30 e^2 + 30 e^3)}{600}\frac{q_i q_i}{\rho^2\theta^3} 
+\frac{(7 - 39 e - 10 e^2 + 10 e^3)}{1680}\frac{m_{ijk} m_{ijk}}{\rho^2\theta^3} 
&\nonumber\\
&+\frac{(113 - 81 e - 30 e^2 + 30 e^3)}{94080}\frac{R_{ij} R_{ij}}{\rho^2\theta^4}
\bigg],&
\end{flalign}
\begin{flalign}
\label{P3}
\mathcal{P}^3=&-\frac{15(1+e)}{16}\nu\,\rho\,\theta^3 \bigg[(1-e)(115+44 e^2+8 e^4)&
\nonumber\\
&+\frac{(8297-6889 e+2852 e^2-2340 e^3+280 e^4-280 e^5)}{240}\Delta 
\nonumber\\
&+\frac{(16841 - 12617 e + 5476 e^2 - 3940 e^3 + 280 e^4 - 280 e^5)}{6720}\Xi
\nonumber\\
& - \frac{(551 + 217 e + 1084 e^2 - 60 e^3 - 280 e^4 + 280 e^5)}{76800} \left(\Delta^2 - \frac{1}{42} \Delta\,\Xi \right) 
&\nonumber\\
&+ \frac{(3113 - 1193 e - 2396 e^2 + 860 e^3 + 280 e^4 - 280 e^5)}{180633600} \Xi^2 
&\nonumber\\
&+\frac{(281 + 423 e + 516 e^2 - 260 e^3 + 280 e^4 - 280 e^5)}{120} \left(\frac{\sigma_{ij}\sigma_{ij}}{\rho^2\theta^2} + \frac{3}{14} \frac{\sigma_{ij} R_{ij}}{\rho^2\theta^3} \right)
&\nonumber\\
&+ \frac{(685 - 45 e + 996 e^2 + 540 e^3 - 840 e^4 + 840 e^5)}{600}\frac{q_i q_i}{\rho^2\theta^3}
&\nonumber\\
&+\frac{(95 - 415 e - 148 e^2 - 620 e^3 - 280 e^4 + 280 e^5)}{1680}\frac{m_{ijk} m_{ijk}}{\rho^2\theta^3}
&\nonumber\\
&+\frac{(1641 + 215 e - 796 e^2 + 540 e^3 + 280 e^4 - 280 e^5)}{31360}\frac{R_{ij} R_{ij}}{\rho^2\theta^4}
\bigg],
\end{flalign}
where
\begin{align}
\nu=\frac{16}{5}\sqrt{\pi}\,n\, d^2\sqrt{\theta}
\end{align}
is the collision frequency. 
\section{Governing equations for perturbed field variables}\label{App:PertSys}
Inserting the field variables from \eqref{perturbations} into the G26 equations \eqref{massBalSimp}--\eqref{eqn:DeltaSimp} (without the underlined term) and neglecting all the nonlinear terms in perturbations, one obtains the following system of linear partial differential equations in the perturbed field variables.
\begin{align}
\label{massBalPert}
\frac{\partial \tilde{n}}{\partial t}+ v_H(t) \frac{\partial \tilde{v}_i}{\partial x_i}=0,
\end{align}
\begin{align}
\label{momentBalPert}
\frac{\partial \tilde{v}_i}{\partial t} + v_H(t) \left[\frac{\partial \tilde{\sigma}_{ij}}{\partial x_j}  + \frac{\partial \tilde{n}}{\partial x_i} + \frac{\partial \tilde{T}}{\partial x_i}\right] - \frac{1}{2}\xi_0 \nu_H(t) \, \tilde{v}_i=0,
\end{align}
\begin{align}
\label{energyBalPert}
&\frac{\partial \tilde{T}}{\partial t}+ \frac{2}{3} v_H(t) \left[\frac{\partial \tilde{q}_i}{\partial x_i} + \frac{\partial \tilde{v}_i}{\partial x_i} \left(1+\frac{\sigma_{ij}^{(H)}(t)}{n_0 \, T_H(t)} \right) \right] + \nu_H(t) \left[\xi_0 \left(\tilde{n} + \frac{1}{2} \tilde{T}\right) + \frac{(1-e^2)}{192} \tilde{\Delta}\right] = 0,
\end{align}
\begin{align}
\label{stressBalPert}
&\frac{\partial \tilde{\sigma}_{ij}}{\partial t} + v_H(t) \! \left[\frac{\partial \tilde{m}_{ijk}}{\partial x_k} + \frac{4}{5} \frac{\partial \tilde{q}_{\langle i}}{\partial x_{j \rangle}} + 2 \frac{\partial \tilde{v}_{\langle i}}{\partial x_{j \rangle}} + \frac{\sigma_{ij}^{(H)}(t)}{n_0 \,  T_H(t)} \frac{\partial \tilde{v}_k}{\partial x_k} + 2\frac{\sigma_{k\langle i}^{(H)}(t)}{n_0 \,  T_H(t)} \frac{\partial \tilde{v}_{j\rangle}}{\partial x_k}\right] \! - \xi_0 \nu_H(t) \, \tilde{\sigma}_{ij} 
\nonumber\\
&= - \nu_H(t) \! \left[\nu_\sigma^\ast \left\{\!\tilde{\sigma}_{ij} + \frac{\sigma_{ij}^{(H)}(t)}{n_0 \,  T_H(t)}  \left(\tilde{n} + \frac{1}{2}\tilde{T}\right) \!\right\}
\! + \! \nu_{\sigma R}^\ast \left\{\!\tilde{R}_{ij} + \frac{R_{ij}^{(H)}(t)}{n_0 \,  T_H(t)\,v_H^2(t)}  \left(\tilde{n} - \frac{1}{2}\tilde{T}\right) \!\right\}
\right]\nonumber\\
&\quad -\frac{(1+e)(3-e)}{4\times 160} \nu_H(t) \left(\frac{1}{28}  \frac{R_{ij}^{(H)}(t)}{n_0 \,  T_H(t)\,v_H^2(t)} - \frac{1}{3} \frac{\sigma_{ij}^{(H)}(t)}{n_0 \,  T_H(t)}\right)\tilde{\Delta},
\end{align}
\begin{align}
\label{HFBalPert}
&\frac{\partial \tilde{q}_i}{\partial t} + v_H(t) \left[\frac{1}{2}\frac{\partial \tilde{R}_{ij}}{\partial x_j} + \frac{1}{6}\frac{\partial \tilde{\Delta}}{\partial x_i} + \frac{5}{2} a_2 \left(\frac{\partial \tilde{n}}{\partial x_i} + 2 \frac{\partial \tilde{T}}{\partial x_i}\right) + \frac{\partial \tilde{\sigma}_{ij}}{\partial x_j} + \frac{5}{2}\frac{\sigma_{ij}^{(H)}(t)}{n_0 \, T_H(t)} \frac{\partial \tilde{T}}{\partial x_j} 
\right.
\nonumber\\
&\left.+ \frac{5}{2} \frac{\partial \tilde{T}}{\partial x_i} +\frac{m_{ijk}^{(H)}(t)}{n_0 \, T_H(t) \, v_H(t)} \frac{\partial \tilde{v}_j}{\partial x_k} - \frac{\sigma_{ij}^{(H)}(t)}{n_0 \, T_H(t)} \left(\frac{\partial \tilde{\sigma}_{jk}}{\partial x_k} + \frac{\partial \tilde{n}}{\partial x_j}\right) +\frac{7}{5} \frac{q_{i}^{(H)}(t)}{n_0 \, T_H(t) \, v_H(t)} \frac{\partial \tilde{v}_j}{\partial x_j} \right.
\nonumber\\
&\left.+ \frac{q_{j}^{(H)}(t)}{n_0 \, T_H(t) \, v_H(t)} \left(\frac{7}{5} \frac{\partial \tilde{v}_i}{\partial x_j} + \frac{2}{5} \frac{\partial \tilde{v}_j}{\partial x_i}\right) \right] - \frac{3}{2} \xi_0 \nu_H(t) \,\tilde{q}_i
\nonumber\\
&= - \nu_H(t)\,\nu_q^\ast \left[\tilde{q}_i + \frac{q_{i}^{(H)}(t)}{n_0 \, T_H(t) \, v_H(t)} \left(\tilde{n} + \frac{1}{2}\tilde{T}\right)\right]
\nonumber\\
&\quad -\frac{(1+e)(19-3e)}{48\times 480} \nu_H(t) 
\frac{q_{i}^{(H)}(t)}{n_0 \, T_H(t) \, v_H(t)}\tilde{\Delta},
\end{align}
\begin{align}
\label{mijkBalPert}
&\frac{\partial \tilde{m}_{ijk}}{\partial t} + v_H(t) \left[\frac{3}{7}\frac{\partial \tilde{R}_{\langle ij}}{\partial x_{k\rangle}}+3\frac{\partial \tilde{\sigma}_{\langle ij}}{\partial x_{k\rangle}} - 3\frac{\sigma_{\langle ij}^{(H)}(t)}{n_0 \, T_H(t)}\left(\frac{\partial \tilde{\sigma}_{k\rangle l}}{\partial x_l} + \frac{\partial \tilde{n}}{\partial x_{k\rangle}}\right) + \frac{m_{ijk}^{(H)}(t)}{n_0 \, T_H(t) \, v_H(t)} \frac{\partial \tilde{v}_l}{\partial x_l}\right.
\nonumber\\
&\left.+ 3 \frac{m_{l \langle ij}^{(H)}(t)}{n_0 \, T_H(t) \, v_H(t)} \frac{\partial \tilde{v}_{k\rangle}}{\partial x_l} + \frac{12}{5} \frac{q_{\langle i}^{(H)}(t)}{n_0 \, T_H(t) \, v_H(t)} \frac{\partial \tilde{v}_j}{\partial x_{k\rangle}}
\right] - \frac{3}{2} \xi_0 \nu_H(t) \, \tilde{m}_{ijk}
\nonumber\\
&
=- \nu_H(t) \, \nu_m^\ast \left[\tilde{m}_{ijk} + \frac{m_{ijk}^{(H)}(t)}{n_0 \, T_H(t) \, v_H(t)} \left(\tilde{n} + \frac{1}{2}\tilde{T}\right)\right]
\nonumber\\
&\quad +\frac{3(1+e)(3-e)}{8 \times 1120} \nu_H(t) \frac{m_{ijk}^{(H)}(t)}{n_0 \, T_H(t) \, v_H(t)}\tilde{\Delta},
\end{align}
\begin{align}
\label{RijBalPert}
&\frac{\partial \tilde{R}_{ij}}{\partial t} + v_H(t) \left[\frac{R_{ij}^{(H)}(t)}{n_0 \,  T_H(t)\,v_H^2(t)} \frac{\partial \tilde{v}_k}{\partial x_k} + \frac{28}{5} \frac{\partial \tilde{q}_{\langle i}}{\partial x_{j\rangle}} + \frac{28}{5} \frac{q_{\langle i}^{(H)}(t)}{n_0 \, T_H(t) \, v_H(t)} \frac{\partial \tilde{T}}{\partial x_{j\rangle}}
\right.
\nonumber\\
&\left. + 4 \frac{\sigma_{k\langle i}^{(H)}(t)}{n_0 \, T_H(t)} \left(\frac{\partial \tilde{v}_k}{\partial x_{j\rangle}} + \frac{\partial \tilde{v}_{j\rangle}}{\partial x_k}\right) - \frac{\sigma_{ij}^{(H)}(t)}{n_0 \, T_H(t)}\left(\frac{8}{3} \frac{\partial \tilde{v}_k}{\partial x_k} + \frac{14}{3} \frac{\partial \tilde{q}_k}{\partial x_k} + \frac{14}{3} \frac{\sigma_{kl}^{(H)}(t)}{n_0 \, T_H(t)}\frac{\partial \tilde{v}_k}{\partial x_l}\right) 
\right.
\nonumber\\
&\left.+ 2\frac{\partial \tilde{m}_{ijk}}{\partial x_k} + \frac{6}{7} \frac{R_{\langle ij}^{(H)}(t)}{n_0 \,  T_H(t)\,v_H^2(t)} \frac{\partial \tilde{v}_{k\rangle}}{\partial x_k} + \frac{R_{k\langle i}^{(H)}(t)}{n_0 \,  T_H(t)\,v_H^2(t)} \left(\frac{4}{5} \frac{\partial \tilde{v}_k}{\partial x_{j\rangle}} + 2 \frac{\partial \tilde{v}_{j\rangle}}{\partial x_k}\right)
\right.
\nonumber\\
&\left.+14\,a_2 \frac{\partial \tilde{v}_{\langle i}}{\partial x_{j\rangle}} + \frac{m_{ijk}^{(H)}(t)}{n_0 \, T_H(t) \, v_H(t)} \left(7\frac{\partial \tilde{T}}{\partial x_k} - 2 \frac{\partial \tilde{n}}{\partial x_k} - 2 \frac{\partial \tilde{\sigma}_{kl}}{\partial x_l}\right)
\right.
\nonumber\\
&\left.- \frac{28}{5} \frac{q_{\langle i}^{(H)}(t)}{n_0 \, T_H(t) \, v_H(t)} \left(\frac{\partial \tilde{\sigma}_{j\rangle k}}{\partial x_k} +  \frac{\partial \tilde{n}}{\partial x_{j\rangle}}\right)\right]-2 \xi_0 \nu_H(t) \, \tilde{R}_{ij}
\nonumber\\
&= -\nu_H(t) \! \left[\nu_R^\ast \left\{\!\tilde{R}_{ij} + \frac{R_{ij}^{(H)}(t)}{n_0 \,  T_H(t)\,v_H^2(t)}  \left(\tilde{n} + \frac{1}{2} \tilde{T}\right)\!\right\}
-\nu_{R\sigma}^\ast
\left\{\!\tilde{\sigma}_{ij} + \frac{\sigma_{ij}^{(H)}(t)}{n_0 \,  T_H(t)}  \left(\tilde{n} + \frac{3}{2} \tilde{T}\right) \!\right\}\right]
&\nonumber\\
&\quad 
+\frac{(1+e)}{12} \nu_H(t) \bigg[\frac{(52-27 e+66 e^2-30 e^3)}{28 \times 480} \frac{R_{ij}^{(H)}(t)}{n_0 \,  T_H(t)\,v_H^2(t)}  \tilde{\Delta}
&\nonumber\\
&\quad+ \frac{(202 - 207 e - 66 e^2 + 30 e^3)}{480} \frac{\sigma_{ij}^{(H)}(t)}{n_0 \,  T_H(t)} \tilde{\Delta}\bigg],
\end{align}
\begin{align}
\label{DeltaBalPert}
&\frac{\partial \tilde{\Delta}}{\partial t} + v_H(t) \left[(8-20\,a_2) \left(\frac{\partial \tilde{q}_i}{\partial x_i} + \frac{\sigma_{ij}^{(H)}(t)}{n_0 \, T_H(t)} \frac{\partial \tilde{v}_i}{\partial x_j}\right) 
\right.
\nonumber\\
&\left.
-8 \frac{q_{i}^{(H)}(t)}{n_0 \, T_H(t) \, v_H(t)} \left(\frac{\partial \tilde{\sigma}_{ij}}{\partial x_j} + \frac{\partial \tilde{n}}{\partial x_i} - \frac{5}{2} \frac{\partial \tilde{T}}{\partial x_i}\right)
+4 \frac{R_{ij}^{(H)}(t)}{n_0 \, T_H(t) \, v_H^2(t)} \frac{\partial \tilde{v}_i}{\partial x_j}\right]
=-\nu_H(t) \, \nu_\Delta^\ast \, \tilde{\Delta},
\end{align}
where
\begin{align*}
\nu_H(t) = \frac{16}{5}\sqrt{\pi}n_0 d^2 \sqrt{\frac{T_H(t)}{m}}
\quad\textrm{and}\quad
\xi_0 = \zeta_0^\ast.
\end{align*}
\bibliography{refer}
\bibliographystyle{jfm}

\end{document}